\documentclass[iop]{emulateapj}

\usepackage{color}

\slugcomment{To appear in Astrophysical J.}

\shorttitle{Jet Feedback in Obscured Radio-Loud Quasars }
\shortauthors{Lonsdale et al.}

\begin{document}

\title{Radio Jet Feedback and Star Formation in Heavily Obscured Quasars at Redshifts$\sim$0.5-3, I:  ALMA Observations} 


\author{Carol J. Lonsdale\altaffilmark{1}, M. Lacy\altaffilmark{1,2}, A. E. Kimball\altaffilmark{1,2,3}, 
A. Blain\altaffilmark{4}, M. Whittle\altaffilmark{5}, B. Wilkes\altaffilmark{6}, D. Stern\altaffilmark{7}, J. Condon\altaffilmark{1}, M. Kim\altaffilmark{8,9}, R. J. Assef\altaffilmark{2,10}, C.-W. Tsai\altaffilmark{7}, A. Efstathiou\altaffilmark{11}, S. Jones\altaffilmark{4}, P. Eisenhardt\altaffilmark{7}, C. Bridge\altaffilmark{12}, J. Wu\altaffilmark{13}, Colin J. Lonsdale\altaffilmark{14}, K. Jones\altaffilmark{5}, T. Jarrett\altaffilmark{15}, R. Smith\altaffilmark{16}}

\altaffiltext{1}{National Radio Astronomy Observatory, Charlottesville, VA 22903, USA; clonsdal@nrao.edu}
\altaffiltext{2}{Visiting Astronomer, Cerro Tololo Inter-American Observatory.
CTIO is operated by AURA, Inc., under contract to the National Science
Foundation.}
\altaffiltext{3}{CSIRO Astronomy and Space Science, Australia Telescope National Facility, PO Box 76, Epping, NSW 1710, Australia} 
\altaffiltext{4}{Department of Physics \& Astronomy, University of Leicester, University Road, Leicester LE1 7RH, UK}
\altaffiltext{5}{Department of Astronomy, University of Virginia, Charlottesville, VA 22903, USA}
\altaffiltext{6}{Harvard-Smithsonian Center for Astrophysics, Cambridge, MA 02138, USA}
\altaffiltext{7}{Jet Propulsion Laboratory, California Institute of Technology, Pasadena, CA 91109, USA}
\altaffiltext{8}{The Observatories of the Carnegie Institution for Science, 813 Santa Barbara Street, Pasadena, CA 91101, USA} 
\altaffiltext{9}{Korea Astronomy and Space Science Institute, Daejeon 305-348, Republic of Korea; KASI-Carnegie Fellow, The Observatories of the Carnegie Institution for Science, Pasadena, CA 91101, USA}
\altaffiltext{10}{N\'ucleo de Astronom\'ia de la Facultad de Ingenier\'ia,
 Universidad Diego Portales, Av. Ej\'ercito Libertador 441, Santiago,  Chile; NASA Postdoctoral Program (NPP)}
\altaffiltext{11}{ School of Sciences, European University Cyprus, Diogenes Street, Engomi, 1516 Nicosia, Cyprus}
\altaffiltext{12}{California Institute of Technology, 249-17, Pasadena, CA 91125, USA}
\altaffiltext{13}{Division of Astronomy \& Astrophysics, University of California, Los Angeles, Physics and Astronomy Building, 430 Portola Plaza, Los Angeles, CA 90095-1547, USA}
\altaffiltext{14}{Massachusetts Institute of Technology, Haystack Observatory, Westford, MA 01886, USA}
\altaffiltext{15}{Astronomy Department, University of Cape Town, Cape Town, South Africa}
\altaffiltext{16}{Department of Physics, 3141 Chestnut Street, Drexel University, Philadelphia, PA 19104, USA}

\begin{abstract}
We present ALMA  870 $\mu$m (345 GHz) data for 49 high redshift (0.47$<z<$2.85), luminous ($11.7<\log (L_{\rm bol}/L_{\odot})<14.2$) radio-powerful AGN, obtained  to constrain cool dust emission from starbursts concurrent with highly obscured radiative-mode black hole (BH) accretion in massive galaxies which possess a small radio jet.   The sample was selected from WISE with extremely steep (red) mid-infrared (MIR) colors and with compact radio emission from NVSS/FIRST.  Twenty-six sources are detected at 870 $\mu$m, and we find that the sample has large mid- to far-infrared luminosity ratios consistent with a dominant and highly  obscured quasar.      The rest-frame 3 GHz radio powers are $24.7<\log (P_{\rm 3.0 GHz}/W Hz^{-1})<27.3$, and all sources are radio-intermediate or radio-loud.  BH mass estimates are $7.7 <\log(M_{\rm BH}/M_{\odot})<10.2$.       The rest frame 1-5 $\mu$m SEDs are very similar to the ``Hot DOGs'' (Hot Dust Obscured Galaxies), and steeper (redder) than almost any other known extragalactic sources.    ISM masses estimated for the  ALMA detected sources are 9.9 $< \log (M_{\rm ISM}/M_{\odot}) < $11.75 assuming a dust temperature of 30K.   The cool dust emission is consistent with star formation rates (SFRs) reaching several thousand $M_{\odot}$ yr$^{-1}$,  depending on the assumed dust temperature, however we cannot rule out the alternative that the AGN powers all the emission in some cases.    Our best constrained source has radiative transfer solutions with $\sim$ equal contributions from an obscured AGN and a young (10-15 Myr) compact starburst.     
\end{abstract}


\section{Introduction}\label{section-introduction}

Central questions concerning coeval galaxy and super-massive black hole (SMBH) evolution include the relative timescales and mechanisms for stellar mass and BH mass building including the roles and duty cycles of (a) secular vs. merger-triggered  mechanisms for driving material into the central regions; (b) ``radiative-mode''  vs. ``jet-mode" BH accretion modes and rates; and (c) ``quasar-mode'' vs. ``radio-mode'' feedback mechanisms, all as a function of epoch, galaxy mass and galaxy environment.   Jet-mode  AGN are low-excitation systems in which the AGN is powered by advection-dominated accretion flows onto the BH with a low accretion rate, as recently reviewed by  \citet{heckmanbest14}.  AGN in jet-mode are expected to have low radiative emission and to be energetically dominated by the jet kinetic outflow.   Jet-mode radio-AGN are thought to be highly effective in maintaining galaxies free of new gas and star formation via ``radio-mode'' kinetic feedback: jet inflation of bubbles in the surrounding hot intergalactic gas \citep{croton06,cattaneo07,fabian10}.   Radiative-mode, or ``quasar-mode'', AGN have higher accretion rates from a thin accretion disk whose radiation powers the narrow and broad line regions, and which is fed through a surrounding dusty torus or ``torus-like'' structure.  Quasar-mode accretion has a lower duty cycle than jet-mode, occurring when large amounts of material are available for accretion onto the supermassive BH.   Quasar-mode AGN are capable of powering efficient feedback into the host galaxy via thermal winds from the accretion disk, disrupting star formation and ejecting gas.   

Powerful jets are also found in a $\sim$10\% of radiatively efficient AGN, the Radio-Loud QSOs (RLQs) (also known as Broad Line Radio Galaxies, BLRGs, and High Excitation Radio Galaxies, HERGs), and therefore jets may also contribute to feedback activity in quasar-mode AGN \citep{holt08,nesvadba08}.      It is often assumed that the high radio power Fanaroff-Riley type II (FRII) jets are too collimated to impact the ISM within host galaxies \citep{deyoung10}.     However, high resolution hydrodynamic models by  \citet{wagner11}, \citet{wagner12} and \citet{wagner13} have demonstrated that the impact of powerful jets on the  ISM within the central regions of AGN host galaxies is highly dependent on the density, the filling factor and the average size of the cool clouds within the ISM.  High porosity leads to the inflation of a cocoon by the overpressured jet, leading to a quasi-spherical bubble being driven into the ISM.   \citet{wagner11} and \citet{wagner12} find that moderate to high power jets can accelerate dense ISM gas from a hundred to several thousand km s$^{-1}$, with wide-angle flows, within 10 to 100 Myr.  

\citet{bestheckman12} and \citet{best14} have shown that the two populations of radio-AGN, quasar-mode (or radiative-mode) and radio-mode (or jet-mode), are both found across all radio luminosities, and that the radiative-mode radio-AGN show much stronger evolution with cosmic time than jet-mode radio-AGN, with an order of magnitude increase in space density out to $z{\sim}1$.    This evolution in space density is similar to that of the star formation rate density and the quasar luminosity function, consistent with the scenario that the radiative-mode AGN are controlled by episodic cold gas supply.     

Gas flows associated with gas-rich mergers are the likely source of episodic cold gas fueling, important for building galaxies and BHs by triggering both starbursts and AGN activity.     Morphological signatures of mergers have been found in a large faction of powerful, $z<0.7$, radio galaxies (RGs) which display spectroscopic signatures of young stellar populations \citep{tadhunter11}, and those with the youngest stellar populations ($<$ 0.1 Gyr) show a correlation with mid- to far-infrared (MFIR) and [\ion{O}{3}] luminosity, indicating the presence of a radiative-mode AGN.     \citet{tadhunter11} also find significant complexity in correlations between the triggering or re-triggering of jets, recent star formation and  the merging of the dual nuclei, implying chaotic gas infall histories during merger events, while \citet{dicken11} find evidence for the strongest correlation between recent star formation and radio jets for the radio-AGN with the most compact jets.   These results support the idea that small (young) radio jets play an important role in the evolution of massive galaxies and SMBHs during merger-driven high accretion rate phases.

In this series of papers we address the impact of young, moderate to powerful, radio jets from luminous, radiatively-efficient, highly obscured, radio-AGN, on the disruption of the ISM and star formation in their hosts at redshifts near the peak of galaxy and BH building, $z{\sim}1-3$.    We also consider the possibility of ISM compression and starburst triggering by jet kinetic energy.   By selecting systems with a high mid-infrared (MIR) luminosity we aim to identify radiatively efficient AGN, and by selecting compact radio sources we favor young radio jets which are confined within the hosts.    By selecting AGN which are very red through the optical-MIR we favor highly obscured systems likely to be in a peak fueling phase.  

A nearby example of such a system is the MIR-bright QSO Mrk 231, which has  a luminous radiative-mode AGN, two small radio jet systems (2 pc and 40 pc) \citep{ulvestad99,lonsdale03}, and powerful molecular outflows \citep{fischer10,gonzalez14,aalto15,feruglio15}.  Mrk 231 also shows a bright optical core, indicating a complex nuclear geometry.

\subsection{Evidence for Radio Jet Interactions with Molecular Gas}
Most early studies of outflows from AGN targeted the ionized gas, which, because of its much lower mass, requires much less energy to disperse than the neutral and molecular gas.
More recent work shows that jet-induced feedback can indeed impact both warm and cold molecular gas in radio galaxies \citep{feruglio10,fischer10,sturm11,dasyra12,combes13,morganti13,garcia14,dasyra14,gonzalez14,tadhunter14a,tadhunter14b,feruglio15}, including turbulence and shock-excited H$_2$ emission in radio galaxies (eg. Morganti et al. 2003; Nesvadba et al. 2008,2011a,2011b; Guillard et al. 2012, 2015).    Molecular hydrogen emission galaxies (MOHEGs) have large mid-IR H$_2$ luminosities which are too high for photo-dissociation regions, and which are most likely generated by jet-generated shocks in the ISM \citep{appleton06,ogle10,guillard12,lanz15}.      Most of these radio galaxies are in a phase of radio-mode accretion without strong evidence for a concurrent radiative mode AGN core.   Quasar winds may also contribute to molecular outflows in radiative mode radio AGN (eg. Veilleux et al. 2013), such as those we study here, therefore our sample is ideal for studying the relative importance of these two feedback modes in early feedback phases of heavily obscured objects.

\subsection{MIR-Selected Highly Obscured Quasars}
A rare class of highly obscured and luminous quasars at redshifts above 1 was revealed in follow-up studies of extremely red (from the optical to the MIR) sources found in $Spitzer$ surveys \citep{lutz05,yan05,polletta06,dey08,lacy11}. These systems can have luminosities over $10^{13} L_{\odot}$, and deep X-ray observations have revealed Compton thick AGN in some $(N_{\rm H} > 1.5 {\times} 10^{24}$  cm$^{-2}$; Polletta et al. 2008).  The reddest of these $Spitzer$-selected systems are sometimes referred to as Dust Obscured Galaxies, ``DOGs" \citep{dey08}.  

The Wide-field Infrared Survey Explorer (WISE), which covers the entire sky at  3.4, 4.6, 12, and 22 $\mu$m (henceforth refereed to as the W1, W2, W3 and W4 bands) \citep{wright10,jarrett11,cutri12}, has opened up the entire MIR sky to obscured QSO searches by their MIR signatures \citep{eisenhardt12,stern12,wu12,assef13,yan13,bridge13,jones14,wu14,tsai14,stern14}.   Although not as deep as the largest $Spitzer$ surveys (e.g. Lonsdale et al. 2004; Ashby 2009; Ashby 2013), WISE is sensitive enough to see the dust thermal emission from the most powerful quasars to redshifts $>$4.   The first results from WISE follow-up of the reddest sources (selected without regard to radio brightness) have indeed revealed an extremely IR-luminous population of high-redshift quasars, exceeding $10^{14} L_{\odot}$ in bolometric luminosity in some cases \citep{eisenhardt12,wu12,bridge13,jones14,wu14,assef14,tsai14}.  Their  bolometric luminosities are MIR-dominated, which led  \citet{wu12} to dub them ``Hot DOGs" (Hot Dust Obscured Galaxies), while \citet{bridge13} investigate Ly$\alpha$ Blobs discovered around a high percentage of their red WISE sample, the WISE Lyman Alpha Blobs (WLABs).   X-ray observations show only faint fluxes,  consistent with highly obscured X-ray AGN \citep{stern14,pinconcelli15}. In this paper we will henceforth refer to the main discovery papers for these WISE Hot DOGs and WLABs, \citep{eisenhardt12,wu12,bridge13}, as EWB12.   

\subsection{This Work}
We present a snapshot survey of 49 sources with the Atacama Large Millimeter/submillimeter Array, ALMA, at 870 $\mu$m, which represents a southern sky subset of our sample of 156 radio-powerful (RP) obscured quasar candidates.  The sample has been selected to be unresolved in the NVSS and FIRST radio surveys \citep{condon98,becker95} and ultra-red in the WISE MIR survey, with similar selection criteria as the Hot DOGs (EWB12), to search for radio-jet dominated feedback from massive, obscured, quasars.     We also present redshifts obtained from optical and/or near-IR (NIR) spectroscopy for 45 of these ALMA-observed quasars, and additional FIR-submillimeter photometry from other facilities.

The sample is described in Section \ref{section-sample} and the observations in Section \ref{section-observations}.    The results are presented in Section \ref{section-results}, followed by SED model fitting and derivation of luminosities and masses in Section \ref{section-fits}.  The discussion is in Section \ref{section-discussion} and conclusions in Section \ref{section-conclusions}.    Seven sources with near-IR spectroscopy from FIRE on Magellan have been discussed by \cite{kim13}.    \citet{jones15} have published deep SCUBA 850 $\mu$m imaging the the James Clerk Maxwell Telescope (JCMT) for 30 northern sources from our overall sample of 156 RP quasars, detecting four, and finding an excess of serendipitously-detected 850$\mu$m sources in the fields on $\sim$1 Mpc scales.  \citet{silva15} have found an excess of serendipitously-detected 870$\mu$m sources in the 49 ALMA images discussed here, in agreement with \citet{jones15}, although on smaller physical size scales of $\sim$150 kpc, perhaps indicating an excess of starbursting SMGs in the fields of these QSOs.  Subsequent papers will address Jansky VLA  8-12 GHz imaging of the full sample (Carol Lonsdale et al., in preparation), and VLBA C-band imaging of 90 sources, including 33 from the ALMA sample (Colin Lonsdale et al., in preparation).    NIR $J$ and $K_s$ imaging for a subset of the ALMA sample from VLT/ISAAC, and VLT/XShooter spectroscopy, will be presented in A. Blain et al. (in preparation).  We adopt a cosmology with $H_0 = 71~{\rm km}~ {\rm s}^{-1}~\rm Mpc^{-1}$, ${\Omega}_{\Lambda} = 0.73$ and ${\Omega}_M = 0.27$.   

\section{Sample Selection}\label{section-sample}
\begin{figure}[!ht]
	\centering
	\includegraphics[trim=100 80 12 370,clip,width=0.6\textwidth]{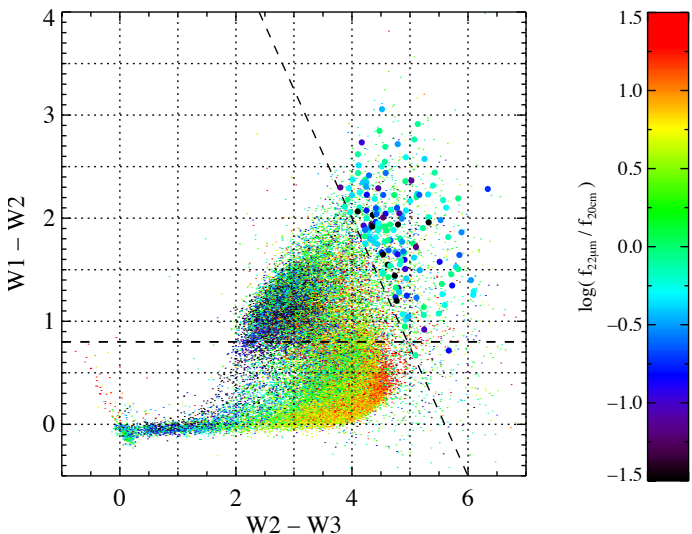}
\caption{WISE $3.4-4.6-12$ $\mu$m ($W1-W2-W3$) color-color plot (using Vega magnitudes) showing the full WISE-NVSS sample of 54,457 sources, color-coded by radio-loudness (see  Figure \ref{fig-radioIR}) as shown by the color bar on the right, radio loudness increasing red to black.   The sequence at the bottom is the sequence of normal spirals and starbursts.  The flux-limited ultra-red sample of 156 sources is highlighted  above the dashed line with larger symbols.   The horizontal dashed line shows  the AGN color-selection criterion used by \citet{stern12}: $(W1-W2) > 0.8$, for comparison to our ultra-red selection criterion.   \label{fig-bubble}}
\end{figure}

For our overall sample of ultra red, radio powerful, sources, we cross-matched the NVSS catalog, which covers the sky north of ${\delta}=-40^{\circ}$, with point sources from the WISE Allsky Catalog.  The ALMA-observed subsample described here was selected at an earlier time when only the WISE Preliminary Catalog was available, as explained further below.   Extended WISE sources were rejected based on the cataloged {\tt ext} flag.  We used positional information from the higher resolution but smaller area FIRST catalog where available, and excluded $\pm 10^{\circ}$ from the Galactic plane. Within this area of 28,443 square degrees of mutual NVSS-WISE coverage are 54,457 WISE sources which have a robust point source detection (SNR $>$ 7) in the WISE 12 $\mu$m and/or 22 $\mu$m bands and an NVSS/FIRST 1.4 GHz match within a separation of $<7\arcsec$, the best compromise between completeness and reliability based on a randomized match analysis.   The entire sample is illustrated in the WISE $3.4-4.6-12$ ${\mu}$m color-color diagram in Figure \ref{fig-bubble}, where we plot the WISE-NVSS sample color-coded by $q_{\rm 22} = \log (f_{\rm 22{\mu}m}/f_{\rm 20 cm})$, the 22 $\mu$m $q$ parameter, which is a measure of radio loudness \citep{appleton04,ibar08,ivison07}.     The sequence of low-redshift normal spirals and starbursts has blue $(W1-W2)$ colors (where $W$X is the Vega magnitude in WISE band number X) with a wide range of $(W2-W3)$ colors, and a cloud of AGN is seen with redder $(W1-W2)$ colors \citep{wright10,jarrett11,yan13}.     The most radio loud systems are seen towards the left of the each of the normal galaxy sequence and the AGN cloud, consisting of the radio mode and the radiative mode AGN, respectively.    For low redshift galaxies the point source fluxes used for this figure may underestimate the total fluxes.

\begin{figure}[!ht]
	\centering
	\includegraphics[trim=80 80 12 370,clip,width=0.65\textwidth]{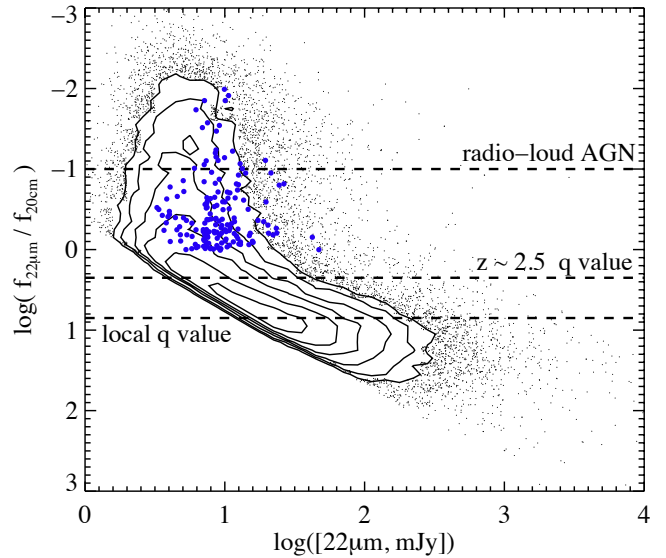}
\caption{The 22$\mu$m, $q$ value, $q_{\rm 22} = \log (f_{\rm 22{\mu}m}/f_{\rm 20cm})$ {\it vs.} WISE 22 $\mu$m flux density, for the 156 ultra-red, radio-powerful sources (blue points), compared to the full WISE/NVSS cross-matched sample of 54,457 sources (dots and contours).    The completeness limit of the NVSS is responsible for the diagonal cutoff at the lower left. Also shown are the mean expected values for star forming galaxies of the infrared-to-radio parameter, $q_{\rm 24} = \log (f_{\rm 24{\mu}m}/f_{\rm 20cm})$ for the $Spitzer$ 24 $\mu$m band by \citet{ibar08} in the local universe and the k-corrected $z{\sim}2.5$ value.  The representative radio-loud SED from \citet{elvis94}, is also shown.   The subset of ultra-red sources observed with ALMA was selected to have $0<\log (f_{\rm 22 {\mu}\rm m}/f_{\rm 22cm})<-1$. \label{fig-radioIR}}	
\end{figure}

Highly obscured luminous AGN are expected to be among the reddest sources in this figure.  Previous authors have used color cuts in MIR color space and/or MIR-optical space to select the reddest extragalactic $\it Spitzer$ and WISE sources (eg. Dey et al. 2006; EWB12).   We have instead chosen to reduce potential bias caused by color cuts, due to the complexity of the observed MIR spectral shape of these sources, which depends strongly on redshift due to the PAH emission features and the 9.7 and 18 $\mu$m silicate features.  Therefore we have included all sources lying significantly redward of the main WISE populations.  This is illustrated in Figure \ref{fig-bubble} by the dashed line, defined empirically as:   $(W2-W3) + 1.25 (W1-W2) > 7$.    The resulting number of sources in the sample redward of our MIR selection threshold in Figure \ref{fig-bubble} is 1858.  To reject radio-quiet systems we added the requirement that the  $\log (f_{\rm 22{\mu}m}/f_{\rm 20cm})<0$ as illustrated in Figure \ref{fig-radioIR}.     For comparison, the color selection criteria for the Hot DOGs of Eisenhardt et al. (2012) and Wu et al. (2012) are either ($W2-W4$)$>$8.2 or ($W2-W3$)$>$5.3, termed by these authors the ``W1W2drop'' criteria.   \citet{bridge13} have used a slightly different color selection method:  ($W2-W3$)$\ge$4.8.  Both studies also use brightness or SNR threshold similar to ours, and neither has a radio flux density criterion.

We inspected all candidates in WISE and DSS images, and SDSS images where available, rejecting low redshift galaxies, artifacts and confused sources.    To avoid rejecting galaxies or quasars that could plausibly be within the redshift range of interest, $1<z<3$, no specific optical magnitude or MIR/optical flux ratio cut was imposed.  Galaxies with a size or brightness large enough to place them at $z < 0.5$ were rejected, while high surface brightness compact sources were retained to brighter limits as potential high redshift quasars.   We are obtaining spectra to reject low redshift sources from our final sample.      Thus, our selection can include objects that are less red (in R-[22]) than the criterion $(R-[24])>14$ used for $Spitzer$ DOGs \citep{dey08} and similar samples, and, for example, may include systems in which a massive galaxy dominates the optical light.  A total of 708 of the NVSS-WISE sample satisfy all our criteria, or $\sim$1.3\% of the entire NVSS-WISE matched sample.    Of the 708 sources, 703 (269) are detected at 12 (22) $\mu$m with SNR$>$7.     

Due to the strongly varying WISE sensitivity across the sky, caused by the varying coverage level, a flux density threshold at 22$\mu$m was applied.  For the subsample observed with ALMA the flux density threshold chosen was 4 mJy.  This was later revised upward to 7mJy for the remainder of the sample when the Allsky Catalog became available.   The final sample is 156 sources, 49 of which belong to the ALMA-observed subsample.

The ALMA subsample fluxes have been revised in the updated WISE Allsky Catalog, some now falling below the original 4 mJy threshold.  These have been retained in the sample.   The declination range for the ALMA subsample was chosen to allow access to many northern facilities as well as ALMA:  $-40^{\circ}<{\delta}<$+1$^{\circ}$.   The ALMA sample is restricted in RA to two regions: $3^{\rm h}<RA<8^{\rm h}30^{\rm m}$ and $13^{\rm h}<RA<21^{\rm h}$, due to the incompleteness of the WISE Preliminary Catalog at that time.   Fifty-five sources met our original selection criteria within these areas.  Six of these have $\log (f_{\rm 22{\mu}m}/f_{\rm 20cm})<$-1 and were excluded to disfavor classical double lobed sources, a criterion which was later dropped for the full sample of 156 sources.  Of the final ALMA subsample of 49 sources, 48 (23) have SNR$>$7 at 12 (22) $\mu$m in the WISE Allsky Catalog, and the minimum 12/22$\mu$m SNR is 6.7/2.6.   Only one source fails the original SNR criterion of SNR$>$7 at 12 or 22 $\mu$m after the Allsky flux revision:  WISE J204049.51-390400.5 which has an SNR at 12/22$\mu$m of 6.7/6.2.   Due to the evolving selection criteria the sample is not complete to a fixed SNR or flux density limit.

\section{Observations}\label{section-observations}

\subsection{FIR and Submillimeter Observations}\label{section-submm}
Twenty-three sources were observed with ALMA in two Band 7 scheduling blocks
on 2011 November 16.  An additional 14 were observed on 2012 May 25 and the final 12 on 2012 August 29.   The central frequency
was set to 345 GHz (870 $\mu$m), with an 8 GHz bandwidth split into two sidebands. Fifteen antennas were used in the ``compact'' Early Science configuration 
for the November 2011 observations, 19 for the May 2012 observations and 23 for the August observations.  The resulting beamsizes are 1$\farcs$2, 0$\farcs$5 and 1$\farcs$2, respectively.

For the November observations, Callisto was used for flux calibration.    Titan was used for 7 sources in May and Neptune for the other 7 sources in May.  Titan was again used for all sources in August.  The objects were observed with the correlator in Time Division Mode, 
which results in 256 channels per sideband with a spectral resolution 
of 14 km s$^{-1}$.  Time on source was 1.5  min per object, resulting in an $rms$ noise of
0.3--0.6 mJy, the lower noise levels generally correspond to the observations with larger antenna numbers.    Flux densities were measured using imfit in CASA, and are reported in Table \ref{tbl-flux}.      

Four sources from our overall sample of 156 were observed at 350 $\mu$m at the Submillimeter High Angular Resolution Camera II (SHARC-II) installed at the 10.4 m Caltech Submillimeter Observatory (CSO) telescope \citep{dowell03}  on UT 2012 March 22.   Two of these, W1343-1136 and W1400-2919, belong to the ALMA subsample.   Our observing and data reduction process follows that described by \citet{wu12}.   

Seventy-seven of our full sample of 156 sources, including all the ALMA targets, were queued for $Herschel${\footnote{\textit{Herschel} is an ESA space observatory with science instruments provided by European-led
Principal Investigator consortia and with important participation from
NASA.} PACS (Poglitsch et al. 2010) and SPIRE (Griffin et al. 2010) photometric observations (PID: OT2\_clonsdal\_1).   In total 15 objects were observed by PACS, and 5 with SPIRE, before the cryogen depletion of the \textit{Herschel} Telescope in 2013 April.  For the PACS observations, two concatenated mini-scans orientated at 70$^\circ$ and 110$^\circ$ were acquired. Each mini-scan has 8 legs, a scan length of 3$\arcmin$ and and a leg separation of 5$\arcsec$, with a time on source of 220 seconds. The SPIRE observations were conducted using the small jiggle map mode with 2 repeats, with 74 seconds on source time.  The data were processed with HIPE 11.1.0 (Ott 2010).    The aperture-corrected flux densities for all 15 {\it Herschel}-observed sources are presented in Table \ref{tbl-herschel}.    

Summarizing, the FIR/submillimeter data available for the 49 ALMA-observed sources discussed in this paper, PACS observations are available for two sources, W1500-0649 and W2059-3541, and SPIRE observations for one, W2059-3541.      Two additional sources have upper limits from the CSO at 350$\mu$m, W1343-1136 and W1400-2919.

\subsection{Optical and Near-IR Photometry and Spectroscopy}\label{section-optical}

Redshifts for 45 of the 49 ALMA sources and $R$-band photometry for 26 were obtained and are presented in Tables \ref{tbl-flux} and \ref{tbl-lum}.   In total 48 of the 49 sources were observed spectroscopically, and three of these yielded no line detections.  Full details of the spectroscopy will be presented in a later publication; here we make use only of the redshifts.
Optical spectra of 23 objects were obtained using the Goodman spectrograph on the SOAR 4.2 m telescope on UT 2012 January 21-24 and UT 2012 December 9-12.   The R-band photometry obtained with SOAR was taken with the spectrograph acquisition camera.
We observed five sources from the WISE-NVSS-ALMA sample, and two from the northern JCMT sample, using the Double Spectrograph on the 5-m Hale telescope at Palomar Observatory between 2012 November and 2013 March.  
As described in more detail by \citet{kim13},  twenty-four sources were observed in the NIR with Magellan on UT 2012 July 27-29.  
Finally, twenty-eight sources were observed with VLT / XShooter on UT 2013 June 04-06 and thirty-one
sources were observed in $K_s$ with VLT/ISAAC over 3 nights from UT 2013 June 01-04, of which 14 were 
also observed in $J$.   

\begin{centering}
\begin{deluxetable*}{llrrrrcrrr}[!ht]
\tabletypesize{\tiny}
\tablecaption{Optical, WISE, NVSS and ALMA Photometry\label{tbl-flux}}
\tablewidth{0pt}
\tablehead{
\colhead{WISE Name}  & \colhead{$R$-band} & \colhead{$f_{\rm 3.4{\mu}m}$}  & \colhead{$f_{\rm 4.6{\mu}m}$} & \colhead{$f_{\rm 12{\mu}m}$}& \colhead{$f_{\rm 22{\mu}m}$} & \colhead{$q_{22}$} & \colhead{$f_{\rm 1.4GHz}$}  &\colhead{$f_{\rm 870{\mu}m}$} & \colhead{log} \\
\colhead{}       & \colhead{app. mag.}&  \colhead{(mJy)}             &   \colhead{(mJy)}      &   \colhead{(mJy)}  &    \colhead{(mJy)}    & \colhead{observed}   & \colhead{(mJy)}  & \colhead{(mJy)} & \colhead{($f_{\rm 870{\mu}m}$}/ \\
\colhead{}       &  \colhead{(Vega)} & \colhead{WISE}             &   \colhead{WISE}      &   \colhead{WISE}  &    \colhead{WISE}    & \colhead{frame}   & \colhead{NVSS}  & \colhead{ALMA} & \colhead{$f_{\rm 22{\mu}m}$)}
}
\startdata
J030427.53-310838 &  22.0$\pm$0.3 & 0.179$\pm$0.0066&    0.693$\pm$0.018&     4.55$\pm$0.12&    10.87$\pm$0.65& -0.70&54.99$\pm$1.72&     2.8$\pm$0.3  & -0.52   \\
J030629.21-335332 &  20.7$\pm$0.2 & 0.194$\pm$0.0070&    0.749$\pm$0.020&     6.19$\pm$0.14&     4.83$\pm$0.65& -0.11& 6.17$\pm$0.52&     1.7$\pm$0.4   & -0.45  \\
J035448.24-330827 &  22.5$\pm$0.2 & 0.086$\pm$0.0063&    0.314$\pm$0.015&     3.34$\pm$0.12&     7.05$\pm$0.78& -0.04& 7.73$\pm$0.54&    $<$1.2 & $<$-1.12 \\
J040403.61-243600 &  20.7$\pm$0.2 & 0.104$\pm$0.0072&    0.326$\pm$0.017&     4.63$\pm$0.15&     4.43$\pm$0.92& -0.40&11.04$\pm$0.58&     3.1$\pm$0.4  & -0.16   \\
J040937.67-183757 &  21.3$\pm$0.2 & 0.517$\pm$0.0147&    2.492$\pm$0.054&    22.97$\pm$0.36&    47.07$\pm$1.68 & 0.00&46.91$\pm$1.48&    $<$1.5 & $<$-1.47 \\
J041754.10-281654 &  23.9$\pm$0.3 & 0.043$\pm$0.0057&    0.143$\pm$0.013&     1.43$\pm$0.10&     3.84$\pm$0.85& -0.63&16.32$\pm$0.68&     4.3$\pm$0.6  & 0.05   \\
J043921.92-315908 &  24.2$\pm$0.3 & 0.034$\pm$0.0051&    0.090$\pm$0.011&     1.19$\pm$0.09&     4.57$\pm$0.77& -0.66&20.88$\pm$0.78&      6 $\pm$0.4   & 0.12  \\
J051905.84-081320 &  $>$24  & 0.059$\pm$0.0071&    0.136$\pm$0.015&     1.76$\pm$0.13&     5.09$\pm$0.85& -0.72&26.79$\pm$0.90&    $<$1.5 & $<$-0.56 \\
J052533.47-361440 &  22.9$\pm$0.3 &  0.013$\pm$0.0053&    0.046$\pm$0.010&     1.38$\pm$0.10&     4.07$\pm$0.71& -0.10& 5.12$\pm$0.56&    $<$1.5 & $<$-0.41 \\
J052624.72-322500 &  22.6$\pm$0.2 & 0.023$\pm$0.0056&    0.106$\pm$0.011&     6.73$\pm$0.17&    26.47$\pm$1.17 & -0.82&173.5$\pm$5.23&     18 $\pm$0.5 & -0.17    \\
J053622.59-270300 &  23.2$\pm$0.3 & 0.065$\pm$0.0060&    0.404$\pm$0.016&     3.26$\pm$0.12&     5.35$\pm$0.80& -0.18& 8.16$\pm$0.56&     2.7$\pm$0.4   & -0.30  \\
J054930.07-373939 &  $>$24 & 0.021$\pm$0.0049&    0.091$\pm$0.010&     1.09$\pm$0.09&     3.37$\pm$0.73& -0.50&10.60$\pm$0.57&      2 $\pm$0.4    & -0.23 \\
J061200.23-062209 &  20.8$\pm$0.2 & 0.330$\pm$0.0118&    0.621$\pm$0.022&     9.52$\pm$0.20&    20.55$\pm$1.01& -0.20&32.79$\pm$1.08&     2.7$\pm$0.6   & -0.88  \\
J061348.08-340728 &  $>$24.6  & 0.065$\pm$0.0065&    0.211$\pm$0.012&     2.62$\pm$0.11&     7.37$\pm$0.71& -0.52&24.15$\pm$1.16&    $<$1.8 & $<$-0.62 \\
J061405.57-093658 &  24.1$\pm$0.3 & 0.017$\pm$0.0073&    0.069$\pm$0.015&     2.16$\pm$0.14&     5.80$\pm$0.84& -0.01& 5.94$\pm$0.51&    $<$1.8 & $<$-0.51\\
J063027.81-212058 &  22.2$\pm$0.2 & 0.020$\pm$0.0070&    0.063$\pm$0.014&     2.62$\pm$0.13&     5.06$\pm$0.90& -0.39&12.55$\pm$0.60&      5 $\pm$1.3 & -0.01    \\
J064228.92-272801 &  21.4$\pm$0.2 &  0.038$\pm$0.0070&    0.108$\pm$0.014&     1.33$\pm$0.16&     3.67$\pm$0.87& -0.24& 6.36$\pm$0.52&     2.2$\pm$0.6    & -0.22 \\
J065215.85-200612 &  $>$23.7 & $<$0.013&    0.045$\pm$0.013&     1.92$\pm$0.13&     4.81$\pm$0.83& -0.18& 7.35$\pm$0.52&     3.2$\pm$0.5   & 0.18  \\
J070257.20-280842 &  21.8$\pm$0.2 & 0.027$\pm$0.0061&    0.140$\pm$0.016&     1.76$\pm$0.12&     4.54$\pm$0.86& -0.25& 8.00$\pm$0.58&    $<$1.8 & $<$-0.40\\
J071433.54-363552 &  22.8$\pm$0.2 & $<$0.012&    0.039$\pm$0.011&     0.99$\pm$0.12&     4.01$\pm$0.84& -0.47&11.95$\pm$0.60&     2.4$\pm$0.3   & -0.22  \\
J071912.58-334944 &  24.1$\pm$0.3 & $<$0.011&    0.081$\pm$0.012&     1.93$\pm$0.12&     4.06$\pm$0.88& -0.78&24.30$\pm$0.87&     5.2$\pm$0.6   & 0.11  \\
J081131.61-222522 &  21.5$\pm$0.2 & 0.132$\pm$0.0086&    0.611$\pm$0.023&     5.62$\pm$0.17&     7.61$\pm$1.17& -0.37&17.84$\pm$0.71&    $<$1.8   & $<$-0.63   \\
J082311.24-062408 &  22.32$\pm$0.16\tablenotemark{a} &  0.118$\pm$0.0078&    0.441$\pm$0.019&     4.08$\pm$0.15&    10.42$\pm$0.97& -0.71&53.99$\pm$1.67&    $<$1.8     & $<$-0.76 \\
J130817.00-344754 &  22.3$\pm$0.2 & 0.086$\pm$0.0056&    0.248$\pm$0.013&     3.36$\pm$0.12&     9.12$\pm$0.73& -0.87&68.10$\pm$2.10&      1.38$\pm$0.34 & -0.81     \\
J134331.37-113609 &  21.7$\pm$0.2 & 0.024$\pm$0.0057&    0.136$\pm$0.013&     1.61$\pm$0.12&     3.81$\pm$0.79& -0.33& 8.18$\pm$0.54&      2.34$\pm$0.31  & -0.22      \\
J140050.13-291924 &  21.7$\pm$0.2 & 0.110$\pm$0.0063&    0.501$\pm$0.018&     5.58$\pm$0.14&    11.85$\pm$0.77& -0.64&51.92$\pm$1.63&      $<$0.90   & $<$-1.37     \\
J141243.15-202011 &  \nodata &  0.092$\pm$0.0063&    0.333$\pm$0.015&     3.39$\pm$0.13&     7.41$\pm$0.78& -0.09& 9.01$\pm$0.55&      2.55$\pm$0.63   & -0.45     \\
J143419.59-023543 &  22.02$\pm$0.18\tablenotemark{a} & 0.058$\pm$0.0056&    0.257$\pm$0.014&     2.13$\pm$0.11&     5.04$\pm$0.71& -0.86&36.15$\pm$1.16&      $<$0.9    & $<$-0.75    \\
J143931.76-372523 & \nodata &  0.027$\pm$0.0071&    0.115$\pm$0.013&     2.34$\pm$0.12&     3.92$\pm$0.83& -0.41&10.02$\pm$0.57&      $<$0.6    & $<$-0.75   \\
J150048.73-064939 & \nodata &   0.068$\pm$0.0065&    0.293$\pm$0.016&     6.26$\pm$0.17&    15.77$\pm$0.94& -0.10&20.01$\pm$0.73&     6.11$\pm$0.28  &   -0.41    \\
J151003.71-220311 & \nodata &  0.143$\pm$0.0095&    0.411$\pm$0.020&     5.34$\pm$0.18&    14.87$\pm$1.09& -0.06&17.14$\pm$0.70&      $<$0.9  &   $<$-1.27   \\
J151310.42-221004 & \nodata &  0.037$\pm$0.0082&    0.214$\pm$0.018&     2.64$\pm$0.16&     9.71$\pm$1.10& -0.50&30.40$\pm$1.03&      4.86$\pm$0.27   & -0.30     \\
J151424.12-341100 & \nodata &  0.076$\pm$0.0091&    0.184$\pm$0.019&     3.12$\pm$0.16&     7.01$\pm$1.03& -0.25&12.39$\pm$0.60&      $<$0.9   & $<$-0.94    \\
J152116.59+001755 & \nodata &   0.039$\pm$0.0046&    0.274$\pm$0.014&     5.41$\pm$0.15&     9.51$\pm$0.70& -0.60&37.89$\pm$1.20&      1.19$\pm$0.28   & -0.90     \\
J154141.64-114409 & \nodata &   0.032$\pm$0.0077&    0.155$\pm$0.017&     2.91$\pm$0.16&    10.74$\pm$1.14& -0.51&34.52$\pm$1.13&      1.2$\pm$0.3  & -1.11   \\
J163426.87-172139 & \nodata &  0.039$\pm$0.0094&    0.101$\pm$0.018&     1.70$\pm$0.17&     3.57$\pm$1.15& -0.42& 9.49$\pm$0.55&   $<$0.84  &  $<$-0.63       \\
J164107.22-054827 & \nodata &   0.086$\pm$0.0083&    0.423$\pm$0.020&     3.14$\pm$0.15&     6.26$\pm$0.89& -0.02& 6.62$\pm$0.48&     2.3$\pm$0.29 & -0.43     \\
J165305.40-010230 & \nodata &   0.083$\pm$0.0074&    0.191$\pm$0.015&     2.56$\pm$0.14&     5.31$\pm$0.93& -0.36&12.21$\pm$0.56&   $<$0.78   & $<$-0.83  \\
J165742.88-174049 & \nodata &   0.073$\pm$0.0102&    0.186$\pm$0.026&     2.82$\pm$0.24&     8.60$\pm$1.01& -0.31&17.48$\pm$0.71&   $<$0.78  & $<$-1.04 \\
J170204.65-081108 & \nodata &  0.021$\pm$0.0690&    0.074$\pm$0.053&     3.05$\pm$0.26&    12.32$\pm$1.40& -0.74&67.59$\pm$2.07&   $<$1.02 & $<$-1.08 \\
J170325.05-051742 & \nodata &   0.021$\pm$0.0082&    0.199$\pm$0.018&     2.35$\pm$0.24&    11.66$\pm$1.42& -0.39&28.77$\pm$0.96&   1.02$\pm$0.27   &-1.05  \\
J170746.08-093916 & \nodata &   0.119$\pm$0.0073&    0.342$\pm$0.020&     3.46$\pm$0.28&     3.27$\pm$1.26& -0.52&10.86$\pm$0.52&   $<$1.02 & $<$-0.51 \\
J193622.58-335420 & \nodata &   0.031$\pm$0.0069&    0.127$\pm$0.016&     2.34$\pm$0.14&     5.27$\pm$0.96& 0.00& 5.27$\pm$0.51&    1.86$\pm$0.36   &-0.45  \\
J195141.22-042024 & \nodata &   0.030$\pm$0.0178&    0.065$\pm$0.036&     2.55$\pm$0.15&     8.56$\pm$1.02& -0.38&20.52$\pm$1.09&   $<$1.05  & $<$-0.91\\
J195801.72-074609 & \nodata &   0.056$\pm$0.0086&    0.203$\pm$0.018&     3.29$\pm$0.16&     7.44$\pm$1.06& -0.64&32.79$\pm$1.06&   $<$0.93  & $<$-0.90\\
J200048.58-280251 & \nodata &   0.027$\pm$0.0169&    0.113$\pm$0.017&     3.21$\pm$0.17&     7.19$\pm$1.20& -0.33&15.33$\pm$0.66&   $<$0.96  & $<$-0.87\\
J202148.06-261159 & \nodata &             $<$0.015&    $<$0.065&     1.03$\pm$0.15&     6.27$\pm$1.01& -0.04& 6.82$\pm$0.55&     4.4$\pm$0.38   & -0.15  \\
J204049.51-390400 & \nodata &   0.070$\pm$0.0077&    0.254$\pm$0.017&     2.75$\pm$0.15&     4.02$\pm$0.91& -0.44&10.95$\pm$0.57&     5.1$\pm$0.43   & 0.10  \\
J205946.93-354134 & \nodata &   0.052$\pm$0.0069&    0.182$\pm$0.015&     2.94$\pm$0.14&     4.75$\pm$0.99& -0.28 & 9.13$\pm$1.07&  $<$0.99  & $<$-0.40   \\
\enddata
\tablenotetext{a}{SDSS $r$-band AB magnitude}
\end{deluxetable*}
\end{centering}

\begin{centering}
\begin{deluxetable*}{lcccccccc}[!ht]
\tabletypesize{\tiny}
\tablecaption{$Herschel$, JCMT and CSO Photometry\label{tbl-herschel}}
\tablewidth{0pt}
\tablehead{
\colhead{WISE Name} & \colhead{ALMA}  & \colhead{Redshift} & \colhead{$f_{\rm 70}$}  & \colhead{$f_{\rm 170}$}  & \colhead{$f_{\rm 250}$} & \colhead{$f_{\rm 350}$} & \colhead{$f_{\rm 500}$} & \colhead{$f_{\rm 850}$}  \\
\colhead{} & \colhead{subsample} & \colhead{} & \multicolumn{6}{c}{(mJy)}   \\
\colhead{} & \colhead{} & \colhead{} & \colhead{PACS}  & \colhead{PACS}  & \colhead{SPIRE} & \colhead{SPIRE/CSO} & \colhead{SPIRE} & \colhead{JCMT/ALMA}  
}
\startdata
W0524+3005  & \nodata &  \nodata &   92.3$\pm$2.1      &  119.8$\pm$7.3     &   38.9$\pm$6.2     &   $<$15.9    &   $<$19.8 & \nodata \\
W0526$+$1259 & \nodata & \nodata & 31.3$\pm$2.0 & 54.5$\pm$7.3 & 34.3$\pm$6.3 & 18.9$\pm$5.2	& $<$19.5	& \nodata			\\
W0537+3947   & \nodata & \nodata &   33.7$\pm$1.7    &  $<$26.1& \nodata     & \nodata     & \nodata & \nodata \\
W0541+1130   & \nodata & \nodata  &   23.6$\pm$1.8    &   $<$30.0    & \nodata    & \nodata     & \nodata  & \nodata  \\
W0844$+$7420	& \nodata	& \nodata	& 106.1$\pm$2.0	& 130.8$\pm$6.9	& 62.0$\pm$6.5	& 44.1$\pm$5.4	& 24.6$\pm$6.2	& \nodata			\\
W1001$-$2141    & \nodata & \nodata &   \nodata    &   \nodata    & \nodata     & $<$45\tablenotemark{c}     & \nodata  & \nodata  \\
W1025+6128    & \nodata & \nodata &   \nodata    &   \nodata    & \nodata     & 30$\pm$13\tablenotemark{c}     & \nodata  & \nodata  \\
W1331$-$3913	& \nodata	& \nodata	& $<$5.7	& $<$24.0	& \nodata	& \nodata	& \nodata	& \nodata			\\
W1332+7907    & \nodata & \nodata &   56.9$\pm$1.5    &   59.6$\pm$7.8    & \nodata   & \nodata  & \nodata  & \nodata  \\
W1343$-$1136 & ALMA & 2.49\tablenotemark{a} & \nodata & \nodata & \nodata  & $<$45\tablenotemark{c} & \nodata & 2.34$\pm$0.31\tablenotemark{d} \\
W1400$-$2919 & ALMA & 1.67\tablenotemark{a} & \nodata & \nodata & \nodata  & $<$45\tablenotemark{c} & \nodata &$<$0.9\tablenotemark{d} \\
W1500$-$0649    & ALMA & 1.500\tablenotemark{a} &  91.0 $\pm$3.3    &  171.6$\pm$7.4    & \nodata   & \nodata  & \nodata  & 6.11$\pm$0.28\tablenotemark{d} \\
W1501+1324   & \nodata & 0.505\tablenotemark{b} &  199.9$\pm$1.7    &  200.2$\pm$6.9    & \nodata   & \nodata  & \nodata  & $<$6.6\tablenotemark{e} \\
W1501+3341   & \nodata & \nodata &    8.1$\pm$1.9    &   49.4$\pm$7.0    & \nodata   & \nodata  & \nodata   & \nodata \\
W1505+0219   & \nodata &  \nodata & 12.2$\pm$ 1.6    &   50.0 $\pm$7.9    & \nodata   & \nodata  & \nodata  & \nodata  \\
W1517+3523   & \nodata & 1.515\tablenotemark{b} &   53.5$\pm$1.6    &   69.8$\pm$8.5    & \nodata   & \nodata  & \nodata  & $<$5.7\tablenotemark{e} \\
W1921+7349   & \nodata & \nodata &   19.4$\pm$2.0    &   49.5$\pm$7.8    & \nodata   & \nodata  & \nodata   & \nodata \\
W2005+0215   & \nodata & \nodata &   14.2 $\pm$1.9    &   $<21.6$     &   $<$18.6    &   $<$16.2         &  $<$19.8 & \nodata   \\
W2059$-$3541    & ALMA & 2.380\tablenotemark{a} &    11.7$\pm$2.7    & $<$19.2       &    $<$19.2       &   $<$16.2         &   $<$19.5     & $<$0.99\tablenotemark{d}  \\
\enddata
\tablenotetext{a}{see Table \ref{tbl-lum}}
\tablenotetext{b}{Palomar 200 inch}
\tablenotetext{c}{CSO}
\tablenotetext{d}{ALMA 870\micron\ flux}
\tablenotetext{e}{Jones et al. (2015)}
\end{deluxetable*}
\end{centering}

\section{Observational Results}\label{section-results}

\subsection{Photometry and Redshifts}
The $R$ \& $r$ magnitudes and the WISE, NVSS and ALMA flux densities for the 49 sources are presented in Table \ref{tbl-flux}.  There are 26 ALMA detections at 3$\sigma$ or above.      None of the sources are resolved.    The two sources from the ALMA subsample observed with {\it Herschel} were both detected at 70$\mu$m, and W1500-0649 also at 170$\mu$m (Table \ref{tbl-herschel}).  Twelve additional sources from the full sample have {\it Herschel} detections.  Neither of the two ALMA sources observed with the CSO, W1343-1136 and W1400-2919, were detected, resulting in 3$\sigma$ upper limits of 45 mJy for each, while one of the other two souces, W1025+6128, has a modest detection, as listed in Table \ref{tbl-herschel}.   There are available $R$-band Vega system magnitudes from SOAR for 16 of the ALMA-detected sources and for 10 of the sources with upper limits.  W0823-0624 and W1434-0235 have Sloan Digital Sky Survey (SDSS) data in one or more bands.  The $R/r$ magnitudes range from 20.7 to 24.1.      The optical photometry is used in this paper only as a  rough constraint on the mass  of the stellar populations; a full analysis of these data and the NIR photometry will be presented by A. Blain et al. (in preparation).   

Redshifts  are available for 25 of the 26 detected sources, and for 20 of the 23  sources with ALMA upper limits.  Six of the 19 sources in Table \ref{tbl-herschel} have known redshifts, including the four sources in the ALMA subsample and the two JCMT-observed sources.  It is beyond the scope of the present paper to analyze the optical/NIR spectroscopy, however we note that many spectra have indications from the ionization levels of an obscured radiative-mode AGN  \citep{baldwin81,kewley13}.  The Magellan/FIRE observations have been published by \citet{kim13} while the VLT/Xshooter observations will be reported by A. Blain et al. (in preparation).   One object was not observed (W1657-0102), one was not detected even in continuum (W0614-0936), and for two objects only a faint continuum was detected (W1707-0811 and W2040-3541).     Additionally the redshifts are uncertain for five sources which have only weak lines, a single line detection or an unresolved line pair.   They are: W0519-0813 (L$\alpha$$\lambda$1216\AA); W0702-2808 ([\ion{O}{2}]$\lambda$3727\AA); W1521+0017(weak H$\beta$,[\ion{O}{3}]$\lambda$4959,5007\AA); W1703-0517 (blended H$\alpha$+[\ion{N}{2}]$\lambda$6584\AA); and W1936-3354 ([\ion{O}{2}]$\lambda$3727\AA).

\begin{center}
\begin{deluxetable*}{llrrrrrrrrrccc}[t]
\tabletypesize{\tiny}
\setlength{\tabcolsep}{0.02in}
\tablecolumns{14}
\tablewidth{0pc}
\tablecaption{Redshifts, Luminosities and Radio Power \label{tbl-lum}}
\tablehead{
\colhead{WISE Name} & \colhead{Redshift} & \multicolumn{2}{c}{log L$_{\rm AGN}$}  & \multicolumn{4}{c}{log L$_{\rm BB}$} &  \multicolumn{3}{c}{log L$_{\rm Total}$} & \colhead{T$_{\rm dust}$} & \colhead{log P$_{\rm 3GHz}$} & \colhead{$q_{\rm 22}$}\\
\colhead{} & \colhead{} & \multicolumn{2}{c}{(L$_{\odot}$)}  & \multicolumn{4}{c}{(L$_{\odot}$)} &  \multicolumn{3}{c}{(L$_{\odot}$)} & \colhead{(K)\tablenotemark{a}} & \colhead{(WHz$^{-1}$)} & \colhead{k-corr'd}\\
\colhead{} & \colhead{} & \colhead{Min} & \colhead{Max} & \colhead{30K} & \colhead{50K} & \colhead{90K} & \colhead{120K} & \colhead{Min} & \colhead{Max} & \colhead{Best} & \colhead{Best model} & \colhead{} & \colhead{} 
}
\startdata
W0304-3108 &  1.54 &  13.01 & 13.16 &    11.49 &    12.46 &    12.92 &    \nodata     &    13.02 &    13.36 &    13.36 & 90 & 26.58 & -1.30 \\
W0306-3353 &  0.78 &  12.19 & 12.23 &    10.98 &    12.07 &    12.28 &    \nodata     &    12.25 &    12.54 &    12.53 & $<$50,90$>$ &        24.92 & -0.55\\
W0354-3308 &  1.37 &  12.70 & 12.72 & $<$11.14 & $<$12.14 & $<$12.59 &    \nodata     & $<$12.71 & $<$12.96 & $<$12.96 & 90 &    25.61 & -1.15 \\
W0404-2436 &  1.26 &  12.58 & 12.63 &    11.46 &    12.48 &    12.93 &   \nodata     &    12.61 &    13.11 &    13.11 & 90 &    25.67 & -1.69 \\
W0409-1837 &  0.67 &  12.73 & 12.74 &    \nodata     & $<$11.99 & $<$12.4  &   \nodata     & $<$12.80 & $<$12.90 & $<$12.90 & 90 &    25.65 & -0.28 \\
W0417-2816 &  0.94 &  11.95 & 12.01 &    11.48 &    12.55 &    12.98 &  \nodata     &    12.12 &    13.02 &    12.47 & $<$30,50$>$ & 25.54 & -1.38 \\
W0439-3159 &  2.82 &  13.44 & 13.45 &    12.04 &    12.84 &    13.33 &    14.17 &    13.47 &    14.24 &    13.70 & 90 &    26.76 & -1.12 \\
W0519-0813 &  2.05\tablenotemark{c} &  13.01 & 13.02 & $<$11.30 & $<$12.2  & $<$12.68 & $<$13.58 & $<$13.03 & $<$13.68 & $<$13.18 & 90 &    26.53 & -1.09 \\
W0525-3614 &  1.69 &  12.51 & 12.60 & $<$11.29 & $<$12.25 & $<$12.71 & $<$13.75 & $<$12.62 & $<$13.77 & $<$12.94 & 90 &    25.64 & -0.55 \\
W0526-3225 &  1.98 &  13.44 & 13.54 &    12.40 &    13.31 &    13.79 &    \nodata     &    13.57 &    13.95 &    13.95 & 90 &    27.33 & -1.18 \\
W0536-2703 &  1.79 &  13.05 & 13.05 &    \nodata     &   \nodata     &    12.94 &    \nodata     &    13.00 &    13.30 &    13.30 & 90 &    25.90  & -0.54 \\
W0549-3739 &  1.71 &  12.56 & 12.56 &    11.39 &    12.34 &    12.81 &    \nodata     &    12.59 &    13.00 &    13.00 & 90 &    25.97 & -0.92 \\
W0612-0622 &  0.47 &  12.03 & 12.04 &    10.89 &    12.05 &    12.20  & \nodata     &    12.07 &    12.43 &    12.43 & 90 &    25.14 & -0.32 \\
W0613-3407 &  2.18 &  13.20 & 13.20 & $<$11.43 & $<$12.31 & $<$12.79 & $<$13.68 & $<$13.21 & $<$13.80 & $<$13.34 & 90 &    26.57 & -0.90 \\
W0614-0936 &  \nodata\tablenotemark{b} &  12.94 & 12.95 & $<$11.40 & $<$12.31 & $<$12.78 & $<$13.69 & $<$12.96 & $<$13.76 & $<$13.17 & 90 &    25.87 & -0.38\\
W0630-2120 &  1.44 &  12.52 & 12.57 &    11.72 &    12.71 &    13.17 &    \nodata     &    12.63 &    13.26 &    12.82 & $<$30,50$>$ & 25.86 & -1.33 \\
W0642-2728 &  1.34 &  12.36 & 12.36 &    11.34 &    12.35 &    \nodata     &    \nodata     &    12.40 &    12.66 &    12.66 & 50 &    25.50 & -1.42 \\
W0652-2006 &  0.60 &  11.58 & 11.59 &    11.11 &    12.24 &    \nodata     &    \nodata     &    11.71 &    12.33 &    12.12 & $<$30,50$>$ & 24.74 & -0.39  \\
W0702-2808 &  0.94\tablenotemark{c} &  12.04 & 12.04 & $<$11.10 & $<$12.17 &    \nodata     &    \nodata     & $<$12.09 & $<$12.41 & $<$12.41 & 50 &    25.23 & -1.00 \\
W0714-3635 &  0.88 &  11.88 & 11.89 &    11.18 &    12.26 &    \nodata    &    \nodata     &    11.97 &    12.41 &    12.41 & 50 &    25.34& -1.09  \\
W0719-3349 &  1.63 &  12.57 & 12.57 &    \nodata     &    12.75 &    13.22 &    \nodata     &    12.97 &    13.31 &    13.01 & $<$50,90$>$ & 26.71 & -1.33 \\
W0811-2225 &  1.11 &  12.63 & 12.63 &    \nodata     &    \nodata     &    $<$12.65 &    \nodata     &    $<$12.94 &    $<$12.94 &    $<$12.94 & 90 &    25.75& -1.53  \\
W0823-0624 &  1.75 &  13.11 & 13.22 & $<$11.36 & $<$12.3  & $<$12.76 &    \nodata     & $<$13.12 & $<$13.35 & $<$13.35 & 90 &    26.70 & -1.10 \\
W1308-3447 &  1.65 &  12.99 & 12.99 &    11.22 &    12.18 &    12.65 &    \nodata     &    13.00 &    13.15 &    13.15 & 90 &    26.74 & -1.40 \\
W1343-1136 &  2.49 &  13.00 & 13.02 &    \nodata     &    12.43 &    12.92 &    13.78 &    13.12 &    13.85 &    13.27 & 90 &    26.23 & -0.75 \\
W1400-2919 &  1.67 &  13.10 & 13.11 & $<$10.78 & $<$11.73 &    \nodata     & $<$13.16 & $<$13.11 & $<$13.43 & $<$13.43 & 120 &   26.63 & -1.13 \\
W1412-2020 &  1.82 &  13.02 & 13.17 &    11.52 &    12.45 &    12.92 &    \nodata     &    13.03 &    13.36 &    13.36 & 90 &    25.95 & -0.43 \\
W1434-0235 &  1.92 &  12.89 & 12.89 &    $<$11.08 &    $<$12.00 &    $<$12.47 &    $<$13.39 &    $<$12.90 &    $<$13.51 &    $<$13.03 & 90 &    26.62 & -1.22 \\
W1439-3725 &  1.19 &  12.29 & 12.29 & $<$10.79 & $<$11.81 & $<$12.26 &    \nodata     & $<$12.30 & $<$12.58 & $<$12.58 & 90 &    25.58 & -1.70 \\
W1500-0649 &  1.50 &  13.07 & 13.50 &    \nodata     &    12.84 &    13.33 &    \nodata     &    13.52 &    13.59 &    13.52 & 90 &    26.10 & -0.89 \\
W1510-2203 &  0.95 &  12.60 & 12.61 & $<$10.74 & $<$11.81 & $<$12.23 &    \nodata     & $<$12.62 & $<$12.76 & $<$12.76 & 90 &    25.57 & -0.84 \\
W1513-2210 &  2.20 &  13.26 & 13.26 &    11.86 &    12.74 &    13.22 &    \nodata     &    13.28 &    13.54 &    13.54 & 90 &    26.68 & -0.89 \\
W1514-3411 &  1.09 &  12.44 & 12.48 & $<$10.81 & $<$11.86 & $<$12.29 &    \nodata     & $<$12.45 & $<$12.70 & $<$12.70 & 90 &    25.57 & -1.36 \\
W1521+0017 &  2.63\tablenotemark{c} &  13.61 & 13.61 &    \nodata     &    \nodata     &    12.63 &    \nodata     &    13.65 &    13.65 &    13.65 & 90 &    25.60 & -1.04 \\
W1541-1144 &  1.58 &  12.81 & 12.94 & 11.13 & 12.10 & 12.57 &    \nodata     & 12.95 & 13.01 & 13.01 & 90 &    26.40 & -1.15 \\
W1634-1721 &  2.08 &  12.83 & 12.83 & $<$11.08 & $<$11.98 & $<$12.46 & $<$13.36 & $<$12.84 & $<$13.47 & $<$12.98 & 90 &    26.11 & -0.80 \\
W1641-0548 &  1.84 &  12.94 & 13.09 &    \nodata     &    12.41 &    12.88 &    \nodata     &    13.05 &    13.30 &    13.30 & 90 &    25.83 & -0.37 \\
W1653-0102 &  2.02 &  13.00 & 13.00 & $<$11.04 & $<$11.94 & $<$12.41 & $<$13.31 & $<$13.00 & $<$13.48 & $<$13.10 & 90 &    26.19 & -0.73 \\
W1657-1740 &  \nodata\tablenotemark{b} &  13.16 & 13.17 & $<$11.03 & $<$11.94 & $<$12.42 & $<$13.32 & $<$13.17 & $<$13.55 & $<$13.60 & $<$90,120$>$ & 26.34 & -0.68 \\
W1702-0811 &  2.85 &  13.60 & 13.60 & $<$11.28 & $<$12.07 &    -\nodata    & $<$13.4  & $<$13.60 & $<$13.81 & $<$13.81 & 120 &   27.26 & -1.20 \\
W1703-0517 &  1.80\tablenotemark{c} &  13.11 & 13.51 & 11.12 & 12.05 & 12.48 & 13.45 & 13.12 &13.61 & 3.60 & $<$90,120$>$ & 26.91 & -0.74 \\
W1707-0939 &  \nodata\tablenotemark{b} &  12.98 & 13.04 & $<$11.15 & $<$12.06 & $<$12.54 & $<$13.43 & $<$12.99 & $<$13.56 & $<$13.56 & 120 &   26.14 & -0.89 \\
W1936-3354 &  2.24\tablenotemark{c} &  13.05 & 13.18 &    11.45 &    12.32 &    \nodata     &    13.69 &    13.06 &    13.78 &    13.78 & 120 &   25.64 & -0.40 \\
W1951-0420 &  1.58 &  12.80 & 12.83 & $<$11.08 & $<$12.05 &    \nodata     & $<$13.45 & $<$12.84 & $<$13.54 & $<$13.54 & 120 &   26.18 & -1.03 \\
W1958-0746 &  1.80 &  12.98 & 12.98 & $<$11.08 & $<$12.01 & $<$12.48 & $<$13.41 & $<$12.99 & $<$13.54 & $<$13.10 & 90 &    26.51 & -0.99 \\
W2000-2802 &  2.28 &  12.75 & 12.90 & $<$10.72 & $<$11.59 & $<$12.07 & $<$12.96 & $<$12.75 & $<$13.17 & $<$12.96 & 90 &    25.99 & -0.73 \\
W2021-2611 &  2.44 &  12.97 & 13.51 &    11.85 &    12.68 &    \nodata     &    14.04 &    13.00 &    14.15 &    14.15 & 120 &   25.93 & -0.45 \\
W2040-3904 &  \nodata\tablenotemark{b} &  12.92 & 13.15 &    11.85 &    12.76 &    13.23 &    \nodata     &    12.96 &    13.49 &    13.49 & 90 &    26.14 & -0.81 \\
W2059-3541 &  2.38 &  13.15 & 13.37 &    \nodata     &    \nodata     & $<$12.46 & $<$12.87 & $<$13.15 & $<$13.42 & $<$13.42 & 90 &    26.26 & -0.69 \\
\enddata
\tablenotetext{a}{$<$T1,T2$>$ denotes an average between these two model fits.}
\tablenotetext{b}{$z$=2 assumed if no spectroscopic redshift exists.}  
\tablenotetext{c}{Uncertain redshift: single line, blended lines or weak lines.}  
\end{deluxetable*}
\end{center}
   
\subsection{Colors and Spectral Energy Distributions}
The ALMA observations were designed to constrain the luminosity of these quasars in the rest-frame FIR-submillimeter compared to the MIR.  We will show that most of our WISE-NVSS-ALMA sample have SEDs dominated by an AGN in the MIR, and possibly through the FIR-submm also.  However, substantial rates of star formation are likely also present.    

We present the main results in Figures \ref{fig-ratioz870}--\ref{fig-lastsed}.    We show these results in two complementary ways:  (1) the ratio of 870 $\mu$m/22 $\mu$m flux density as a function of redshift is shown in Figure \ref{fig-ratioz870}; and (2) we use the range in redshifts for the sample to construct an ``ensemble" rest-frame SED for our sample in Figure \ref{fig-seds}.  We show ensemble SEDs of several comparison samples in Figure \ref{fig-lastsed}.  In all the plots we also show the tracks for several templates of nearby well-studied sources.    The ensemble SEDs must be interpreted carefully because the choice of normalization wavelength affects the relative appearance of dispersion between the points at other wavelengths.    We have chosen to normalize the templates and data at rest frame 4.6 $\mu$m.    The selection of 4.6$\mu$m has the disadvantage that this spectral region may suffer significant extinction from a thick torus or other nuclear dusty structure, however it is the longest wavelength (and hence has the lowest optical depth) which both avoids the PAH and silicate features and which still lies within the WISE rest-frame wavelength range for all of the ALMA sources.   

The  source templates in all five figures are from \citet{polletta07}, and include the cool starburst-dominated ULIRG Arp 220, the starburst M82, and the broad-line dusty QSO Mrk 231.   The torus model in Figures \ref{fig-seds}--\ref{fig-lastsed} is based on the tapered disk models of \citet{efstathiou95} and has been fitted to one of our best sampled SEDs in this paper (see Section \ref{rt}). This torus model has an opening angle of 45$^{\circ}$, an inclination angle of 54$^{\circ}$ and a UV equatorial optical depth $\tau_{\rm {\nu}}$=500.  The intrinsic AGN SEDs, from  \citet{shang11} and \citet{mullaney11}, are empirical SEDs of nearby AGN from which the host galaxy light has been subtracted, and they are quite similar to each other and to other published intrinsic AGN SEDs \citep{elvis94,richards06,netzer07,assef10}.     In Figure \ref{fig-ratioz870}  we include the modeled SED of a $\sim$ 4 pc radius dust torus in the nearby AGN NGC 3081 that \citet{ramos11}  derived based on $Herschel$ data and subarcsecond MIR imaging with Gemini T-ReCS.      

\begin{figure}[!ht]
	\centering
	\includegraphics[trim=80 70 12 365,clip,width=0.55\textwidth]{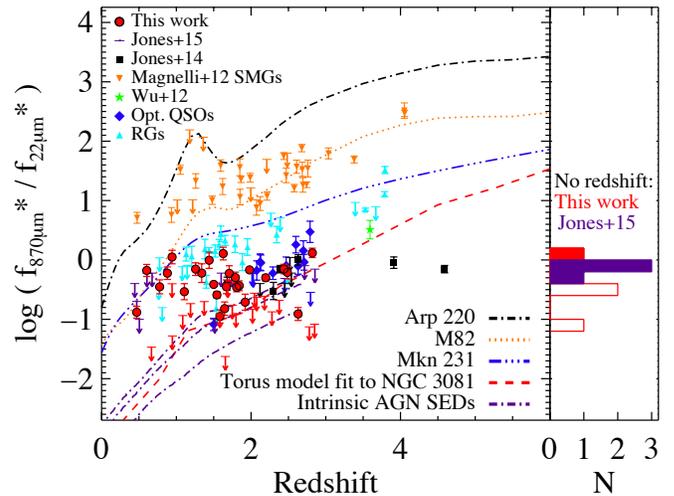}
\caption{Ratio of ALMA 870 $\mu$m to WISE 22 $\mu$m flux density {\it vs.} redshift for the ALMA sample (red points and limits).     Histogram: sources without  redshift (submillimeter-detected sources:  filled bars;  upper limits: empty bars).  The comparison templates and source samples are described in the text.       The three intrinsic AGN curves (purple dash-dot lnes) are the mean [$\log (L_{\rm 2-10kev}/{\rm erg}~{\rm  s}^{-1}) < 42.9$], mean[all], mean [$\log (L_{\rm 2-10kev}/{\rm erg}~\rm{s}^{-1}) > 42.9$]  SEDs of \citet{mullaney11} in descending order which cover 6--1000 $\mu$m rest frame; the full dispersion of intrinsic AGN SEDs reported in that paper is significantly wider.   The three ALMA sources lying on or above the Mkn 231 template are W0417-2816, W0652-2006 and W0714-3635.   $^{*}$: The comparison samples include data at 850 $\mu$m and at 24 $\mu$m.   \label{fig-ratioz870}}
\end{figure}

Most of our WISE-NVSS-ALMA sample is much more strongly MIR-dominated than the starburst templates (M82 and Arp 220) in Figure \ref{fig-ratioz870}, and also compared to the $1<z< 3$ Submillimeter Galaxies (SMGs) \citep{magnelli12}, which we illustrate to provide a comparison to high redshift starburst-dominated systems.     The WISE radio-blind Hot DOG samples (EWB12; Jones et al. 2014), and our northern sample of RP sources \citep{jones15} are similar to the ALMA sample in showing a low submm/MIR flux density ratio compared to the galaxy templates.   Generally speaking, the WISE sources lie between the intrinsic SEDs and the Mkn 231 template, though some of the upper limits lie below even the intrinsic SEDs shown.    We also show some representative redshift $\sim$1--3 radio galaxies \citep{archibald01,grimes05,seymour07} and broad-line (optically selected) quasars \citep{priddey03,priddey07} with available 850$\mu$m data in the plot, however a detailed comparison to the FIR-submillimeter properties of well selected samples of AGN of various types is beyond the scope of this paper.    The broad-line QSOs shown tend to have similar colors to our sample, while the RGs shown tend to be a bit more FIR-strong relative to the MIR, however these trends may be dominated by selection effects and should be used only as a very general comparison to our sample.    

\begin{figure*}[!ht]
	\centering
	\includegraphics[trim=60 110 12 350,clip,width=0.8\textwidth]{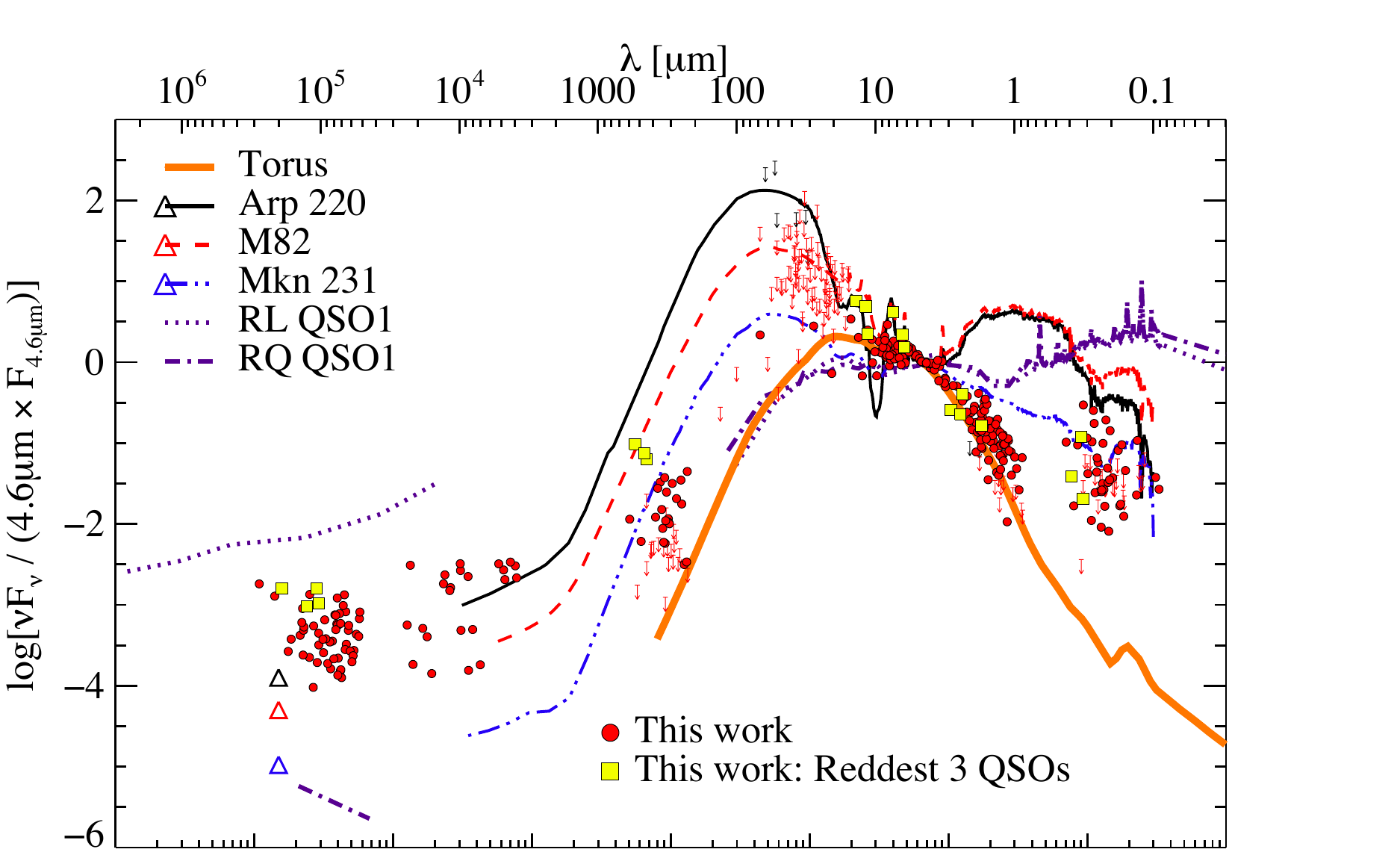}
	\includegraphics[trim=60 75 12 390,clip,width=0.8\textwidth]{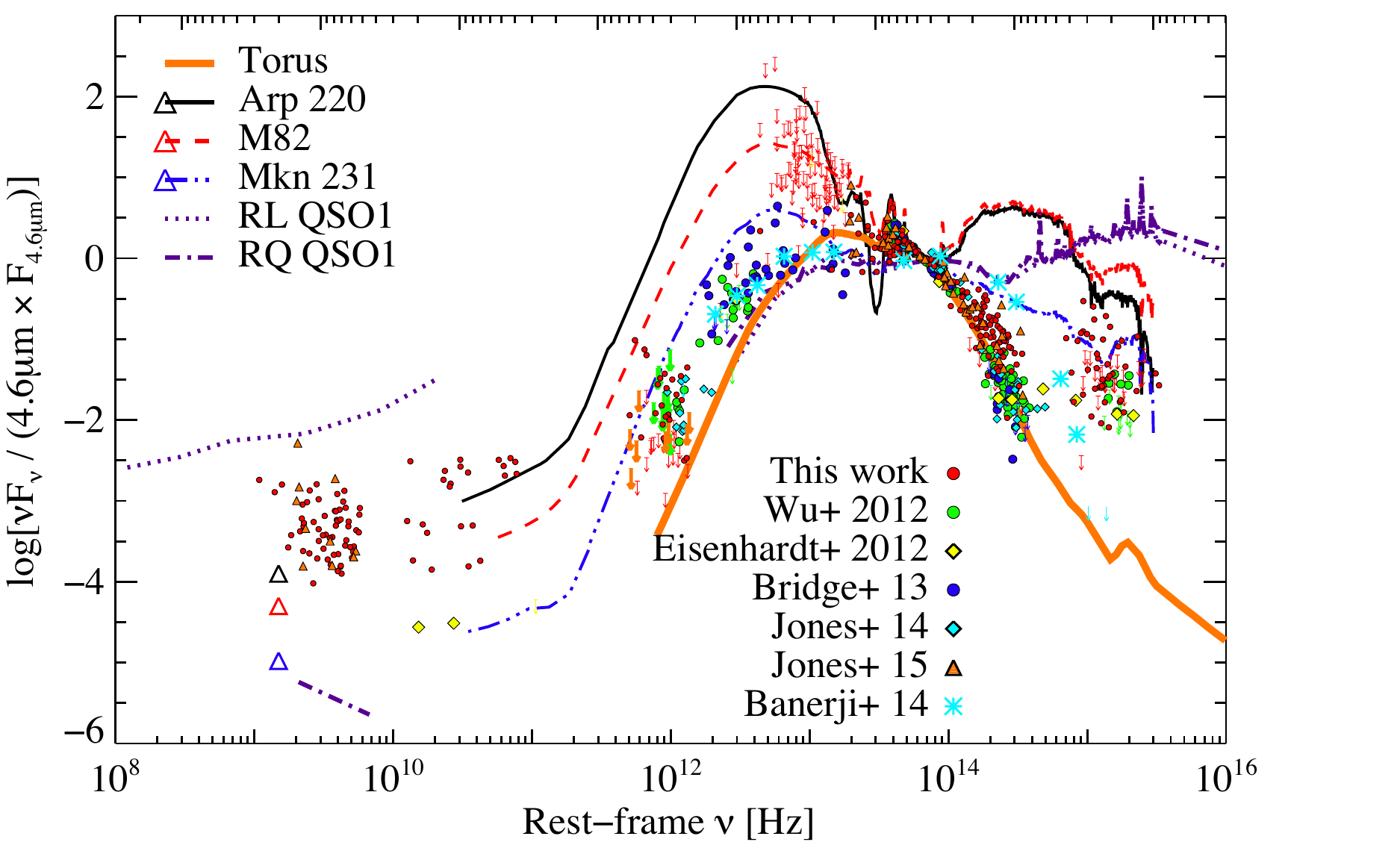}
	\caption{Rest-frame ensemble SEDs for WISE-selected ultra red systems.   {\bf Top:} the 45 sources in our ALMA sample with known redshift, normalized at 4.6 $\mu$m, compared with the templates from the library of \citet{polletta07},  the torus model fitted to one of our sources in Section \ref{rt}, and the intrinsic AGN SEDs of \citet{shang11} (labelled RQ QSO1 and RL QSO1).    The colored triangles in the radio correspond to the matched color template.   The three reddest objects, relative to Mkn 231, from Figure \ref{fig-ratioz870} are highlighted in yellow.   The upper limits in the rest $\sim$ 12--100 $\mu$m range are largely from IRAS, with a few from $Herschel$ (see text for details). Several sources lie at redshifts where the 22 $\mu$m band coincides with the 9.7 $\mu$m silicate absorption feature, which may partially account for the turnover in their spectral shapes.   Note the intermediate radio power compared to classical (evolved) RQ and RL QSOs, when normalized to 4.6 $\mu$m power.  {\bf Bottom}: Rest-frame SEDs for the WISE-selected red Hot DOG sources from \citet{wu12} (green circles), \citet{eisenhardt12} (yellow diamonds), \citet{jones14} (cyan diamonds) and the WLABs from \citet{bridge13} (blue circles) compared to our samples (red circles; orange triangles (Jones et al. 2015).  Also shown is the reddest of the dust-reddened type 1 quasars from \citet{banerji14} (cyan asterisks).      \label{fig-seds}}
\end{figure*}

\begin{figure*}[!ht]
	\centering
	\includegraphics[trim=65 110 12 340,clip,width=0.8\textwidth]{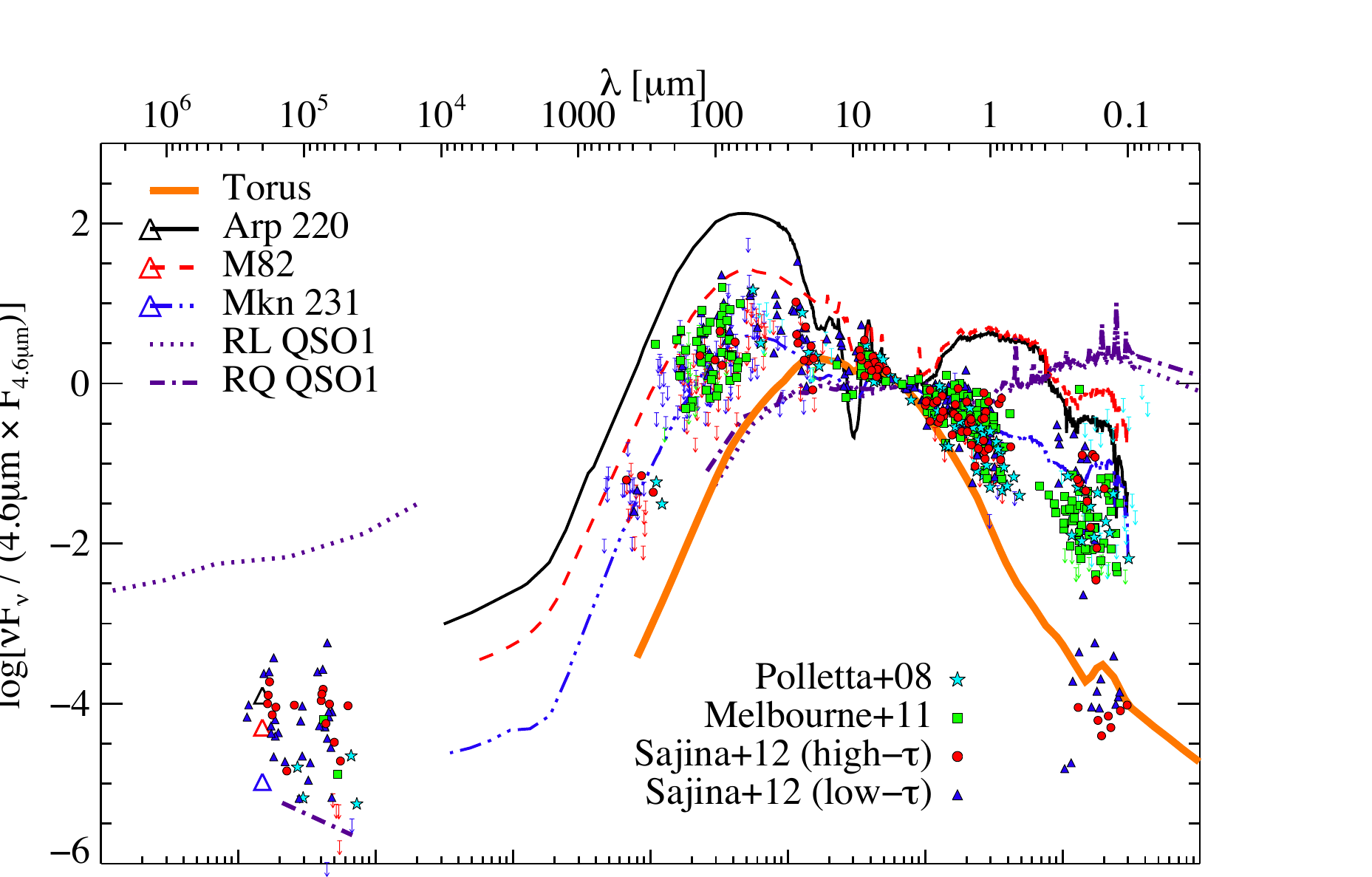}
	\includegraphics[trim=60 75 12 390,clip,width=0.8\textwidth]{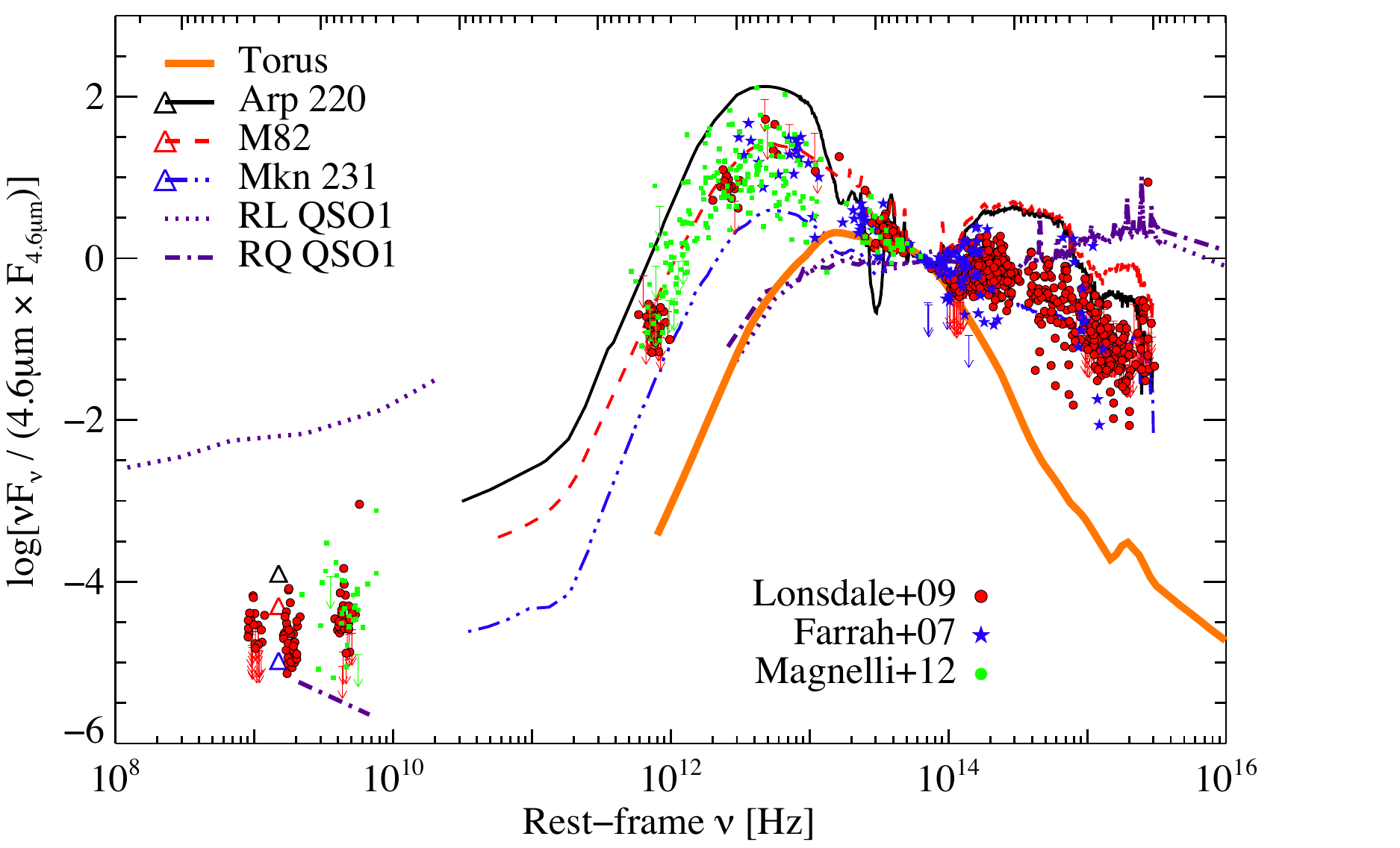}
	\caption{Rest-frame SEDs for several comparison {\it Spitzer} and {\it Herschel} samples.  {\bf Top}:Rest-frame SEDs for 152  $Spitzer$ DOGs which were selected in a similar manner to our sample \citep{polletta08,melbourne12,sajina12}.    The extreme redness in the UV--optical for many of the Sajina et al. (2012) sources is due to their selection criteria.  {\bf Bottom}: Rest-frame SEDs for 61 high-redshift MIR-selected starburst-dominated ULIRGs from \citet{lonsdale09} and \citet{fiolet09} (red symbols), 16 $Spitzer$ 70 $\mu$m-selected galaxies from \citet{farrah07} (blue symbols) and 61 SMGs with {\it Herschel} data from  \cite{magnelli12} (green symbols).   \label{fig-lastsed}}
\end{figure*}

The conclusions from Figure \ref{fig-ratioz870} are emphasized in the ensemble SED (Figure \ref{fig-seds}-top), where we see a strong similarity amongst the MIR-submm SEDs of the ALMA sample.  They resemble the high optical depth torus model in the rest frame NIR-MIR, as expected since this model was designed to fit one of our sources, however they are systematically {\it steeper} than any of the galaxy or intrinsic AGN templates in this wavelength range, when normalized at 4.6 $\mu$m.}    There is a lack of observational data available in the rest frame FIR, although the IRAS limits (red limits at $100<{\lambda}_{\rm rest}<10$ $\mu$m)  help to constrain the flux of many of our WISE-NVSS sample, ruling out SEDs like Arp 220 and M82.    

The three sources with highest submillimeter/MIR flux ratios (relative to the templates) from Figure \ref{fig-ratioz870},  W0417-2816, W0652-2006 and W0714-3635 (yellow points in Figure \ref{fig-seds}) are the strongest candidates for possessing significant ongoing star formation.   Their presence in our sample may be the result of a selection effect due to strong 6.7--7.7 $\mu$m PAH features falling in the W3 filter, or an 11.7 $\mu$m PAH feature falling in the W4 filter, as can be seen in Figure \ref{fig-seds}.   

In Figure \ref{fig-seds}-bottom we add the 10 sources with known redshift from \citet{jones15} (orange triangles), which are drawn from our northern WISE-NVSS sample.  They all have upper limits at 850 $\mu$m from JCMT and they show very similar SEDs to the ALMA subsample.  We also compare our radio-selected samples to the radio-blind WISE Hot DOG samples of EWB12 and of \citet{jones14}.   We see a very close similarity between the radio-selected (red and orange symbols) and radio-blind samples (green, blue, cyan and yellow filled symbols).  The radio-blind Hot DOG  samples have systematically larger redshifts than our sample, displacing the two sets of SEDs from each other in rest wavelength somewhat, however the radio-selected and radio-blind samples fall within a continuous band.  The different redshift selection function for the two samples may be unrelated to radio power, instead being due at least in part to the redder ($W2-W3$) threshold of EWB12.  This probably eliminates sources with silicate absorption that falls into the 22$\mu$m filter, and therefore favors sources with $z\gtrsim$1.5-2.    \citet{jones15} have suggested that the northern sources from our NVSS-WISE sample that they observed at 850 $\mu$m at the JCMT may show slightly less steep rest-frame SEDs than the EWB12 samples.   This is an interesting possibility that requires further study.

The WISE samples of EWB12 have more data from $Herschel$ in the rest MIR-FIR than we do for our sample, and their ensemble FIR SEDs tend to fall in the region occupied by the intrinsic SEDs, the torus template and the Mkn 231 template (Figure \ref{fig-seds}).    The two WISE-NVSS sources for which we have $Herschel$ data fall in the same region.     All of these sources lie well below both M82 and Arp 220 in this wavelength region, emphasizing the result from Figure \ref{fig-ratioz870}.  The limited $Herschel$ data which we do have for our sample, and the IRAS limits, support a picture in which our WISE-NVSS sample have similar FIR-submm SEDs to the radio-blind WISE samples.  The one Hot DOG with published radio data \citep{eisenhardt12} has lower radio power than our sample, when normalized to 4.6 $\mu$m power. 

It is interesting to compare the WISE-NVSS objects to the reddest known broad-line type 1 quasar from \citet{banerji14}, ULAS J1234+0907 ($z = 2.50$), shown as the cyan asterisks in Figure \ref{fig-seds} bottom.       The SED shape is very similar to the Hot DOGs through the MIR-submm but is much less red than any of the WISE sample in the rest NIR 1--5 $\mu$m region.  Unlike the WISE red sample, ULAS J1234+0907 follows the Mkn 231 template through $\sim$ 1 $\mu$m, only dropping steeply at shorter wavelengths.       The best fit model by \citet{banerji14} has $A_V = 6$.    This result emphasizes the likelihood that the steep red WISE-optical SEDs for the WISE-selected samples are caused by heavy obscuration.

In Figure \ref{fig-lastsed}-top we turn to a comparison to the $Spitzer$-selected ``power-law'' DOGs  \citep{polletta08,melbourne12,sajina12}, including the two $z > 3$ Compton-thick quasars discussed by \citet{polletta08}.   As a class the WISE samples show significantly redder ensemble rest NIR slopes than most the $Spitzer$ DOG samples.  The WISE sources also appear to turn over into the FIR/submm at systematically shorter wavelength than the DOGs, as can be seen by comparing Figures \ref{fig-seds}-bottom and \ref{fig-lastsed}-top, and as previously discussed by EWB12 and \citet{jones14,jones15}.    To first order, this is likely to indicate a larger ratio of star formation to AGN accretion in the $Spitzer$ DOGs than in the Hot DOGs.     

Lastly we illustrate in Figure \ref{fig-lastsed}-bottom the ensembles SEDs of several samples of {\it Spitzer}-selected starburst-dominated Ultra Luminous InfraRed Galaxies (ULIRGs) \citep{farrah07,lonsdale09,fiolet09}, and SMGs observed with {\it Herschel} \citep{magnelli12}.   These samples have more marked FIR emission than either the WISE sources or the {\it Spitzer}-selected DOGs, relative to MIR emission, and bluer optical-MIR SEDs.    

Turning to the cm-radio emission there is a steady decrease in the average radio/4.6 $\mu$m flux density ratio from our sample through the {\it Spitzer} DOGs to the SMGs / starbursts, although there is a lot of overlap between the latter two samples.    This in entirely consistent with the selection in favor of bright radio sources in our sample, and the dominance of star formation in the SMGs and starbursts.

In summary, the WISE-selected ultra red samples have very similar SED shapes from the rest-frame NIR through the FIR, with no obvious difference between the radio-selected samples (this paper and Jones et al. 2015) and the radio-blind Hot DOGs (EWB12; Jones 14) except that the radio-blind samples have a larger mean redshift, which may be caused by the different selection functions.   Together, these sources are redder than any other known source type in the NIR-MIR, and most of them turnover into the FIR at higher frequencies than the Spitzer DOGs, starbursts and SMGs.   

\subsection{Synchrotron Contribution to the ALMA fluxes}\label{synch}
Before addressing the possible range of star formation rates in these sources, we first consider the possibility that a fraction of the 345 GHz flux is due to synchrotron emission.  Since we have selected compact radio sources in radiatively efficient AGN there is the possibility that the radio emission is beamed and that some of our sources are blazers.   We briefly address the possibility that some of the 345 GHz emission arises from non-thermal synchrotron emission associated with the radio sources here, however we defer detailed discussion of this topic to the next paper in our series, in which we present the high resolution X-band (8-12 GHz) imaging from the VLA.   

We can make some preliminary conclusions from the measured spectral indices across the 8-12 GHz VLA X-band, which are more reliable than indices derived from non-contemporaneous and non-beam matched 1.4 GHz NVSS data and the much later X-band imaging.    We find that the majority of the sample has steep spectral indices between 8-12 GHz; 42 of them have ${\alpha}^{8}_{12}<-0.8$ (27 have indices steeper than -1.0),   characteristic of optically thin synchrotron emission and potentially consistent with being Gigahertz Peaked Sources (GPS; $<$ 1kpc in size with a synchrotron peak $\sim$ 1 GHz) or Compact Steep Spectrum sources (CSS; $<$ 20 kpc in size with a synchrotron peak below 1 GHz) \citep{odea98}.  Several sources are also resolved or multiple on scales of 1-10 kpc.  For most of these steep spectrum sources the synchrotron contribution to the 345 GHz flux is likely to be $<$10\%.

The remaining seven sources have flat or inverted 8-12 GHz spectral indices, however two of these show a steepening spectral index between 12 and 20 GHz (from our limited VLA K-band imaging) thus they are also likely to be dominated by optically thin synchrotron emission (W0526-3225 and W0823-0642).   Two of the remaining 5 sources (W0642-2728 and W1434-0235) have an ALMA measurement (a detection and a limit, respectively) which is well in excess (by a factor of 5 or more) of the extrapolated flat radio SED.  The other three sources (W0536-2703, W1412-2020 and W1634-1721)  require a synchrotron peak beyond 12 GHz to avoid exceeding the ALMA flux density detection or limit, and these are also probably not blazers unless they are exceptionally variable.   They are more likely to be High Frequency Peakers (small $<$ 100 pc sources with synchrotron peaks above 4 GHz; Dallacasa et al. 2000) and it is possible that their 345 GHz flux has a significant contribution from optically thin synchrotron emission.         None of the flat or inverted spectrum sources has a plausible SED that can explain the WISE data as synchrotron emission from a blazar.  

\section{ISM Mass, SED Fits and Derived Parameters}\label{section-fits}
As noted in the previous section, the simplest description for the NIR-FIR SED shapes of the majority of our sample is that they resemble the intrinsic shapes of local AGN samples, derived by subtracting the host galaxy emission \citep{elvis94,richards06,netzer07,assef10,mullaney11,shang11}.    If we interpret the SEDs in this fashion then we could conclude that {\it all} of the MIR-submillimeter dust emission stems from re-radiation of accretion disk energy by a nuclear torus or other dusty structure.    In this section we explore, as an alternative, the range of plausible contributions from star formation that may be permitted by reasonable models of the SEDs.

\subsection{Interstellar Medium Mass}\label{ism}
We derive the ISM masses assuming that 100\% of the 345 GHz flux is thermal dust emission.  As noted in the previous section the thermal fluxes could be overestimated by $<$10\% for most of the sample due to possible contributions from synchrotron emission, and by up to 100\% for 3 sources.    We do not correct the data for this contamination because it is highly uncertain at this point in time.

\citet{scoville14} derive the gaseous ISM mass ($M_{\rm HI} + M_{\rm H_2}$) of populations of distant galaxies in the COSMOS field using ALMA 870$\mu$m data as a measure of cool dust mass.  They  
use a local sample of well studied galaxies to show that the observed ratio of 850 $\mu$m specific luminosity to ISM mass $ L_{{\nu}\rm 850 {\mu}m} / M_{\rm ISM} = 1 {\pm} 0.23 {\times} 10^{20}$ ${\rm ergs}$ ${\rm s^{-1}}$ ${\rm Hz^{-1}}$ $M_{\odot}^{-1}$ for low redshift spirals, with a dispersion of a factor of about 5.     Using this empirical calibration, \citet{scoville14} derive the following relation for a flux density measured at observed frequency $\nu_{\rm obs}$ (valid for $\lambda_{\rm rest} > 250 {\mu}$m on the Rayleigh-Jeans tail where the emission will be optically thin) to derive ISM masses from ALMA data for SMGs (their equation 12,  valid for a dust temperature of 25K):  

\small
\[\frac{f_{\nu_{obs}}}{mJy} \ = 0.83 \ \frac{M_{\rm ISM}}{10^{10} M_{\odot}} \ (1+z)^{4.8} \left(\frac{\nu_{\rm obs}}{\nu_{850{\mu}m}}\right)^{3.8} \frac{\Gamma_{\rm RJ}}{\Gamma_0} \left(\frac{\rm Gpc}{d_L}\right)^2 \ (1)\]
\normalsize

Here ${\Gamma_{\rm RJ}}/{\Gamma_0}$ corrects for the departure of the dust emission spectrum from the Rayleigh-Jeans tail as the redshift increases from 0 (and the rest frequency approaches the peak of the spectrum) and $d_L$ is the luminosity distance. \citep{scoville14} adopt 250 ${\mu}m$ as the minimum acceptable rest-frame wavelength for derivations of ISM mass, thus we expect our results to provide reasonable estimates for sources with z$<$2.5, which represents 94\% of our sample, with only small inaccuracies for the 3 sources with 2.5$<$z$<2.85$.   $\Gamma_{\rm RJ}$ is a function of assumed dust temperature, which was taken to be 25K by \citet{scoville14}.  The constant 0.83 in the equation is also proportional to $T_d$.  We derive ISM masses (Table \ref{tbl-mass}, columns 13 \& 14), for dust temperatures of 30 K and 90 K, using the appropriate values for the constant,  $\Gamma_{\rm 0}$, and ${\Gamma_{\rm RJ}}$.   The majority of the dust in these systems is unlikely to be as warm as 90 K, therefore these values provide fairly strong lower limits to the ISM masses when we have an 870 $\mu$m detection.   We estimate the overall uncertainty to be a factor of 5, based on the unknown dust temperature and the overall dispersion for local starburst systems in Figure 1 of Scoville et al. (2014).

The ISM masses for $T_d$ = 30 K range from 0.8 to 56 ${\times}10^{10} M_{\odot}$, with a median for the 26 ALMA-detected sources of 5.9 ${\times}10^{10} M_{\odot}$.  This may be compared to  the values derived by \citet{scoville14} for their mid- plus high-redshft sample galaxy stacks from the COSMOS field, which cover a similar redshift range as our sources.     Their ``IR Bright'' sample has a median ISM mass of 11.91$\pm$0.77 ${\times} 10^{10} M_{\odot}$, about twice that of our sample.          

\subsection{Spectral Energy Distribution Fits}
The SEDs of most of the red WISE-NVSS sources are dominated by warm dust in the MIR with a strong decline into the submm, and many of them display spectroscopic evidence of an obscured, high excitation, AGN.      The minimum AGN luminosity can be reasonably well determined from the warm dust emission which dominates the WISE data.  The total 1-000$\mu$m luminosity is less well constrained, however, for most sources, because we lack data in the rest-frame FIR.     We can model the emission by assuming a dominant dust temperature, $T_d$, for the cooler dust that peaks in the FIR wavelength range, however the luminosity of such a component depends on the 4$^{th}$ power of $T_d$, and is uncertain by 2-3 dex without measurements at the FIR peak of the dust SED.   

Two sources have available $Herschel$ data, W1500-0649 and W2059-3541, and for these we find that the SED shapes fall well below the SEDs of M82 and Arp 220 longward of the MIR, resembling the radio blind-selected HotDOGs (Figure \ref{fig-seds}).       We have constructed radiative transfer (RT) models for W1500-0649 (z=1.50), which has the most detections (7 bands in total) that include good constraints on the peak of the SED,  in order to provide some insights into the nature of these sources.   For the remaining sample there are insufficient data points to justify RT models, therefore we  use a parametric ``torus'' model plus modified blackbody (BB) fits for the cooler dust.    

\subsubsection{Radiative Transfer Model for WISE 1500-0649}\label{rt}

We have constructed example models for W1500-0649 using both a tapered disk model (Efstathiou \& Rowan-Robinson 1995)
for the torus (see also Efstathiou et al. 2013) and the two-phase clumpy torus models of Stalevski et al (2012).   For the cooler dust component radiating in the FIR we use the starburst models of
Efstathiou, Rowan-Robinson \& Siebenmorgen (2000) which were revised by 
Efstathiou \& Siebenmorgen (2009).

\begin{figure}[bp]
	\centering
	\includegraphics[trim=90 55 0 360,clip,width=0.5\textwidth]{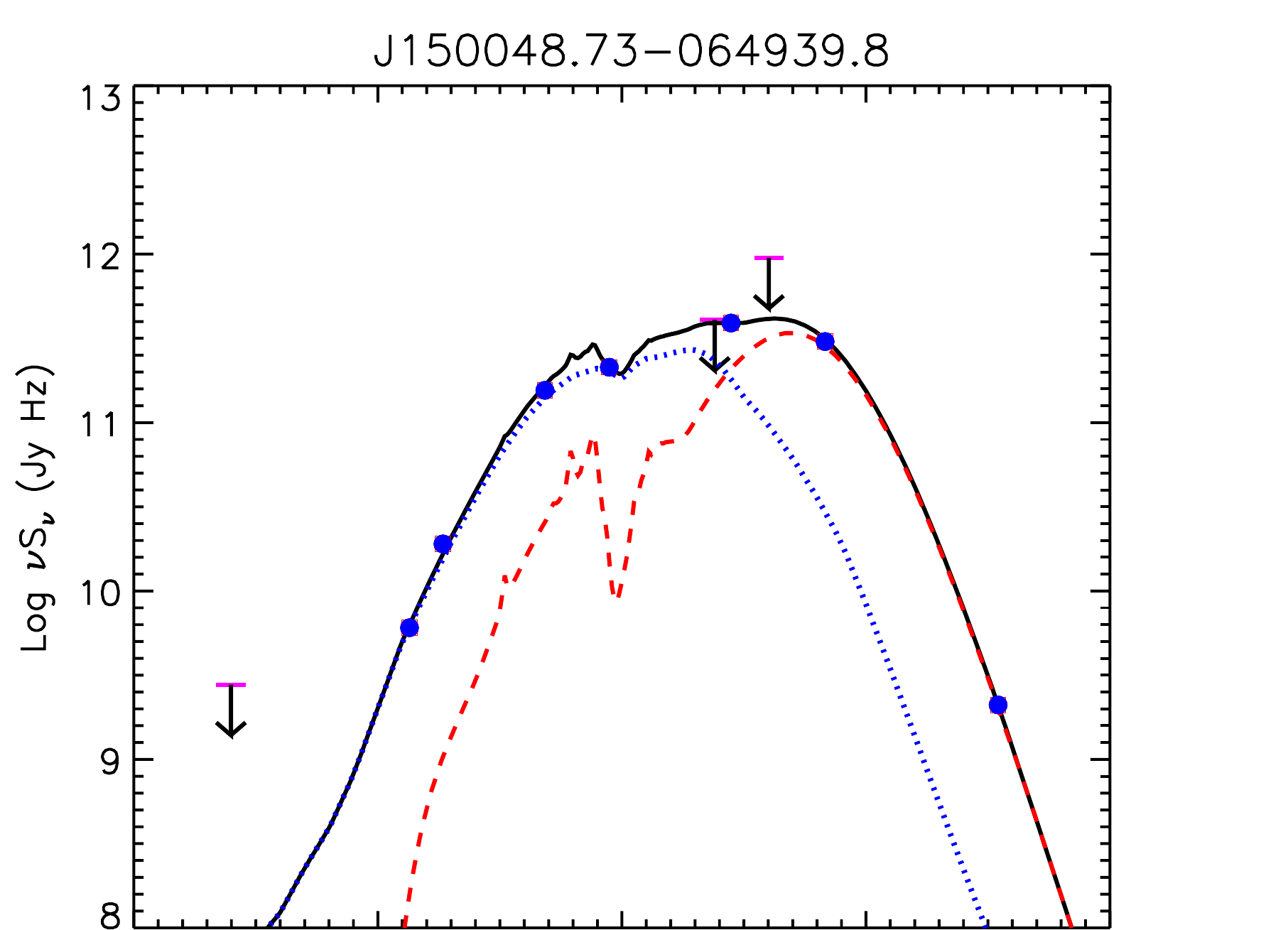}	
	\includegraphics[trim=90 10 0 390,clip,width=0.5\textwidth]{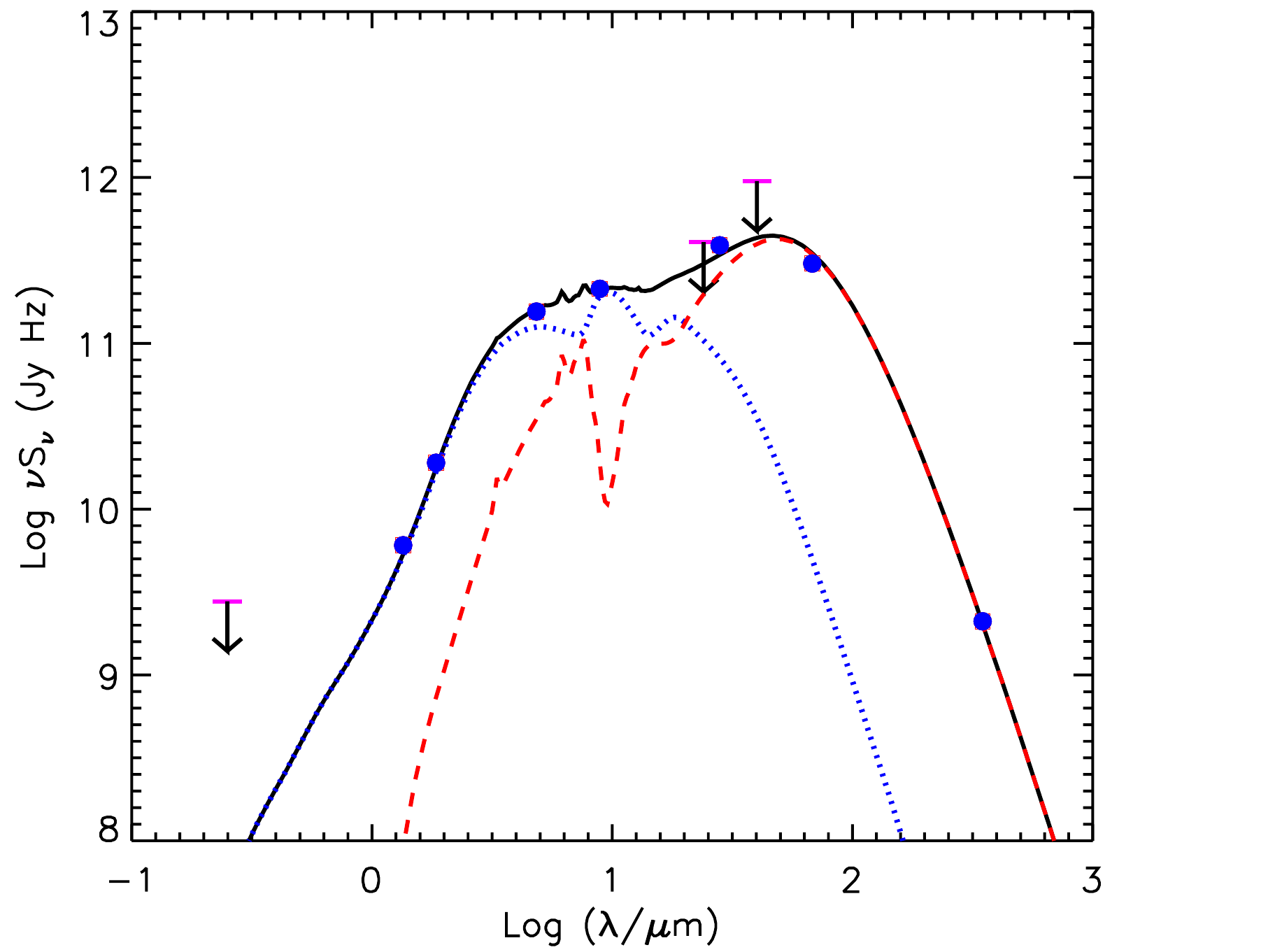}
	\caption{Best fit radiative transfer models for WISEJ150048.73-064939.8.  {\bf Top}: AGN tapered disk model of \citet{efstathiou95} which has 
$\tau_{uv}^{eq}=500$, $r_2/r_1=160$, $i=54^o$ and $t_*=15$ Myr.;
{\bf Bottom:}  AGN Clumpy torus model of \citet{stalevski12} (blue dotted lines).  
The starburst models  (red dashed lines)  are from \citet{efstathiou09}.  \label{fig-AEmodel}}
\end{figure}

The  tapered disk models have 4 parameters plus a normalization factor
and the starburst models have 3 parameters plus a different
normalization factor. In the tapered disc AGN models we fix the
opening angle of the torus $\Theta_0$ at 45$^o$ and vary the equatorial
optical depth of the torus $\tau_{uv}^{eq}$ in the range 250--1250 ($A_V
 \approx 50-250$), the inclination $i$ in the range 45 to 90$^o$ and the
ratio of outer torus radius to inner radius $r_2 / r_1$ from 20 to 160. In
 the starburst models we fix the initial optical depth of the molecular 
clouds $\tau_V$ at 75 (which is the average value in the model grid) and the
e-folding time of the starburst at 20 Myr which from previous work appears
to be a reasonable timescale for starbursts (e.g. Efstathiou et al. 2000). 
In the starburst models we only vary the age of the
 starburst $t_*$. 

We use a standard $\chi^2$ minimization technique to find the model parameters
that best fit the data. =The best tapered disk fit  is shown in Figure \ref{fig-AEmodel}-top (the
AGN torus model is plotted with a blue dotted line and
the starburst model with a red dashed line) assumes the following parameters:
$\tau_{uv}^{eq}=500$, $r_2/r_1=160$, $i=54^o$ and $t_*=15$ Myr.    

The AGN torus has a derived luminosity of $1.9 \times 10^{13} L_\odot$. This
needs to be multiplied by the anisotropy correction factor 
$A$ (Efstathiou 2006) which for this particular combination
of parameters is 0.84 to give an AGN luminosity of
$1.6 \times 10^{13} L_\odot$. The starburst luminosity is
$1.8 \times 10^{13} L_\odot$ so the total
luminosity of the system is predicted to be $3.4 \times 10^{13} L_\odot$. 

In Figure \ref{fig-AEmodel}-bottom we show the SED of
W1500-0649 fitted with the two-phase clumpy torus models of Stalevski et al (2012)
in combination with the starburst models of Efstathiou et al. (2009). We find that a good fit
can be obtained with a torus that assumes a half-opening angle of 50 and an inclination
of 90 degrees. The AGN luminosity is predicted to be $1.2 \times 10^{13} L_\odot$    and the starburst luminosity $2.4 \times 10^{13} L_\odot$.
The starburst is still predicted to be a young system with an age of 10 Myr. We conclude that
irrespective of the uncertain dust geometry, in  W1500-0649 the
AGN and the starburst emit comparable luminosity.

\subsubsection{Three-Component SED Fits} \label{threefit}
We model the full sample by fitting the minimum number of simple spectral components to the SEDs that will define a reasonable {\it maximum} star formation rate that is consistent with the observed data, and to provide minimum and maximum estimates of the luminosity of the AGN-heated, dust.     A third SED component in the optical is used to estimate stellar mass, and we assume the AGN is completely obscured in the optical-UV.   The results are presented in Tables \ref{tbl-lum}--\ref{tbl-sfr} and illustrated in Figure \ref{fig-fits1} (which is continued in the Appendix).     A more thorough analysis of the SEDs using radiative transfer models for the full sample is in preparation by A. Efstathiou et al.

The method fits a parametrized ``torus'' model to the MIR data and a modified blackbody (BB) to the longer wavelength SED.    The ``torus'' model could describe a classical torus, or some other dusty structure heated by the AGN,  including a spherical ``cocoon'' with 100\% covering factor of warm dust, ${\Omega}_{\rm WD}$.  For the BB component the characteristic dust temperature is undefined for most sources, therefore we construct four models with different fixed dust temperatures.  Modeling a range of dust temperatures within an individual source is not justified by the available data points.  We do not expect all sources to be fitted well for each of these temperature choices, particularly the higher values, and we carry forward into the analysis only those SED fits that are viable.   

The methodology follows \citet{sajina12} but is simplified from their four dust components to only two, due to the limited SED data available for our sample.   The AGN MIR emission is modeled with a parameterization which is consistent with the clumpy torus models of \citet{nenkova08}.  It has the functional form: 

\[ f_{\nu} = \frac{\nu}{(\frac{\nu}{\nu_0})^{\alpha}e^{0.5\nu} + (\frac{\nu}{\nu_0})^{-0.5}+(\frac{\nu}{\nu_0})^{-3}} \ \ (2)\]

where $\alpha$ and $\nu_0$ are free parameters.   The MIR fit is phenomenological and so it is  not characterized by specific values for the dust temperature range, orientation, torus size or optical depth.     We refer to this component henceforce as the ``AGN'' component.  We emphasize that  it could represent a structure of different shape than a classical torus, such as a more spherical cocoon, or a dusty NLR or polar wind.  We also do not rule out in our later discussion that some of this warm emission could be contributed by a young compact starburst.   

The modified blackbody component has a fixed dust temperature and emissivity, $\beta$.    With only the single ALMA data point long ward of 22 $\mu$m for most sources, it is impossible to fit the long wavelength portion of the SED using free parameters, therefore a single greybody is the best approximation. This component may represent dust heated by star formation, or dust heated by the AGN that is cooler than the ``AGN component'', and of course would in reality it have a wide range of dust temperatures compared to the single value used here.   The fit to the MIR region of the spectrum with the AGN component varies very little as the BB component temperature is changed, because the AGN component it is very well defined by the WISE data on the short wavelength side.   For the four sources without measured redshift we assume $z = 2$.

For the stellar luminosity and mass we use either a 100 or 600 Myr stellar population from \citet{maraston05}.   
This is constrained only by the $R$-band data point in most cases, except for the two SDSS-detected sources, and the shape of the MIR SED at the shortest wavelengths.  

We have made four fits to each source, each with a different temperature for the BB component.  We have allowed the temperatures to take on a wide range so that we can interpret the luminosity as arising from disk-like distributed star formation (the coolest dust temperature, 30 K); a starburst similar to those found in local LIRGs and ULIRGs (50 K) \citep{melbourne12,bendo15}; and two additional warmer temperatures that might be appropriate for very young and compact starbursts and/or for additional AGN-heated dust \citep{wilson14}.    For $\beta$ we have selected  a value of 1.5, consistent with the range of values found in the literature.  For the 50K model only, we instead used a value of $\beta$=2, consistent with the largest values found in the literature, in order to illustrate that the uncertainties due to the unknown dust temperature distribution in each source far outweigh the effect of the choice of $\beta$.  This may be seen by comparing the 30 K   and 50 K model fits in Figure \ref{fig-fits1}.  We do not expect any of these fits to be unique for any of our sources; instead we use them to constrain the range of plausible AGN and starburst luminosities.

\begin{figure*}[!ht]
	\centering
	\includegraphics[trim=35 255 40 240,clip,width=0.9\textwidth]{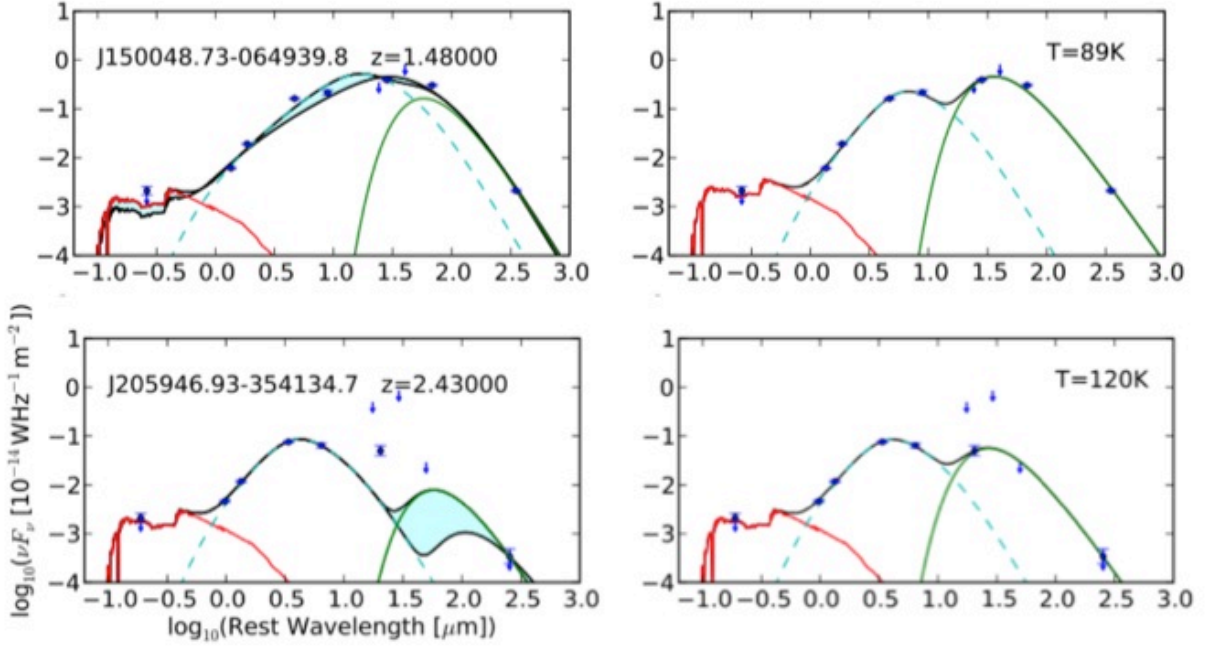}
\caption{SED models for the two sources for which we have sufficient data near the peak of the SED, from $Herschel$, to obtain well constrained fits across the MIR and FIR: W1500-0649 and W2059-3541.    ``AGN'' (dashed cyan line); modified blackbody ``BB'' dust component (green lines); stellar population (red lines); summed model (black line).  For the AGN and stellar components, and for the total fits, only one fit is shown in each panel, for clarity.   The ALMA data point is the right-most point, and the limits near the peak of the cool component are 60 $\mu$m and 100 $\mu$m IRAS data.  {\bf Left}: models with $T_d$ = 50 K, $\beta$ = 2 (upper) and $T_d$ = 30 K, $\beta$ = 1.5 (lower).   {\bf Right}:  models in which the temperature of the BB dust component is allowed to float, for these two sources only, with $\beta$ = 1.5.   \label{fig-tfree}} 
\end{figure*}

To select appropriate dust temperatures for the two warmer BB models, we ran a set of models where $T_d$ was allowed to be a free parameter for the two sources which have $Herschel$ data, and the results are shown in Figure \ref{fig-tfree}-{\it right}, compared to the 30 K and 50 K fixed-T fits in the left panels.      For W1500-0649, which was fitted in the previous section with radiative transfer models, there are two plausible fits.  The 50 K BB model is reasonable, although the IRAS 60 $\mu$m limit is slightly exceeded.  This fit requires a large fraction of the FIR emission to be explained by the warm AGN component (dashed cyan line).   The best fit free-$T_d$ model has $T_d$ = 89 K and provides a total MIR-submm  luminosity of  $\log (L_{\rm Total}/L_{\odot}) = 13.52$, and a division between the AGN and BB components of 1.17 and 2.14 $\times 10^{13} L_{\odot}$.    This is very similar result to the radiative transfer model.   For W2059-3541 a dust temperature greater than 50 K is required to match the 70 $\mu$m PACS data.   The best-fit $T_d$--free model has $T_d = 120$ K, with a total MIR-submm luminosity of log $L_{\rm bol}/L_{\odot} <$ 13.33 (the 870$\mu$m measurement is a non-detection for this source), assuming $\beta$=1.5.   

For the remainder of the sample, for which free-$T_d$ fits are insufficiently constrained, we  have adopted the two dust temperature values which have fitted W1500-0649 and W2059-3541 successfully: 90 K (rounding up from 89 K) and 120 K, for the two warmest of the 4 BB models.   The two fits with the cooler dust temperatures, 30 K and 50 K, are shown in the left hand panels of Figure \ref{fig-fits1}, while the fits with the two warmer temperatures, 90 K and 120 K, are shown in the right hand panels.  The results are tabulated in Table \ref{tbl-lum}.    Since the AGN luminosities do not vary much between the four models, we list only the minimum and maximum AGN component luminosities in columns 3 and 4 of Table \ref{tbl-lum}: $L_{\rm AGN-Min}$ and L$_{\rm AGN-Max}$.  

Many of the BB fits are not successful in fitting all the data points; in particular the 120 K fits exceed the IRAS limits for many sources.  Also the stellar component is not always fitted well by the code.  We have not attempted to refine the fitting procedures to improve this situation because the current optical photometry is inadequate for this purpose.    Only fits judged to be acceptable are tabulated and used in the subsequent analysis.    

\begin{figure}[!ht]
	\centering
	\includegraphics[trim=35 130 50 150,clip,width=0.5\textwidth]{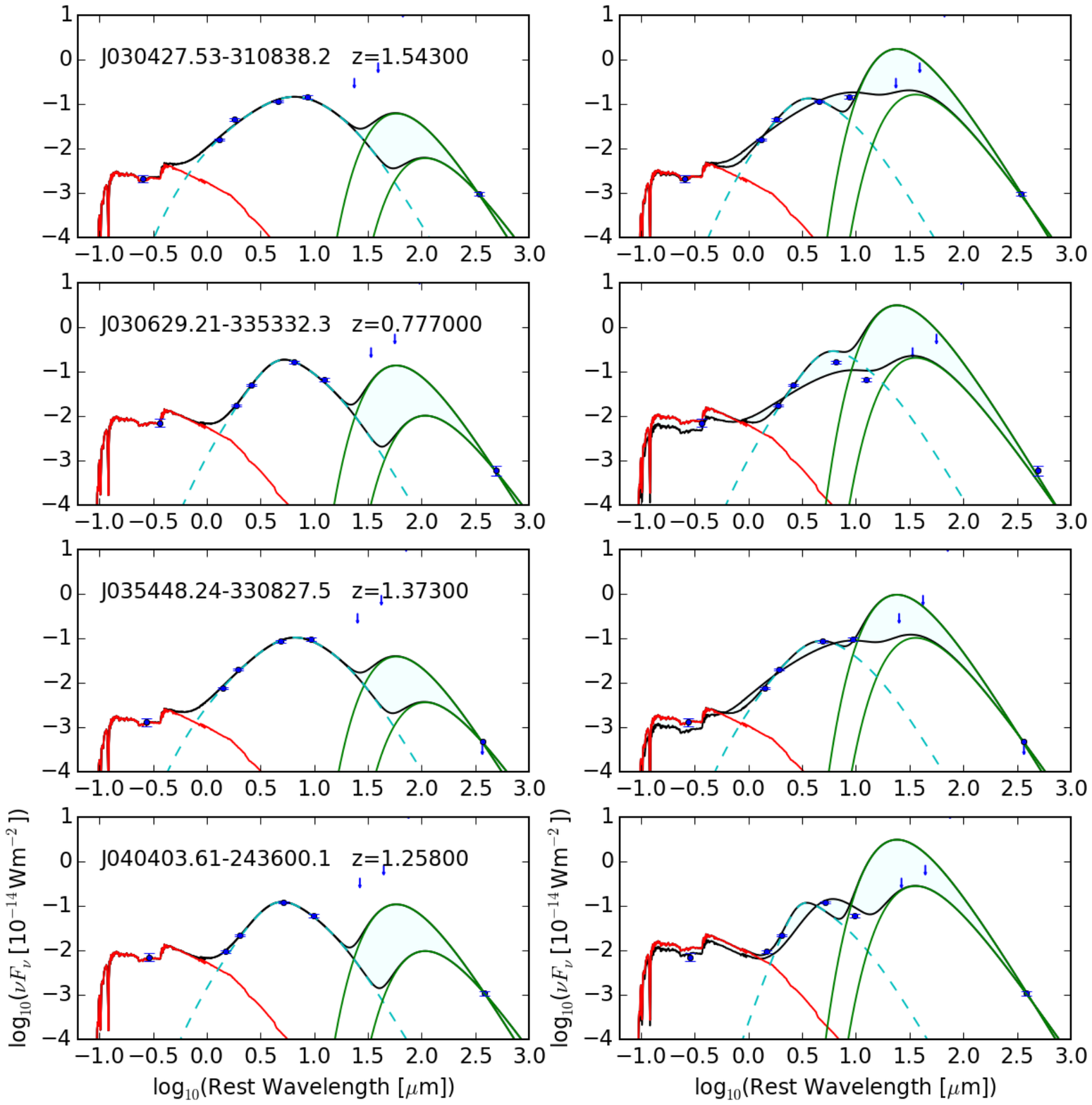}
	\caption{Example of the SED models for each source: data and fits as in Figure \ref{fig-tfree}.    The full set of SED fits is shown in the appendix.  Many of the BB models do not provide good fits, and these are not considered further in our analysis.  In some cases no consistent fit is found for the stellar component.  {\bf Left}: models with $T_d$ = 50 K, $\beta$ = 2 (upper) and $T_d$ = 30 K, $\beta$ = 1.5 (lower).   {\bf Right}:   models with 90 K (lower) and 120 K (upper), with $\beta$ = 1.5; temperature choices based on the results for W1500-0649 and W2059-3541 in Figure \ref{fig-tfree} (continued in Appendix). \label{fig-fits1}}
\end{figure}

For the two warmer BB models, the IRAS 60 $\mu$m limit (and occasionally the 100 $\mu$m limit) is helpful in constraining the dust temperature of the fits, for many sources.       Four sources fail to find a reasonable fit for the IRAS limits with $T_d$ = 90 K, while for 29 sources, the IRAS 60 $\mu$m limit rules out the 120 K model.    In one case, W1500-0649, the 120 K model is ruled out not by IRAS but by ALMA and $Herschel$.    In some cases the low temperature BB models fail to allow the overall model to fit the WISE data; these sources are better fit with one or both of the two warmer models.  

We derive star formation rates from the BB components of the models using the Kennicutt  (1998) conversion from far-infrared luminosity to star formation rate: 
\[ {\rm SFR} = 4.4 \ 10^{-37} L_{\rm BB} {\rm (W)} \ M_{\odot} \ {\rm yr^{-1}} \ \ (3)\]   The SFR results are  given in Table \ref{tbl-sfr}, columns 3--6, for those models which achieved successful fits.      
 
We also list in Table \ref{tbl-lum} columns 9 and 10 the minimum and maximum summed luminosity from the up to 4 viable $L_{\rm Total}$ ($= L_{\rm AGN} + L_{\rm BB}$) models, $L_{\rm Total-Min}$ and $L_{\rm Total-Max}$, for each source.   In addition we list $L_{\rm Total-Best}$ in column 11, which represents the $L_{\rm Total}$ fit that best resembles the SED shapes of the two sources with well fitted SEDs: W1500-0649 and W2059-3541, and the Hot DOGs with well-sampled SEDs (Figure \ref{fig-seds}).   The best fit model is indicated in column 12.    In some cases the best match is derived from the average of two of the $L_{\rm Total}$ models.   

The luminosities derived for the AGN from the AGN model component range from $ \log (L_{\rm AGN-Min}/L_{\odot}) = 11.58$ to $ \log( L_{\rm AGN-Max}/L_{\odot}) = 13.61$.      We also consider a maximum AGN luminosity derived from the Total-Best models:  $\log (L_{\rm Total-Best} / L_{\odot}) = 12.01$--14.15.   The BB component luminosity depends strongly on assumed dust temperature.  Considering all acceptable fits to the SEDs for all of the 870 $\mu$m-detected sources, the acceptable range in $\log (L_{\rm BB} / L_{\odot})$  is 10.89--14.17, and the corresponding SFR range is 13.5--25700 $M_{\odot}$ ${\rm yr}^{-1}$.   The total summed AGN+BB luminosity range is $\log (L_{\rm Total} / L_{\odot}) = 11.71$--14.24.  These results are summarized in Table \ref{tbl-range}.

\subsection{Stellar Masses}
The stellar population fit can be used to constrain the host galaxy mass from the rest frame H-band absolute magnitude of the fitted stellar component.    The host galaxy mass has only a small dependence on the stellar population selected by the fitting code.    The masses do not depend on the dust temperature assumed for the  BB dust component because the optical fit is dominated by the AGN component fitted to the WISE data.    In a few cases, a self-consistent fit could not be obtained, indicating some possible confusion in the $R$-band data point.   The H-band absolute magnitudes are upper limits in more than half the sample since there is no $R$-band measurement or detection, therefore most of the stellar mass estimates in Columns 11--12 of Table \ref{tbl-mass} are also upper limits.   The stellar masses derived from the models range from $\log (M_{\rm stars}/M_{\odot})=9.83-11.09$ for $R$-band detected sources.  The upper limits ranges from $\log (M_{\rm stars}/M_{\odot})<9.82$ to $<$11.34.

The modeled intrinsic H-band absolute magnitudes and the derived stellar masses are highly uncertain and should be viewed as indicative only.   We have no measurement of extinction for the optical or WISE data points, and no color information to help constrain the stellar populations.  We also have little information on morphology, and therefore on the stellar components included in the flux density.    It is also possible that some of the emission is scattered AGN light.    The fitting of the stellar component will be greatly improved by the use of the J and K$_s$ data therefore we limit discussion of these results in this paper.  

\begin{centering}
\begin{deluxetable*}{lccccccccc}[t]
\tablecolumns{10}
\tablewidth{0pc}
\tabletypesize{\tiny}
\tablecaption{Minimum and Maximum Model Range in Luminosity, Mass, SFR \& Accretion Rate ($\dot{M}$)\label{tbl-range}}
\tablehead{
\colhead{}		&	 \colhead{AGN}	&		\colhead{BB}	&	\colhead{Total}	&	\colhead{Total-Best}	&	\colhead{BB}	&	\colhead{Total}	&	\colhead{Total-Best}		\\			
\colhead{Parameter}	& & \multicolumn{3}{c}{----------------870$\mu$m Detected---------------}	&	\multicolumn{3}{c}{--------------870$\mu$m Not  Detected----------------} 	
}
\startdata
Min Value  &W0652-2006&W0612-0622&W0652-2006	&	W0612-0622	&	W2000-2802	&	W0702-2808	&	W0702-2808		\\	
Max Value  	&	W1521+0017	&		W0439-2159	&	W0439-2159	&	W2021-2611	&	W0525-3225	&	W1702-3225	&	W1702-3225		\\
Redshift   	&	0.60--2.63	&	0.47--2.82	&	0.60--2.82	&	0.47--2.28	&	2.28--1.69	&	0.94--2.85	&	0.94 --2.85	\\		
log $L$ (L$_{\odot}$)  &	11.58--13.61	&		10.89--14.17	&	11.71--14.24&12.07--14.15&$<$10.72--$<$13.75&$<$12.09--$<$12.46&$<$12.41--$<$13.81		\\		
SFR (M$_{\odot}$/yr)&	\nodata	&		13.5--25700	&	22--25700 	&	\nodata\tablenotemark{a}	&	$<$9--$<$9800	&	$<$22--$<$4360	&	\nodata\tablenotemark{a}		\\		
log $M_{\rm BH}$ (M$_{\odot}$)\tablenotemark{c}  & 7.66--9.69 &  \nodata & \nodata\tablenotemark{b} & 8.11--10.23 & \nodata & \nodata\tablenotemark{b} & $<$8.49--$<$9.89 \\
$\dot{M}$ (M$_{\odot}$/yr)\tablenotemark{c} & 0.24--26 & \nodata & \nodata\tablenotemark{b} & 0.7--90 & \nodata & \nodata\tablenotemark{b} & $<$1.6--$<$41 \\
\enddata
\tablenotetext{a}{No star formation is present in the Total-Best model by definition, since it is designed to resemble the intrinsic AGN SEDs.}
\tablenotetext{b}{Some Total=AGN+BB fits are inconsistent with a torus-like or intrinsic AGN SED shape; the Total-Best model is the preferable maximal fit for an AGN.}
\tablenotetext{c}{$M_{\rm BH}$ and accretion rate are directly proportional to the luminosity since we have assumed a fixed Eddington ratio and accretion efficiency.}
\end{deluxetable*}
\end{centering}

\subsection{Radio Powers, Radio Loudness and the q Parameter}\label{section-radio}
The rest frame 3 GHz radio powers in column 13 of Table \ref{tbl-lum} are found to lie in the range $\log$  $(P_{\rm 3 GHz}/W Hz^{-1})$ = 24.74--27.33, adopting a power law spectral index $\alpha = -1.0$ ($f_\nu \propto \nu^{\alpha}$) for the $k$-correction to 3 GHz rest frequency.    The median/mean value is 25.97/26.05.  We also list the rest-frame (i.e. $k$-corrected) $q_{\rm 22}=\log(f_{\rm 22{\mu}m}/f_{\rm 1.4GHz })$  values in column 14.    The MIR $k$-correction for $q$ obviously depends strongly on the assumed rest frame SED.  In our situation this is particularly tricky because the unknown depth of the silicate feature will affect the observed $f_{\rm 22}$ flux density strongly at redshifts near 1.5 (e.g. see Figure \ref{fig-ratioz870}).  We have used the QSO2 (``Torus'') template of \citet{polletta07} because it was successfully fitted by them (their Figure 9) to a very red, Compton thick, Spitzer-selected obscured QSO that is similar in NIR-MIR spectral shape to our sources.     To this we have grafted on the silicate absorption feature from the Arp 220 template of \citet{polletta07}.  We also derived the $k$-correction without the added silicate feature.  All of the templates have been convolved with the WISE 22 $\mu$m filter as a function of redshift by \citet{polletta07}.    Our quasars have comparable $k$-corrected $q_{\rm 22}$ values to the most radio-powerful and most radio-loud sources in the large Spitzer SXDF (Subarau X-ray Deep Field) sample of \citet{ibar08}, and all of our sources would be considered to be moderately to very radio-loud after the $k$-correction, based on this criterion.  This is the case even if the large silicate optical depth is omitted from the template.  This conclusion is also apparent from Figure \ref{fig-seds} in which our quasars lie between the radio-quiet and radio-loud quasar templates, when normalized at 4.6 $\mu$m rest.

\subsection{Black Hole Masses and Accretion Rates} \label{bhmass}
Lacking a high quality spectroscopic indicator of BH masses such as the Mg II emission line width, we derive BH masses and accretion rates from the AGN luminosities.  These three parameters are therefore directly proportional to each other in this work.  We assume an Eddington ratio, ${\lambda}_{\rm Edd}$, of 0.25, which is typical of $z{\sim}2$ quasars \citep{kormendyho13}:

\[L_{\rm Edd} = (4 \pi Gm_pc/{\sigma}_T) M_{\rm BH} = 3.3 \ 10^4 \ M_{\rm BH} \ \ (4)\] 

We assume a covering factor of unity and spherical symmetry, consistent with late stage mergers with heavily obscured nuclei.   The covering factor cannot actually be this high in most sources,  since emission lines from the NLR are visible in many cases; therefore our AGN luminosities and BH masses  may be under-estimates due to missing emission in the X-ray to optical range which is not absorbed by the dust.    We present the results  in Table \ref{tbl-mass}, columns 3--5, using our $L_{\rm AGN-Min}$, $L_{\rm AGN-Max}$, and $L_{\rm Total-Best}$ estimators.    The median/mean BH mass values are 1.0/1.15, 1.07/1.45, and 1.55/2.75 ${\times}10{^9} M_{\odot}$ respectively.   In columns 6-7 we also list the BH masses derived by \citet{kim13}, based on the [\ion{O}{3}]${\lambda} 5007\AA$ line luminosity.

The accretion rates are derived from the AGN luminosities assuming an efficiency for the conversion of matter into radiant energy of $\epsilon$ = 0.1 \citep{heckmanbest14}:

\[L_{\rm bol}=\epsilon \dot{M}c^2 \ \ (5)\]

The rates are listed in columns 8-10 of Table \ref{tbl-mass}, and range from 0.24 to 25.3 $M_{\odot}{\rm yr}^{-1}$, with median/mean values of 6.1/5.6, 7.6/6.2 and 14.5/9.1 M$_{\odot}{\rm yr}^{-1}$ respectively for the AGN-Min, AGN-Max and Total-Best values.   The overall ranges in BH mass and accretion rate are listed in Table \ref{tbl-range}.

\begin{centering}
\begin{deluxetable*}{lcllrccllrrrrr}[t]
\tabletypesize{\tiny}
\tablecolumns{14}
\tablewidth{0pc}
\tablecaption{Masses and Accretion Rates \label{tbl-mass}}
\tablehead{
\colhead{WISE Name} & \colhead{Redshift} & \multicolumn{3}{c}{log M$_{\rm BH}$}  & \multicolumn{2}{c}{log M$_{\rm BH:Kim,{\it etal.}}$} &  \multicolumn{3}{c}{Accretion Rates} & \colhead{Absolute} & \colhead{log M$_{\rm stars}$} & \multicolumn{2}{c}{log M$_{\rm ISM}$} \\
\colhead{}   & \colhead{} & \multicolumn{3}{c}{(M$_{\odot}$)} & \multicolumn{2}{c}{(M$_{\odot}$)}&  \multicolumn{3}{c}{(M$_{\odot}/yr$)} & \colhead{H-} & \colhead{(M$_{\odot}$)} & \multicolumn{2}{c}{(M$_{\odot}$)}  \\
\colhead{} & \colhead{} & \colhead{Min} & \colhead{Max} & \colhead{Total-} & \colhead{Lower} & \colhead{Upper} & \colhead{Min} & \colhead{Max} & \colhead{Total-} & \colhead{Magnitude} & \colhead{} & \colhead{30K} & \colhead{90K} \\
\colhead{} & \colhead{} & \colhead{} & \colhead{} & \colhead{Best} & \colhead{} & \colhead{} & \colhead{} & \colhead{} & \colhead{Best} & \colhead{} & \colhead{} & \colhead{} & \colhead{} 
}
\startdata
W0304-3108 &  1.54&       9.09&       9.24&           9.44&     \nodata & \nodata &       6.52   &     9.20&          14.50&   -24.5 &       10.6&      10.83   &     10.20   \\
W0306-3353 &  0.78&       8.27&       8.31&           8.61&    \nodata & \nodata     &       0.99    &    1.08&          2.18&    -23.9 &      10.4&      10.51   &     9.96    \\
W0354-3308 &  1.37&       8.78&       8.80&        $<$9.04&     \nodata & \nodata     &       3.19     &   3.34&        $<$5.82&   -23.6 &       10.2&   $<$10.52   &  $<$9.91    \\
W0404-2436 &  1.26&       8.66&       8.71&           9.19&    \nodata & \nodata     &       2.42     &   2.72&          8.14&    -25.1 &      10.8&      10.86   &     10.26    \\
W0409-1837 &  0.67&       8.81&       8.82&        $<$8.98&    \nodata & \nodata     &       3.42      &  3.50&        $<$5.10&    -22.6 &       9.8&      $<$10.44   &     $<$9.90    \\
W0417-2816 &  0.94&       8.03&       8.09&           8.55&       9.06&       9.72&       0.57   &     0.65&          1.87&  -22.9 &         9.9&      10.63   &     10.06    \\
W0439-3159 &  2.82&       9.52&       9.53&           9.78&    \nodata & \nodata     &       17.54     &  17.95&         31.56&    -23.6 &      10.3&      11.15   &     10.38    \\
W0519-0813 &  2.05&       9.09&       9.10&        $<$9.26&    \nodata & \nodata     &       6.52      &  6.67&        $<$9.72&    $>$-23.1& $<$10.0&   $<$10.53   &  $<$9.85    \\
W0525-3614 &  1.69&       8.59&       8.68&        $<$9.02&    \nodata & \nodata     &       2.06     &   2.54&        $<$5.58&    -23.5 &    10.2&   $<$10.59   &  $<$9.94    \\
W0526-3225 &  1.98&       9.52&       9.62&           10.03&     \nodata & \nodata     &      17.54    &   22.08&         56.80&  -24.1 &      10.4&      11.64   &     10.96    \\
W0536-2703 &  1.79&       9.13&       9.13&           9.38& \nodata & \nodata    &       7.14      &  7.14&          12.69&    -23.7 &    10.3&      10.82   &     10.16    \\
W0549-3739 &  1.71&       8.64&       8.64&           9.08&     \nodata & \nodata     &       2.31     &   2.31&          6.42&   $>$-22.7 &  $<$9.9&      10.69   &     10.04    \\
W0612-0622 &  0.47&       8.11&       8.12&           8.51&     \nodata & \nodata     &       0.68      &  0.70&          1.71&    -23.8 &      10.2&     10.53   &    10.01    \\
W0613-3407 &  2.18&       9.28&       9.28&        $<$9.42&       9.07&       9.46&       10.09   &    10.09&       $<$14.02&   $>$-22.6 & $<$9.8&   $<$10.64   &  $<$9.94    \\
W0614-0936 &  \nodata\tablenotemark{a}&       9.02&       9.03&        $<$9.25&    \nodata & \nodata     &       5.55    &    5.68&        $<$9.51&   -22.9 &     $<$9.9&   $<$10.64   &  $<$9.96    \\
W0630-2120 &  1.44&       8.60&       8.65&           8.90&     \nodata & \nodata     &       2.11      &  2.37&          4.17&   -24.0 &       10.4&      11.08   &     10.46    \\
W0642-2728 &  1.34&       8.44&       8.44&           8.74&     \nodata & \nodata     &       1.46      &  1.46&          2.88&    -24.1 &      10.4&      10.72   &     10.11    \\
W0652-2006 &  0.60&       7.66&       7.67&           8.20&       8.85&       9.61&       0.24   &     0.25&          0.84&  $>$-21.3 &    $<$9.2&      10.70   &     10.16    \\
W0702-2808 &  0.94&       8.12&       8.12&        $<$8.49&       8.92&       9.73&       0.70    &    0.70&        $<$1.64&  -21.9 &         9.5&   $<$10.58   &  $<$10.01    \\
W0714-3635 &  0.88&       7.96&       7.97&           8.49&       9.26&       9.34&       0.48     &   0.49&          1.64&   -21.7 &        9.5&      10.69   &     10.13    \\
W0719-3349 &  1.63&       8.65&       8.65&           9.09&    \nodata & \nodata     &       2.37    &    2.37&          6.47&   -22.5 &        9.8&      10.51   &     9.74    \\
W0811-2225 &  1.11&       8.71&       8.71&           $<$9.02&     \nodata & \nodata     &       2.72     &   2.72&          $<$5.56&   -24.0 &       10.4&      $<$10.61   &     $<$10.02    \\
W0823-0624 &  1.75&       9.19&       9.30&        $<$9.43&     \nodata & \nodata     &       8.20      &  10.57&       $<$14.23&  -24.2 &        10.5&   $<$10.64   &  $<$9.99    \\
W1308-3447 &  1.65&       9.07&       9.07&           9.23&    \nodata & \nodata     &       6.22&        6.22&          9.07&  -24.4 &        10.6&      10.53   &     9.89    \\
W1343-1136 &  2.49&       9.08&       9.10&           9.35&    \nodata & \nodata     &       6.37   &     6.67&          11.81&   -25.7 &       11.1&      10.74   &     10.00    \\
W1400-2919 &  1.67&       9.18&       9.19&        $<$9.51&     \nodata & \nodata     &       8.02     &   8.20&        $<$17.22&   -24.8 &       10.7&   $<$10.09   &  $<$9.44    \\
W1412-2020 &  1.82&       9.10&       9.25&           9.44&     \nodata & \nodata     &       6.67       & 9.42&          14.72&  $>$-24.7  &  $<$10.7&      10.80   &     10.14    \\
W1434-0235 &  1.92&       8.97&       8.97&           $<$9.11&     \nodata & \nodata     &       4.94&        4.94&          $<$6.82&   -24.5 &       10.6&      $<$10.34   &     $<$9.67   \\
W1439-3725 &  1.19&       8.37&       8.37&        $<$8.66&     \nodata & \nodata     &       1.24  &      1.24&        $<$2.40&   $>$-23.2 &   $<$10.1&   $<$10.21   &  $<$9.61    \\
W1500-0649 &  1.50&       9.15&       9.58&           9.60&    \nodata & \nodata     &       7.48    &    20.14&         21.10&   $>$-23.2 &  $<$10.1&     11.18   &     10.56    \\
W1510-2203 &  0.95&       8.68&       8.69&        $<$8.84&     \nodata & \nodata     &       2.54      &  2.59&        $<$3.68&   $>$-23.1 &  $<$10.0&  $<$10.23   &  $<$9.66    \\
W1513-2210 &  2.20&       9.34&       9.34&           9.62&     \nodata & \nodata     &       11.59&       11.59&         22.16&   $>$-23.9 &  $<$10.4&     11.07   &     10.36    \\
W1514-3411 &  1.09&       8.52&       8.56&        $<$8.78&     \nodata & \nodata     &       1.75   &     1.92&        $<$3.16&    $>$-23.6&  $<$10.2&  $<$10.25   &  $<$9.67    \\
W1521+0017 &  2.63&       9.69&       9.69&           9.73&    \nodata & \nodata     &       25.94   &    25.94&         28.66&   $>$-26.1 &  $<$11.2&     11.75   &     11.20    \\
W1541-1144 &  1.58&       8.89&       9.02&        9.09&    \nodata & \nodata     &       4.11&        5.55&        6.48&  $>$-23.5 &   $<$10.2&  10.46   &  9.83    \\
W1634-1721 &  2.08&       8.91&       8.91&        $<$9.06&     \nodata & \nodata     &       4.31  &      4.31&        $<$6.14&   $>$-25.1 &  $<$10.8&  $<$10.31   &  $<$9.62    \\
W1641-0548 &  1.84&       9.02&       9.17&           9.38&     \nodata & \nodata     &       5.55    &    7.83&          12.66&  $>$-26.4 &   $<$11.3&     10.75   &     10.09    \\
W1653-0102 &  2.02&       9.08&       9.08&        $<$9.18&     \nodata & \nodata     &       6.37      &  6.37&        $<$8.00&   $>$-25.4 &  $<$11.0&  $<$10.28   &  $<$9.60    \\
W1657-1740 &  \nodata\tablenotemark{a}&       9.24&       9.25&        $<$9.68&     \nodata & \nodata    &       9.20&        9.42&        $<$25.42&   $>$-25.1 &  $<$10.8&  $<$10.28   &  $<$9.60    \\
W1702-0811 &  2.85&       9.68&       9.68&        $<$9.89&     \nodata & \nodata     &       25.35  &     25.35&       $<$41.35&  $>$-25.2 &   $<$10.9&  $<$10.38   &  $<$9.61    \\
W1703-0517 &  1.80&       9.19&       9.59&        9.68&     \nodata & \nodata    &       8.20     &   20.61&       25.62&  $>$-25.1 &   $<$10.8&  9.92   &  9.16     \\
W1707-0939 &  \nodata\tablenotemark{a}&       9.06&       9.12&        $<$9.64&   \nodata & \nodata     &       6.08&        6.98&        $<$23.22&   $>$-25.1 &  $<$10.8&  $<$10.39   &  $<$9.71    \\
W1936-3354 &  2.24&       9.13&       9.26&           9.86&    \nodata & \nodata     &       7.14  &      9.64&          38.33&   $>$-24.5 &   $<$10.6&     10.97   &     10.32    \\
W1951-0420 &  1.58&       8.88&       8.91&        $<$9.62&     \nodata & \nodata     &       4.02    &    4.31&        $<$21.97&   $>$-24.9 &  $<$10.7&  $<$10.40   &  $<$9.76    \\
W1958-0746 &  1.80&       9.06&       9.06&        $<$9.18&       9.53&     \nodata     &       6.08    &    6.08&        $<$8.00&   $>$-24.7 &  $<$10.7&  $<$10.36   &  $<$9.70    \\
W2000-2802 &  2.28&       8.83&       8.98&        $<$9.04&     \nodata & \nodata     &       3.58 &       5.06&        $<$5.81&    $>$-23.6&   $<$10.2&  $<$10.36   &  $<$9.65    \\
W2021-2611 &  2.44&       9.05&       9.59&           10.23&    \nodata & \nodata     &      5.94   &     20.61&         89.96&   $>$-25.4 &  $<$10.9&     11.24   &     10.56    \\
W2040-3904 &  \nodata\tablenotemark{a}&       9.00&       9.23&           9.57&    \nodata & \nodata     &       5.30      &  8.99&          19.81&    $>$-24.8&   $<$10.7&     11.09   &     10.41    \\
W2059-3541 &  2.38&       9.23&       9.45&        $<$9.50&       9.07&       9.10&       8.99   &     14.93&       $<$16.76&   $>$-25.2 &  $<$10.9&  $<$10.35   &  $<$9.62    \\
\enddata
\tablenotetext{a}{Redshift assumed to be 2 if no spectroscopic redshift available.}.
\end{deluxetable*}
\end{centering}

\begin{centering}
\begin{deluxetable*}{llrrrrrrrrrrr}[t]
\tabletypesize{\tiny}
\tablecolumns{11}
\tablewidth{0pt}
\tablecaption{Star Formation Rates and Gas Depletion Timescales \label{tbl-sfr}}
\tablehead{
\colhead{WISE Name} & \colhead{Redshift} & \multicolumn{4}{c}{log SFR} & \multicolumn{3}{c}{log (Gas Depletion Time)} \\\colhead{} & \colhead{} & \multicolumn{4}{c}{(L$_{\odot}$/yr)} &  \multicolumn{3}{c}{(years)} \\
\colhead{}	& \colhead{} & \colhead{30K} & \colhead{50K} & \colhead{90K} & \colhead{120K} &  \colhead{30K} & \colhead{50K} & \colhead{90K}	
}
\startdata
W0304-3108&1.54	&   1.73&   2.70&   3.16&  \nodata &  	   9.10	&	   7.81	&	   7.04	\\				
W0306-3353&0.78	&   1.22&   2.31&   2.52&   \nodata	&	   9.29	&	   7.93	&	   7.44	\\				
W0354-3308&1.37	&$<$1.38&$<$2.38&$<$2.83&   \nodata		&	$<$9.14	&	$<$7.83	&	$<$7.08	\\				
W0404-2436&1.26	&   1.70&   2.72&   3.17&   \nodata	&   	   9.16	&	   7.84	&	   7.09	\\				
W0409-1837&0.67	&   \nodata	&$<$2.23&$<$2.64&   \nodata		&	   \nodata	&	$<$7.94	&	$<$7.26	\\				
W0417-2816&0.94	&   1.72&   2.79&   3.22&   \nodata	&   	   8.91	&	   7.56	&	   6.84	\\				
W0439-3159&2.82	&   2.28&   3.08&   3.57&   4.41&   8.87	&	   7.66	&	   6.81	\\				
W0519-0813&2.05	&$<$1.54&$<$2.44&$<$2.92&$<$3.82&	$<$8.99	&	$<$7.74	&	$<$6.93	\\				
W0525-3614&1.69	&$<$1.53&$<$2.49&$<$2.95&$<$3.99&	$<$9.06	&	$<$7.77	&	$<$6.99	\\				
W0526-3225&1.98	&   2.64&   3.55&   4.03&   \nodata	&   9.00 	&	   7.74	&	   6.93	\\				
W0536-2703&1.79	&   \nodata  	&   \nodata	&   3.18&   \nodata	&    	   \nodata	&	   \nodata	&	   6.98	\\				
W0549-3739&1.71	&   \nodata	&   2.58&   3.05&   \nodata	&   	   9.06	&	   7.78	&	   6.99	\\				
W0612-0622&0.47 &   1.13&   2.29&   2.00&\nodata    &             9.40 &          7.99	&          7.57 \\
W0613-3407&2.18	&$<$1.67&$<$2.55&$<$3.03&$<$3.92&	$<$8.97	&	$<$7.73	&	$<$6.91	\\				
W0614-0936&\nodata \tablenotemark{a}&$<$1.64&$<$2.55&$<$3.02&$<$3.93&	$<$9.00	&	$<$7.74	&	$<$6.94	\\		
W0630-2120&1.44	&   1.96&   2.95&   3.41&   \nodata	&   9.12	&	   7.82	&	   7.05	\\				
W0642-2728&1.34	&   1.58&   2.59&   \nodata	&   \nodata	&  	   9.14	&	   7.82	&	   \nodata	\\				
W0652-2006&0.60	&   1.35&   2.48&   \nodata	&   \nodata		&	   9.35	&	   7.96	&	   \nodata	\\				
W0702-2808&0.94	&    $<$1.34 & $<$2.41 &  \nodata   	&   \nodata	&	$<$9.24	&	$<$7.89	&          \nodata	\\				
W0714-3635&0.88	&   1.42&   2.50&   \nodata	&   \nodata	&	   9.27	&	   7.91	&	   \nodata	\\				
W0719-3349&1.63	&   \nodata	&   2.99&   3.46&   \nodata		&	   \nodata	&	   7.11	&	   6.28	\\			
W0811-2225&1.11	&   \nodata	&   \nodata	&  $<$2.89&   \nodata	&	   \nodata	&	   \nodata	&	  $<$7.13	\\				
W0823-0624&1.75	&$<$1.60&$<$2.54&$<$3.00&   \nodata		&	$<$9.04	&	$<$7.77	&	$<$6.99	\\				
W1308-3447&1.65	&   1.46&   2.42&   2.89&   \nodata		&	   9.07	&	   7.78	&	   7.00	\\				
W1343-1136&2.49	&   \nodata	&   2.67&   3.16&   4.02&     \nodata 	 &  7.69	&	   6.84	\\				
W1400-2919&1.67	&$<$1.02&$<$1.97&   \nodata	&$<$3.40&	$<$9.07	&	$<$7.79	&	   \nodata	\\				
W1412-2020&1.82	&   1.76&   2.69&   3.16&   \nodata	&   	   9.04	&	   7.77	&	   6.98	\\				
W1434-0235&1.92	&   $<$1.32&   $<$2.24&   $<$2.71&   $<$3.63&   	   $<$9.02	&	   $<$7.75	&	   $<$6.96	\\				
W1439-3725&1.19	&$<$1.03&$<$2.05&$<$2.50&   \nodata	&	$<$9.18	&	$<$7.86	&	$<$7.11	\\				
W1500-0649&1.50	&   \nodata	&   3.08&   3.57&   \nodata	&     \nodata	&	   7.79	&	   6.99	\\				
W1510-2203&0.95	&$<$0.98&$<$2.05&$<$2.47&   \nodata	&	$<$9.25	&	$<$7.90	&	$<$7.19	\\			
W1513-2210&2.20	&   2.10&   2.98&   3.46&   \nodata	&   	   8.97	&	   7.72	&	   6.90	\\				
W1514-3411&1.09	&$<$1.05&$<$2.10&$<$2.53&   \nodata	&	$<$9.20	&	$<$7.86	&	$<$7.14	\\				
W1521{\tiny +}0017&2.63	&   \nodata	&   \nodata	&   2.87&   \nodata	&    	   \nodata	&	   \nodata	&	   8.33	\\				
W1541-1144&1.58	&1.37&2.34&2.81&   \nodata	&	9.09	&	7.80	&	7.02	\\				
W1634-1721&2.08	&$<$1.32&$<$2.22&$<$2.70&$<$3.60&	$<$8.99	&	$<$7.73	&	$<$6.92	\\				
W1641-0548&1.84	&   \nodata	&   2.65&   3.12&   \nodata		&	   \nodata	&	   7.76	&	   6.97	\\				
W1653-0102&2.02	&$<$1.28&$<$2.18&$<$2.65&$<$3.55&	$<$9.00	&	$<$7.75	&	$<$6.95	\\				
W1657-1740&\nodata \tablenotemark{a}	&$<$1.27&$<$2.18&$<$2.66&$<$3.56&	$<$9.01	&	$<$7.75	&	$<$6.94	\\		
W1702-0811&2.85	&$<$1.52&$<$2.31&   \nodata	&$<$3.64&	$<$8.86	&	$<$7.66	&	   \nodata	\\				
W1703-0517&1.80	&1.36&2.29&2.72&3.69&	8.56	&	7.23	&	6.44	\\				
W1707-0939&\nodata \tablenotemark{a}	&$<$1.39&$<$2.30&$<$2.78&$<$3.67&	$<$9.00	&	$<$7.74	&	$<$6.93	\\		
W1936-3354&2.24	&   1.69&   2.56&   \nodata	&   3.93&   	   9.28	&	   8.08	&	   \nodata	\\				
W1951-0420&1.58	&$<$1.32&$<$2.29&   \nodata	&	$>$5.41	&	$<$9.08	&	$<$7.79	&	   \nodata	\\				
W1958-0746&1.80	&$<$1.32&$<$2.25&$<$2.72&$<$3.65&	$<$9.04	&	$<$7.77	&	$<$6.98	\\				
W2000-2802&2.28	&$<$0.96&$<$1.83&$<$2.31&$<$3.20&	$<$9.40	&	$<$8.16	&	$<$7.34	\\				
W2021-2611&2.44	&   2.09&   2.92&   \nodata	&   4.28&   	   9.15	&	   7.97	&	   0.00	\\				
W2040-3904&\nodata \tablenotemark{a}	&   2.09&   3.00&   3.47&   \nodata	&   	   9.00	&	   6.94	&	   \nodata	\\				
W2059-3541&2.38	&   \nodata	&   \nodata	&$<$2.70&$<$3.11&	   \nodata	&	   \nodata	&	$<$6.92		
\enddata
\tablenotetext{a}{$z$=2 assumed for sources without a spectroscopic redshift.}
\end{deluxetable*}
\end{centering}

\section{Discussion}\label{section-discussion}

Our goals are to search for the most luminous obscured quasars at redshifts $\sim$1--3, the peak epoch of massive BH building, which are in the process of quenching star formation, and to investigate the role of radio jets in that process.  We are specifically interested in the kinetic role of moderate to high power jets on the ISM within the galaxy host, while the AGN is still accreting strongly in ``quasar-mode''.  This is in contrast to the role of jets in typical ``radio-mode'' AGN, which are thought to be accreting at low rates and to have a role in maintaining galaxies free of infalling gas.   As outlined in Section \ref{section-introduction}, models which take the porosity of the ISM into account show that high power jets could be quite effective in a dense dusty environment, such as found the central regions of major mergers \citep{wagner11,wagner12}.

We have shown that WISE has found sources that are steeper (redder)  in the rest-frame 1-10$\mu$m range than most known $Spitzer$-selected red sources, including the so-called DOGs.   The two other samples of very red WISE sources, the Hot DOGs and the WISE Lyman Alpha Blobs (WLABs) (EWB12) which were selected without regard for radio emission, have extremely red rest-frame 1-10 $\mu$m SEDs which are very similar to those our our radio-selected sample.  The radio-blind samples have higher average redshifts than our radio powerful sample, probably as a result of the different WISE-optical color selection criteria.  We attribute our ability to identify these extremely red sources to the much larger survey volume of WISE compared to $Spitzer$.      These systems are candidate luminous quasar-mode AGN which are highly obscured, and the radio-bright ones selected here potentially have young jet activity.     Recent X-ray observations of a few Hot DOGs confirm the likelihood of highly buried AGN \citep{stern14,pinconcelli15}.  The hydrogen column for W1835$+$4355 was found to be $N_{\rm H} \ge 10^{\rm 23} \ \rm cm^{-2}$ by  \citet{pinconcelli15}.

We also find that the WISE-NVSS-ALMA sample sources are more strongly dominated by AGN than the $Spitzer$ DOGs.  Star formation rates of hundreds up to a few thousand $M_{\odot}$  yr$^{-1}$ could also be present in some of the systems,  although the  IR-submm SEDs of some of the ALMA sources, in particular those with only upper limits at 345 GHz,  may be consistent  with the AGN torus model in Figure \ref{fig-seds} without any SF contribution.     Therefore these sources are indeed objects in which it is likely that the accretion rates are still very high but the star formation rates are low {\it relative to the accretion power}, and could thus be ideal sources for investigating recent and ongoing quenching by jet-powered AGN feedback.    

Some of our systems could be High Excitation Radio Galaxies (HERGs) seen at high inclination through an optically thick torus, and it is possible that we have included some with lobes that are unresolved by  the 45{\arcsec} NVSS beam.   We will show in the next paper (Carol Lonsdale et al. 2015, in preparation) that our 8--12 GHz VLA data rule out this scenario for most sources, although a small subset of the VLA sample of 156 red obscured quasars does indeed turn out to have large (several Mpc scale), double lobes ($\sim$7\%).   Another small percentage shows evidence of small scale double lobes on scales $\sim$2--10 kpc.

\subsection{The AGN-Heated Source} 
In this work we have derived a range in the plausible AGN luminosities depending on whether we assume that any of the far-infrared luminosity component is AGN-heated:
\[ L_{\rm bol, AGN} = L_{\rm AGN, MIR} + f_{\rm AGN} L_{\rm BB} \]
where $L_{\rm bol, AGN}$ is the bolometric luminosity of the AGN, $L_{\rm AGN, MIR}$ is the mid-infrared AGN luminosity from the fit to Equation (2),  $f_{\rm AGN}$ is the fraction of the far-infrared emission contributed by the AGN and $L_{\rm BB}$ is the far-infrared luminosity.   $f_{\rm AGN}$ is assumed to be 0 for the $L_{\rm AGN}$ models and to be 1 for the scenario in which we assume that $L_{\rm Total-Best}$ is completely AGN-powered.     
The warm AGN component luminosity is very insensitive to the dust temperature of the BB component, due to the strong constraints placed on the AGN model shape by the four WISE data points.   The range of values for the ratio of the maximum plausible AGN luminosity to the minimum estimate is $1.1<L_{\rm Total-Best}/L_{\rm AGN-Min}<5.5$, with a median value of 2.0.        

We have assumed a covering factor of warm ($\sim 300$K) dust that emits in the mid-infrared, $\Omega_{\rm WD}$:

\[ L_{\rm bol, AGN} = L_{\rm AGN, MIR}/\Omega_{\rm WD} \]

of unity and spherical symmetry, consistent with late-stage mergers with heavily-obscured nuclei.    This may not be a valid assumption in which case we may have under-estimated the AGN luminosities and BH masses.    In particular narrow emission lines photoionized by the AGN are seen in many cases, therefore these luminosity estimators may be excluding flux emitted by the AGN that does not intercept the dusty structure.     A further complication is that the covering factor of the cold ($\sim 50$K) dust is likely to be different from that of the hot dust, and its optical depth to the mid-infrared emission is unknown.

An added complication is the possibility that a non-spherical source emits non-isotropically because it is optically thick, as is often the case for torus models (eg. Efstathiou et al. 2013).  In that case a further correction for anisotropy is required, and this depends on the particular torus model and also on the inclination of the torus.   Generally speaking, for edge-on viewing the luminosity will be underestimated whereas for face-on viewing it may be overestimated.    Face-on viewing is ruled out by the red MIR-optical SEDs and therefore we can conclude that $L_{\rm AGN, MIR}$ provides a firm lower limit to the bolometric luminosity of the AGN.

Support for the high covering factor interpretation for most of our WISE-NVSS-ALMA sample comes from the overall relative numbers of obscured and unobscured AGN amongst the highest luminosity radiative mode AGN at high redshift ($z{\sim}$1--3), which are roughly equal \citep{lacy15,assef14,tsai14}.    Taken as face value this would imply an average covering fraction of $\sim$50\%, if these populations differ only by orientation (ignoring the differing selection functions for type 1 and type 2 AGN, Elitzur 2012).

\subsubsection{A Torus?}

Although we have commented above on the likelihood that the covering factor may be lower than 1.0 for many sources, it might be questionable to conclude that a classical smooth torus could be responsible for the very high luminosities found for some sources, because they would require very large tori (several hundred pc to over a kpc in diameter), especially if the $L_{\rm Total-Best}$ values are interpreted as fully AGN-powered.   Such large thin structures would be unstable.      A clumpy torus would be more plausible, as it could achieve a wider range of dust temperatures than a smooth torus of the same diameter.    

A torus-like structure would imply that a large fraction of the total AGN luminosity escapes dust absorption and would be easily visible at optical wavelengths.  Since the line-of-sight optical continuum emission is very faint compared to type 1 AGN (Figure \ref{fig-seds}) the tori would all have to be inclined closely to our line of sight.   Our selection function of course has favored the selection of highly obscured systems, and we do indeed find a fraction of our sample to have (small) double radio lobes in the plane of the sky (Section \ref{section-radio}),  as might be expected for a radio quasar or radio galaxy interpreted as viewed through an edge-on disk in the standard unification picture.   The majority of our sample does not display extended radio lobes however.  

\subsubsection{Extended NLRs, Polar Winds and Lyman-$\alpha$ Blobs}
High resolution observations of nearby low luminosity AGN show that the MIR emission can lie in the polar direction and may be associated with NLR clouds or with thermal winds from the AGN (Zhang et al. 2013 and references therein).    It is also known that a significant fraction of FRII radio AGN have extended emission line regions (EELRs) up to tens of kpc in size \citep{fustockton07}, which are outflowing \citep{shihstockton14}.   The Compact-Steep-Spectrum (CSS) sources, which are younger versions of the FR IIs, have smaller EELRs which are better aligned with the radio structures than in FR IIs \citep{axon00}.      Therefore it is possible that some of the MIR emission in our sample is associated with NLR or EELR clouds which are heated by the central AGN, or shock-heated by the radio jet interactions \citep{mullaney13}.    \citet{efstathiou13} have found that a dusty NLR component is needed to fit the MIR-FIR SED of the hyperluminous galaxy IRAS10214+4724 at $z=2.285$.

\citet{bridge13} have discovered that a subset of the Hot DOGs  which have extended faint optical emission also possess Lyman Alpha Blobs (LABs; defined to have Ly$\alpha$ emission  extended on scales $>$ 30 kpc).  The WISE LABs are non-symmetric and there is evidence for large outflow velocities.   
The presence of extended ionized gas suggests significant shock heating or that a significant fraction of the nuclear ionizing radiation must be able to find its way out of the galaxy, as in the radio galaxy EELRs, even in these highly obscured MIR-dominated systems.    
Few of our sources currently have spectra that cover Ly$\alpha$.   W0613-3407 has Ly$\alpha$ extended spatially on a scale of ~3\arcsec, so qualifies as an LAB by the size definition.    W1343-1136 appears to also have extended emission (about 2.5\arcsec; 25 kpc) but to  just miss the usual LAB definition of $>$30 kpc.   Although it remains to be seen whether a large fraction of our WISE-NVSS-ALMA sources possess LABs, we have some evidence that some of them possess broad forbidden emission lines that might  indicate substantial outflows.    \citet{kim13} find that the [\ion{O}{3}] lines are exceptionally broad for six of our quasars, with full width at half maximum $\sim$1300 to 2100 km s$^{-1}$, significantly larger than that of typical distant quasars.

\subsection{Star Formation}
The contribution to the 1-1000$\mu$m luminosity from star formation has larger overall uncertainty than the AGN contribution because star formation can produce a larger range in observed dust temperature, varying by $\sim$2 dex between the viable 30 K to 120 K BB models (Table \ref{tbl-sfr}, columns 4-6).     The maximum star formation luminosity, L$_{\rm BB-Max}$, is comparable to the the AGN luminosity estimators, while the minimum starburst luminosity, coming from the 30 K model in all but one case, can be well over an order of magnitude smaller than the AGN luminosity.            A minimum SFR of 0 is also possible if L$_{\rm Total-Best}$ is interpreted as completely AGN-powered.  

If we adopt 50K dust for the BB component, the SFRs lie between $\sim$200 and 3500 $M_{\odot}$ yr$^{-1}$. The lower values  are consistent with  main sequence galaxies at these redshifts while the higher values may require a starburst \citep{delvecchio15}.

Our radiative transfer modeling of our best observed source favored the presence of a young (10-15 Myr), presumably compact, starburst contributing about 50\% of the 1--1000$\mu$m luminosity.  Compact starbursts have been investigated in several local galaxies.    The nuclear structures around the twin nuclei of Arp 220 are thought to be powered by compact starbursts with relatively high dust temperatures (over 100 K) and to be highly optically thick out to  $\gg$100 $\mu$m \citep{wilson14,scoville15,barcos15}.  \citet{tsai14} have considered in more detail the possibility that a compact starburst could contribute a significant luminosity to their most luminous Hot DOGs, those with $L_{\rm bol} > 10^{14} L_{\odot}$.   Using He 2-10 as a local analog and the STARBURST99 code \citep{leithererchen09}, they find SFRs $>5{\times}10^{3-4} M_{\odot}$ $\rm yr^{-1}$ for the conservative case of a top heavy IMF.  

\citet{tsai14} conclude that there is insufficient CO in these systems  to support very large amounts of star formation.   In our case we find mean ISM masses of between $\sim$ 2--9 ${\times} 10^{10} M_{\odot}$, depending on the assumed dust temperature.   For the range of estimated SFRs across our models the gas would be depleted in $\sim$ 2 Myr -- 2 Gyr.    The lower depletion times (corresponding to the higher SFRs) are not insupportable, in the scenario of a late-stage, violent, gas-rich merger.  Therefore it is quite possible that a vigorous starburst is present in some of our systems, and that a compact young starburst may contribute to the warmest dust emission.

\subsection{High Accretion Rates Relative to Star Formation}
\begin{figure}[!ht]
	\includegraphics[trim=0 20 0 20,clip,width=0.5\textwidth]{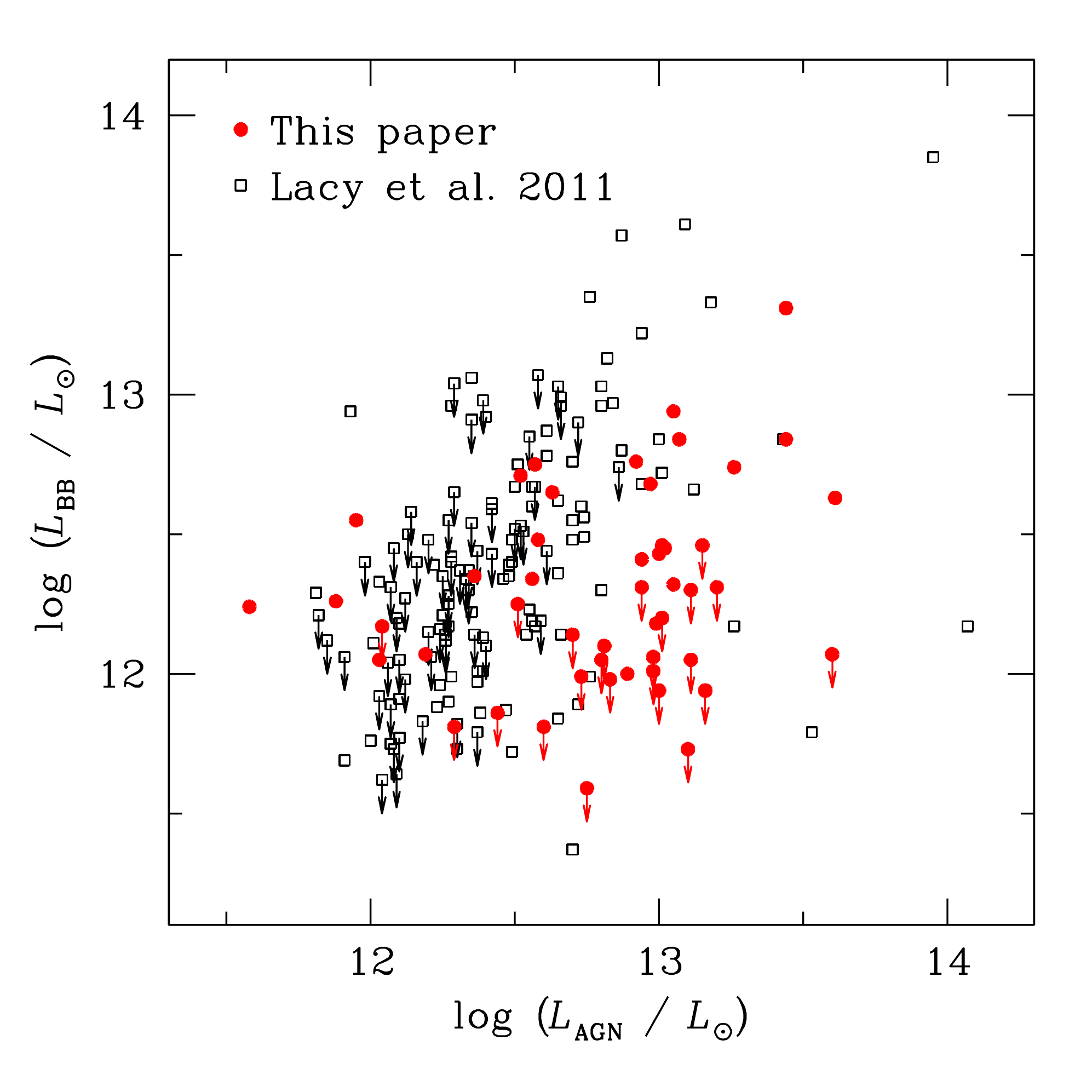}
\caption{Comparison of the $Spitzer$-selected sample of Lacy11 with our sample in the $L_{\rm AGN}$ {\it vs.} $L_{\rm BB}$ plane.   For this figure we show the BB luminosity from our 50K model to match the method used by Lacy11.      \label{fig-lacycomp}}
\end{figure}

In Figure \ref{fig-lacycomp} we plot $L_{\rm AGN-Min}$ vs. $L_{\rm BB-50K}$: no correlation is apparent.  We compare our sample to the radio-quiet $Spitzer$-IRAC-selected quasar sample of \citet{lacy11} (Lacy11), for which fits have been done using the same formalism as here.    The Lacy11 sample has a wider range in mid-IR color selection than our sample, including Type 1 quasars, reddened Type 1 quasars and Type 2 quasars, and is therefore representative of the IR-bright AGN population as a whole.   There is some evidence that the heavily obscured quasars in our sample have similar far-infrared luminosities to the Lacy11 sample, ie. similar star formation rates, but systematically higher AGN luminosity.  This is consistent with them being in a  systematically higher accretion rate phase relative to star formation. 

\subsection{ISM Masses }
The masses of the BH, the ISM and the stellar component are tabulated in Table \ref{tbl-mass}.    The ISM mass is directly proportional to the 870 $\mu$m luminosity, but depends much less strongly on dust temperature than does $L_{\rm BB}$ and SFR.   

We found in Section \ref{ism} that the ISM masses are comparable to those of the ``IR-bright'' high redshift sources in the COSMOS field \citep{scoville14}.          We can also compare our sample to the compilation of all known molecular masses for z$>$1 systems of \citet{carilli13}, their Figure 9.  Assuming a CO to H$_2$ conversion factor appropriate for starbursts \citep{downessolomon98,bolatto13} of ${\alpha}_{\rm CO} {\sim}$ 0.8 M$_{\odot}$ / (K km s$^{-1}$ pc$^2$), and assuming 50\% of the gaseous ISM is in molecular form \citep{carilli13}, our median detected ISM mass implies a median molecular gas mass of 2.95${\times} 10^{10} M_{\odot}$ and a median CO line luminosity of 3.7 ${\times} 10^{10}$ (K km s$^{-1}$ pc$^2$).  This lies at about the 80$^{th}$ percentile of the high redshift QSOs in Figure 9 of \citet{carilli13}, and about the 50$^{th}$ percentile of the SMGs.   It is lower than four of the RGs in this figure and comparable to the two others.

In summary, if we assume no contribution to the measured ALMA fluxes from non-thermal synchrotron emission, we find that the implied ISM masses are quite high, comparable to those of typical ``IR Bright'' star forming systems in the COSMOS field at these redshifts, to the most gas rich galaxies in the local CO survey of \citet{leroy09}, and consistent with the large CO masses of z$>$1 QSOs, SMGs and RGs.     This gas and dust could exist in nuclear, AGN-heated structures, or it could be powered by star formation somewhere within the host system.    

\subsection{BH Masses}
The BH mass estimates derived by \citet{kim13} for 3 of the 6 quasars with available [\ion{O}{3}]$\lambda5007\AA$ line luminosities are significantly larger that the values derived from our MIR data.     Our masses may underestimated due to extreme extinction in the MIR, (although we might expect to recover such dust-absorbed energy in the FIR-submm), and our assumption of $\sim$ 100\% covering factor may be incorrect.   Another possibility is that the [\ion{O}{3}] line strengths of \citet{kim13} are boosted by shocks and outflows.   We will address the relationship between $L_{\rm MIR}$ and $L_{\rm [OIII]}$ for these systems (cf. Mullaney et al. 2013) in a forthcoming paper presenting the spectroscopy.

\begin{figure*}[!ht]
	\centering
	\includegraphics[trim=30 385 0 130,clip,width=1.0\textwidth]{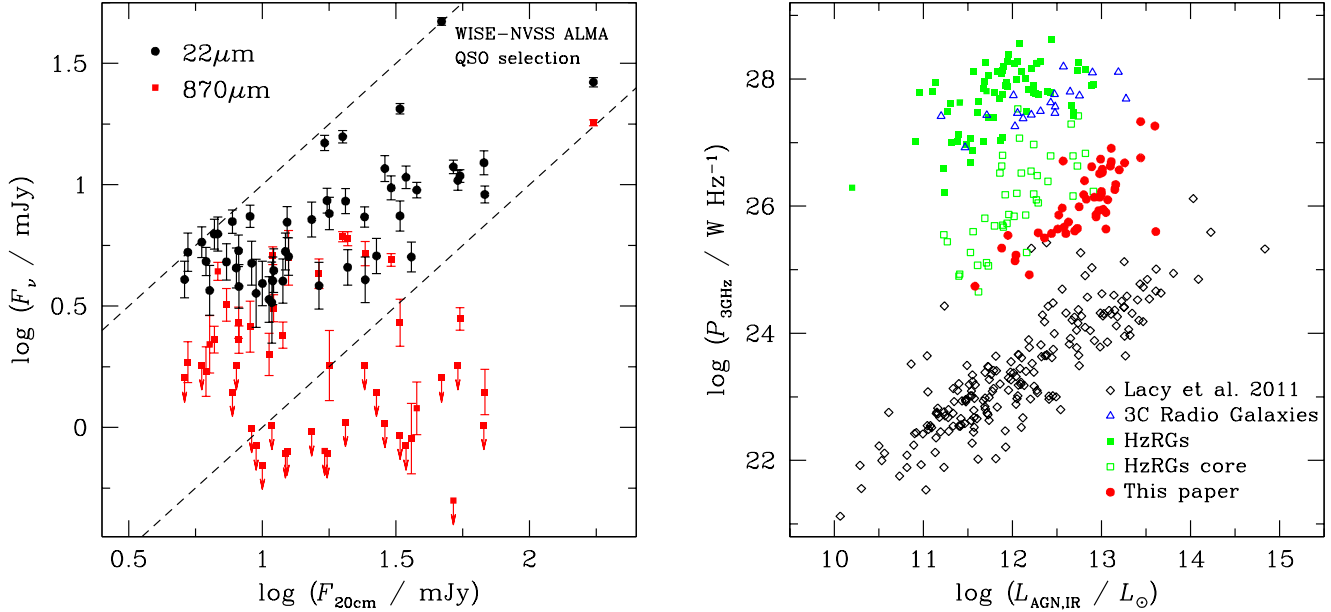}
	\caption{{\bf Left}:  Selection criteria for the ALMA quasar sample at 22 $\mu$m and 20 cm.   An apparent weak correlation between 22 $\mu$m and 20 cm flux density may be due to the selection boundaries (dashed lines).  Also shown is the 870 $\mu$m flux density {\it vs.} 20 cm flux density, showing no trend.     {\bf Right}: AGN luminosity {\it vs.} radio power for our WISE sample and for the HzRGs  of \citet{debreuck10} (both total emission and core emission) the 3C quasars from \citet{cleary07} and the MIR-selected AGN of Lacy11.  \label{fig-rgs_selecn}}
\end{figure*}

\subsection{Nature of the Radio Sources} 
We will address the morphology of the radio sources in our paper presenting the VLA results, where we will show that  the majority of the sample are compact on $\sim$1--3 kpc scales.  The rest frame 3 GHz radio powers lie in the range $\log$  $(P_{\rm 3.0 GHz}$  $/ W Hz^{-1}) =$ 24.74 -- 27.33.      The radio power of both RQ and RL systems evolves with redshift (Best et al. 2014) and some radio sources at $z=$1--3 with radio powers in this range are found to radio quiet based on the 24 $\mu$m $q$ value, $q_{\rm 24} = \log (f_{\rm 24{\mu}m}/f_{\rm 20cm})$ \citep{simpson12}.     Our sources have values of $q_{\rm 22}$ that are significantly too low for them to be considered radio-quiet by this criterion.    

In Figure \ref{fig-rgs_selecn}$\tiny-right$ we compare $L_{\rm AGN-Min}$ with $P_{\rm Radio-3GHz}$, and also include the high-redshift (1$<z<$5.2) radio galaxy (HzRG) sample of \citet{debreuck10}, and the 3C radio galaxy sample of \citet{cleary07} at 0.4$<z<$1.2.  For the HzRG sample we also include the estimated core flux, using the 20 GHz core fractions given by \citet{debreuck10} and assuming the same fraction is appropriate at 3 GHz.  There is a large disparity between our sample, the Lacy11 sample and the two high redshift samples.  The HzRGs and 3C RGs have 2--3 orders of magnitude more radio power than our sample, for a given mid-IR AGN luminosity, which is expected given that the power of the HzRGs is dominated by the extended lobes.   The core radio powers for the classical HzRGs have a similar range as our sample but our sources are significantly more luminous in the infrared, consistent with a higher accretion rate. The radio-blind MIR-selected sample of Lacy11 has on average 2--3 orders of magnitude less radio power than our sample, as expected for an RQ-dominated sample.

All four samples in this figure display an apparent correlation between $L_{\rm AGN}$ and $P_{\rm Radio-3GHz}$.   Given the rarity of these sources and the large redshift range involved, the correlation for our sample may be an artifact of Malmquist bias resulting from the 22 $\mu$m and the 1.4 GHz flux density thresholds.  We show in Figure \ref{fig-rgs_selecn}-$\tiny left$  that there exists an apparent correlation between 22 $\mu$m and 20 cm flux density that may be caused by the limited dynamic range in the flux ratio selection thresholds for the WISE-NVSS-ALMA sample:   $-1<log(f_{\rm 22{\mu}m}/f_{\rm 20cm})<0$ (dashed lines).   Therefore the apparent correlation seen between $L_{\rm AGN}$ and $P_{\rm Radio-3GHz}$ in Figure \ref{fig-rgs_selecn}-$\tiny right$ may not be real, at least for our sample.

\section{Conclusions}\label{section-conclusions}
We have selected a sample of extremely red, luminous, radio-powerful sources in the $0.5<z<3$ redshift range using WISE MIR colors and the NVSS \& FIRST 20 cm radio surveys.   We present ALMA 870 $\mu$m photometry for 49 southern sources from the total sample of 156 red sources, and redshifts for 45 of them from a combination of optical and near-infrared spectroscopy.    JCMT imaging at 850 $\mu$m has been presented for 30 of the northern sample by \citet{jones15}.   Combined, 30 sources have a detection at 850 or 870 $\mu$m, 25 of them with a known redshift.  We also have $R$-band imaging for 27 sources, $Herschel$ photometry for fifteen sources, including two of the ALMA sample, and CSO 350 $\mu$m data four additional sources, including two from the ALMA sample.

Having compared the SEDs of the red WISE selected sources with other samples and template SEDs, we conclude that the rest frame MIR-submm SEDs of our WISE-NVSS sources are dominated by AGN emission in the MIR. They have extremely red optical-MIR colors and high bolometric luminosities in the ULIRG and HyLIRG regime, In some cases  approaching or exceeding 10$^{14} L_{\odot}$.  They are redder in the NIR-MIR than previous samples selected from $Spitzer$ surveys, including almost all of the $Spitzer$ DOGs.  They also display systematically warmer overall MIR-submm SEDs, and probably have higher levels of accretion, relative to star formation, than the DOGs.   BH mass estimates for our sample are $7.7 <\log(M_{\rm BH}/M_{\odot})<10.2$.   We conclude that these sources are best labeled as obscured radio-powerful QSOs.       The rest-frame 3 GHz radio powers are $24.7<\log (P_{\rm 3.0 GHz}/W Hz^{-1})<27.3$, and all sources are radio-intermediate or radio-loud.    The ability of WISE to find this rare and unique sample is due to the large volume accessible to WISE.      

Our best constrained source has radiative transfer solutions with $\sim$ equal contributions for an obscured AGN and a young (10--15 Myr) compact starburst.   Simpler two-component fits to the whole sample find that the SFRs of the sample could be in the hundreds to thousands of solar masses per year range, but it also possible to fit a significant fraction of the ALMA data for some sources with a centrally-heated dusty structure, in which case most of the entire bolometric luminosity could be attributed to the obscured AGN.    In that scenario it is likely that the emission is dominated by a small high covering factor cocoon and/or an extended NLR.     

Our sample is similar in MIR selection method to \citet{eisenhardt12}, \citet{wu12} and \citet{bridge13}, who did not use radio flux density as a selection criterion.  Their Hot DOG samples exhibit similar SED shapes to our radio-selected sample. The MIR SEDs of the EWB12 samples may be steeper on average than those of our radio-selected sample, however this may be due to different selection effects between the MIR-optical color selection criteria for the two samples, resulting in an average higher redshift for the radio-blind sample.     
  
\acknowledgments
This paper makes use of the following ALMA data: ADS/JAO.ALMA\#2011.0.00397.S. ALMA is a partnership of ESO (representing its member states), NSF (USA) and NINS (Japan), together with NRC (Canada), NSC and ASIAA (Taiwan), and KASI (Republic of Korea), in cooperation with the Republic of Chile. The Joint ALMA Observatory is operated by ESO, AUI/NRAO and NAOJ.  The National Radio Astronomy Observatory is a facility of the National Science Foundation operated under cooperative agreement by Associated Universities, Inc.  This publication makes use of data products from the Wide-field Infrared Survey Explorer, which is a joint project of the University of California, Los Angeles, and the Jet Propulsion Laboratory/California Institute of Technology, funded by the National Aeronautics and Space Administration. This work is based on observations made with the Caltech Submillimeter Observatory, which is operated by the California Institute of Technology under funding from the National Science Foundation, contract AST 90-15755.   This paper uses data from SDSS (DR 8). Funding for SDSS-III has been provided by the Alfred P. Sloan Foundation, the Participating Institutions, the National Science Foundation, and the U.S. Department of Energy Office of Science. The SDSS-III web site is http://www.sdss3.org/.  RJA was supported by Gemini-CONICYT grant number 32120009.  We thank the anonymous referee for comments which helped improve the paper.

{\it Facilities:} \facility{WISE}, \facility{VLA}, \facility{ALMA}, \facility{CTIO (SOAR--Goodman)}, \facility{Palomar 200 inch}, \facility{VLT}, \facility{Herschel}, \facility {CSO}, \facility{Magellan}
{}

\addtocounter{figure}{-3}
\begin{figure}[!ht]
	\centering
	\includegraphics[trim=35 130 40 100,clip,width=0.6\textwidth]{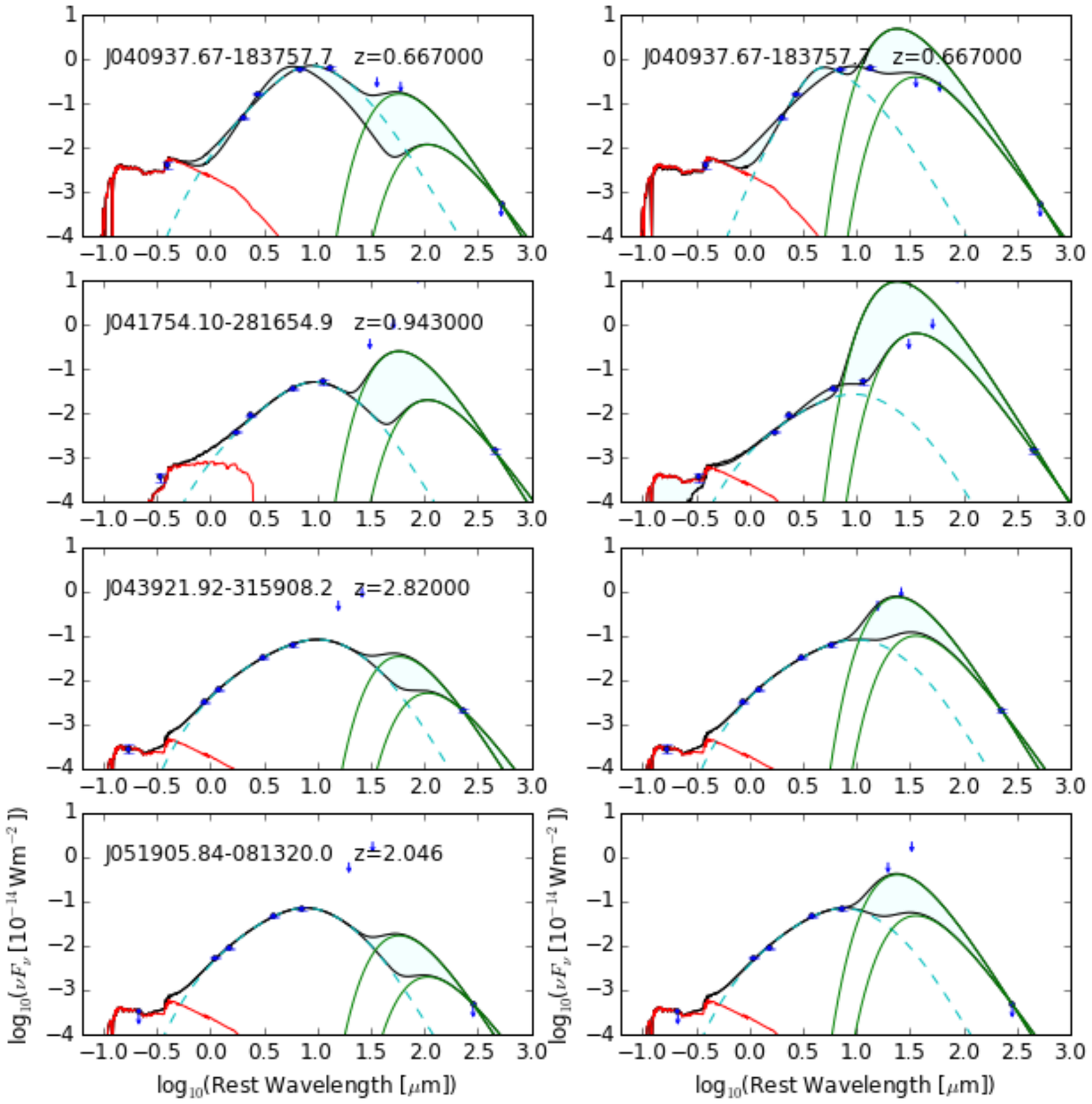}
	\includegraphics[trim=35 130 40 100,clip,width=0.6\textwidth]{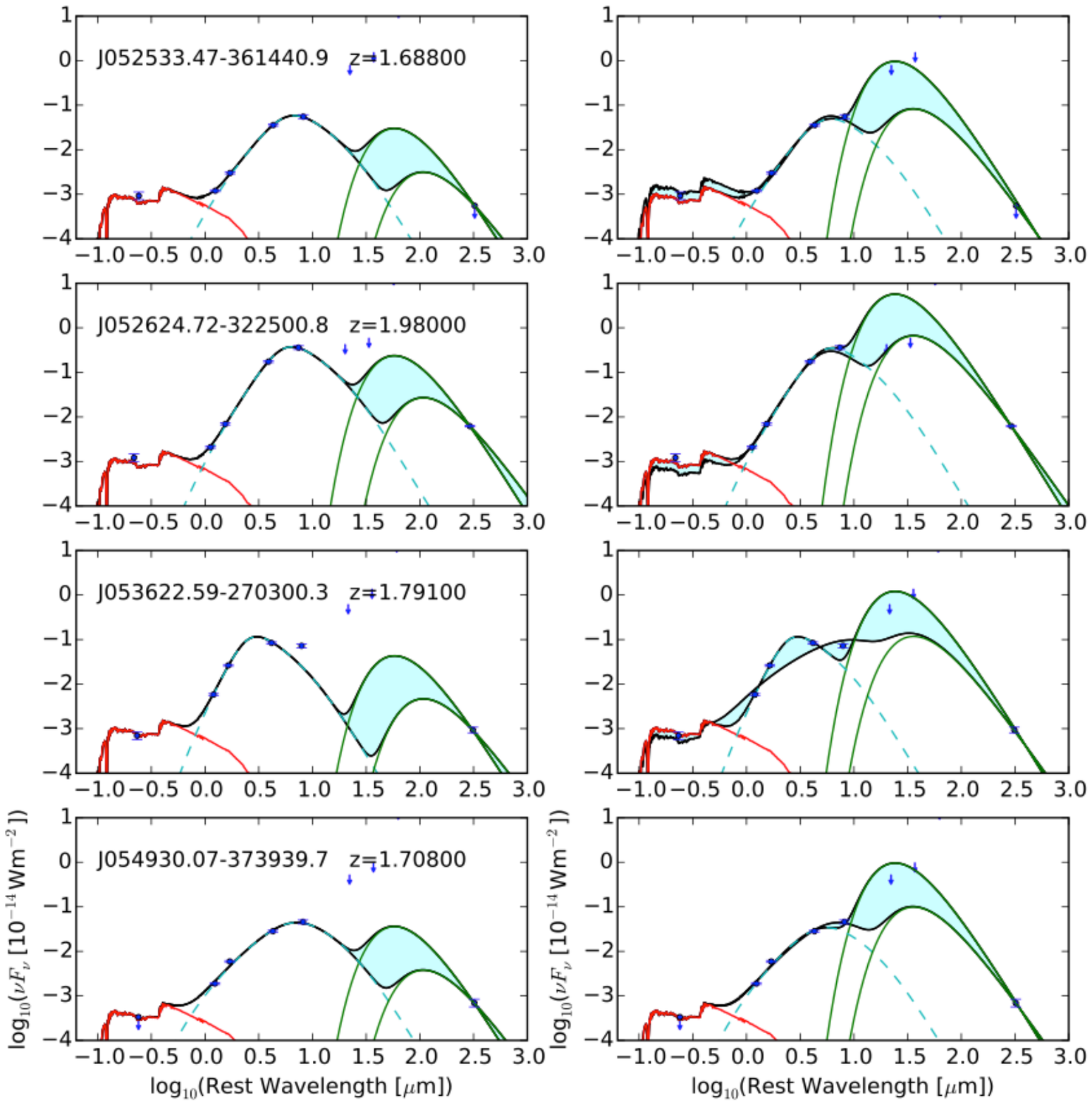}	
	\caption{continued.   SED plots.}
\end{figure}
\addtocounter{figure}{-1}
\begin{figure}[!ht]
	\centering
	\includegraphics[trim=35 130 40 100,clip,width=0.6\textwidth]{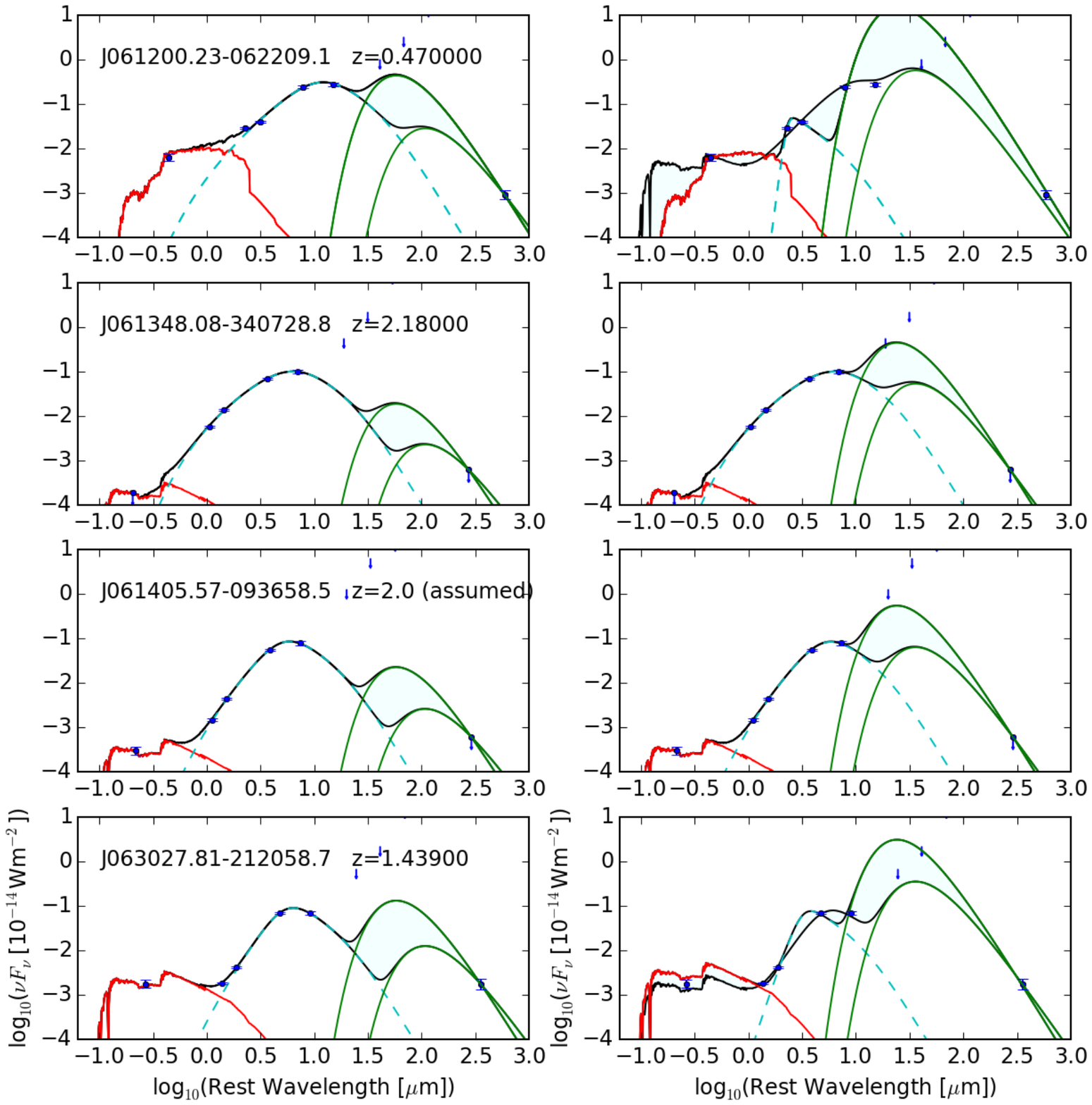}	
	\includegraphics[trim=35 130 50 100,clip,width=0.6\textwidth]{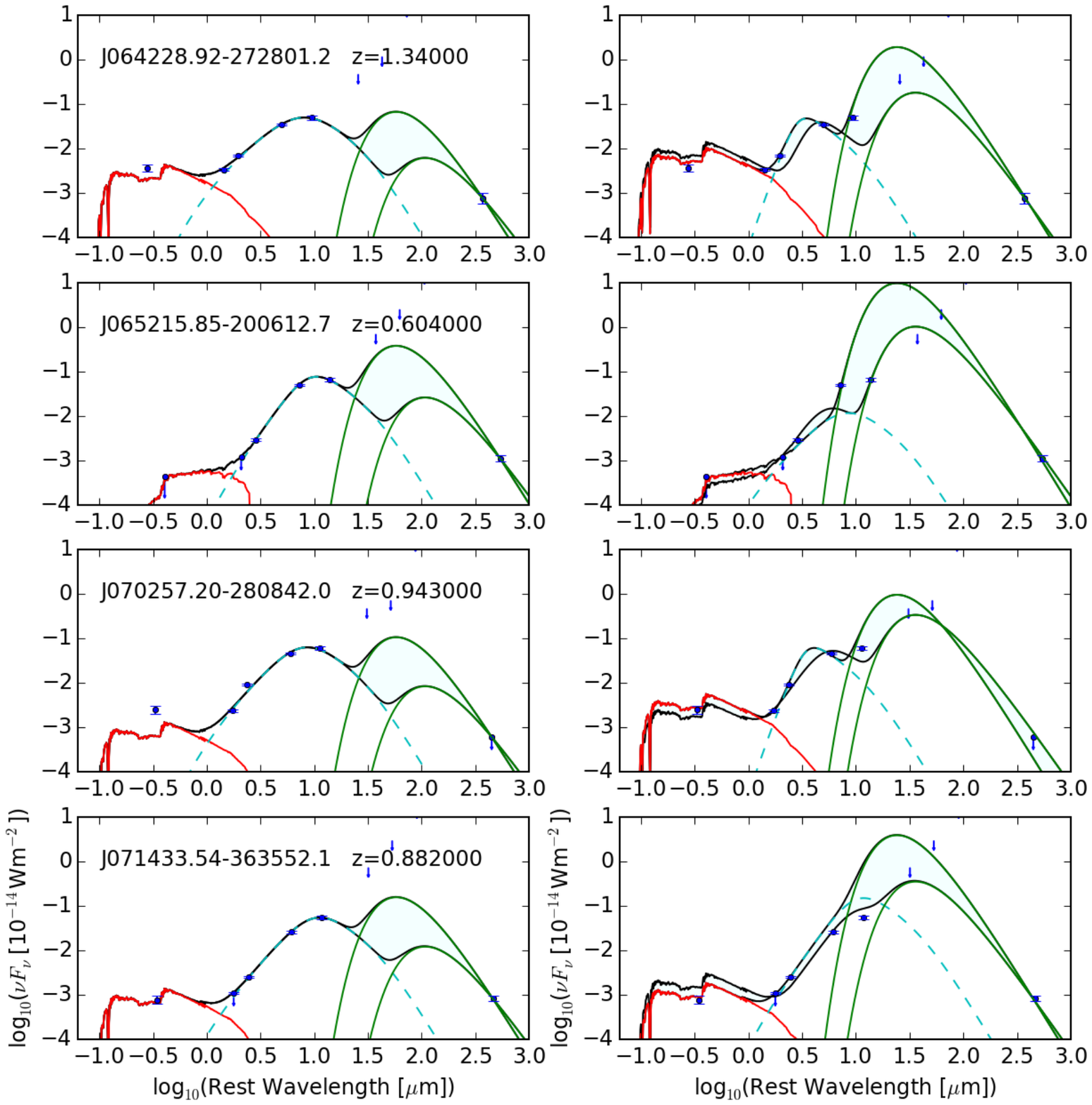}
	\caption{continued.   SED plots.}
\end{figure}
\addtocounter{figure}{-1}
\begin{figure}[!ht]
	\centering
	\includegraphics[trim=35 130 50 100,clip,width=0.6\textwidth]{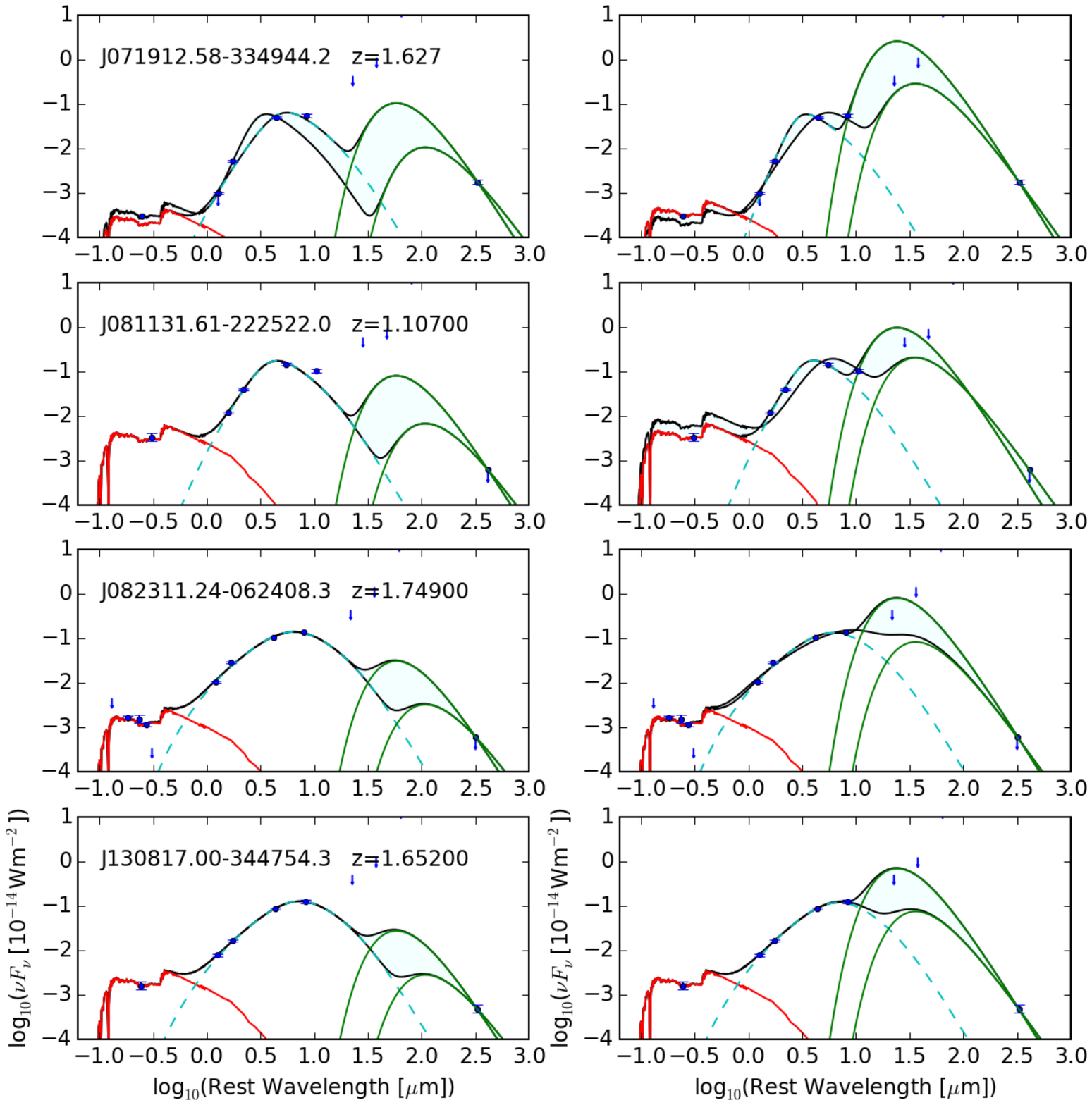}
	\includegraphics[trim=35 130 50 100,clip,width=0.6\textwidth]{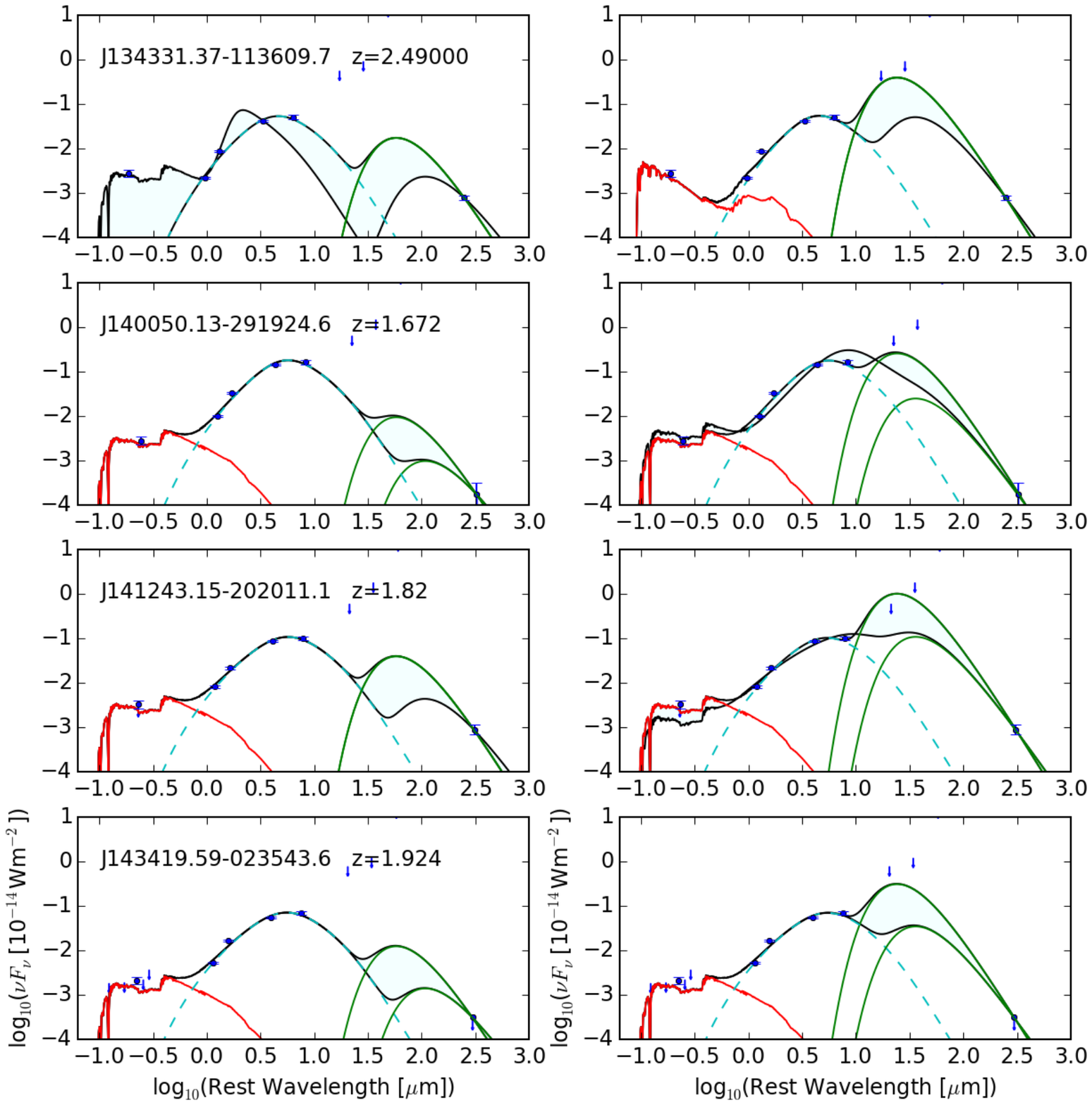}
	\caption{continued.   SED plots.}
\end{figure}
\addtocounter{figure}{-1}
\begin{figure}[!ht]
	\centering
	\includegraphics[trim=35 130 50 100,clip,width=0.6\textwidth]{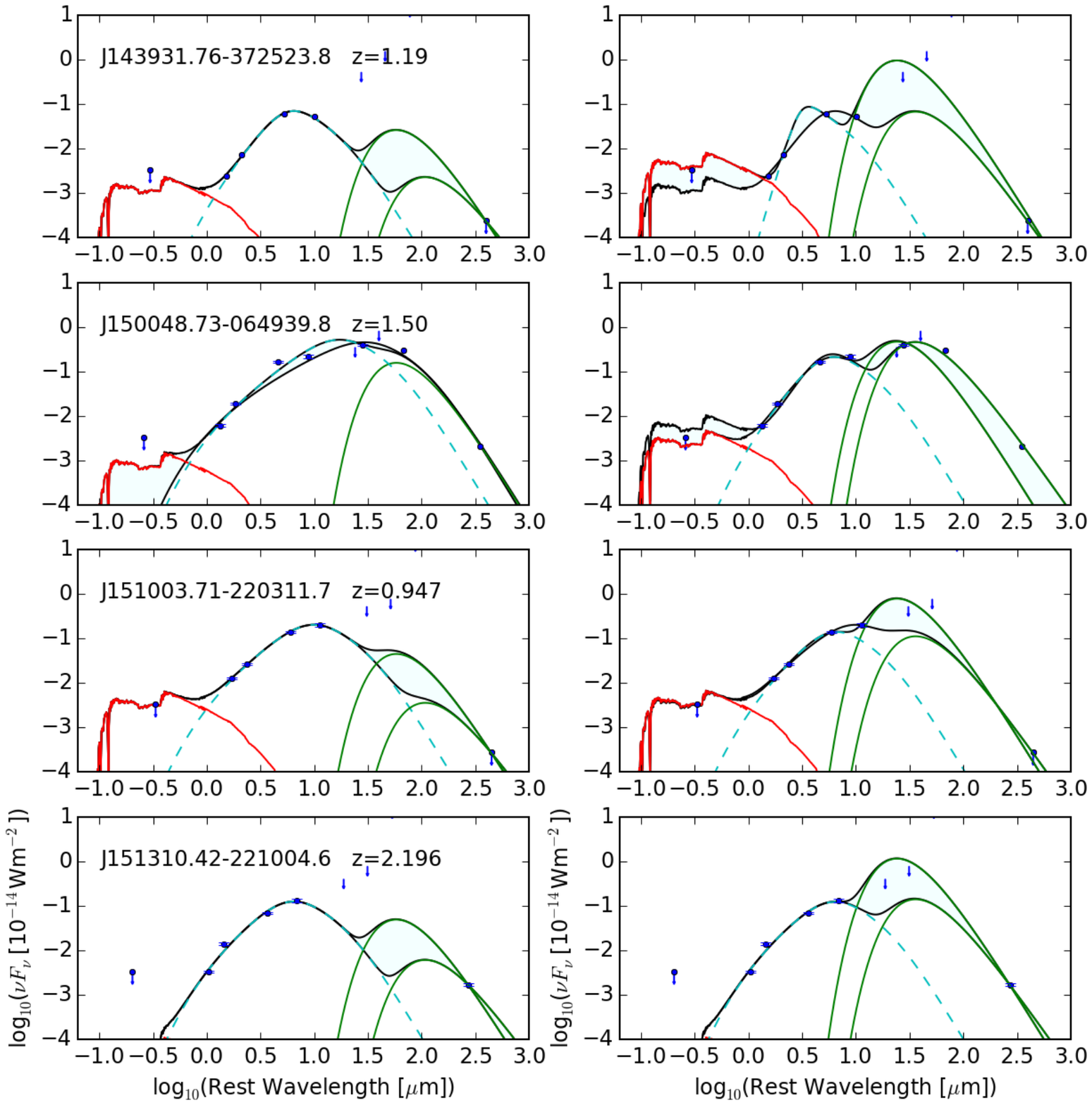}
	\includegraphics[trim=35 130 50 100,clip,width=0.6\textwidth]{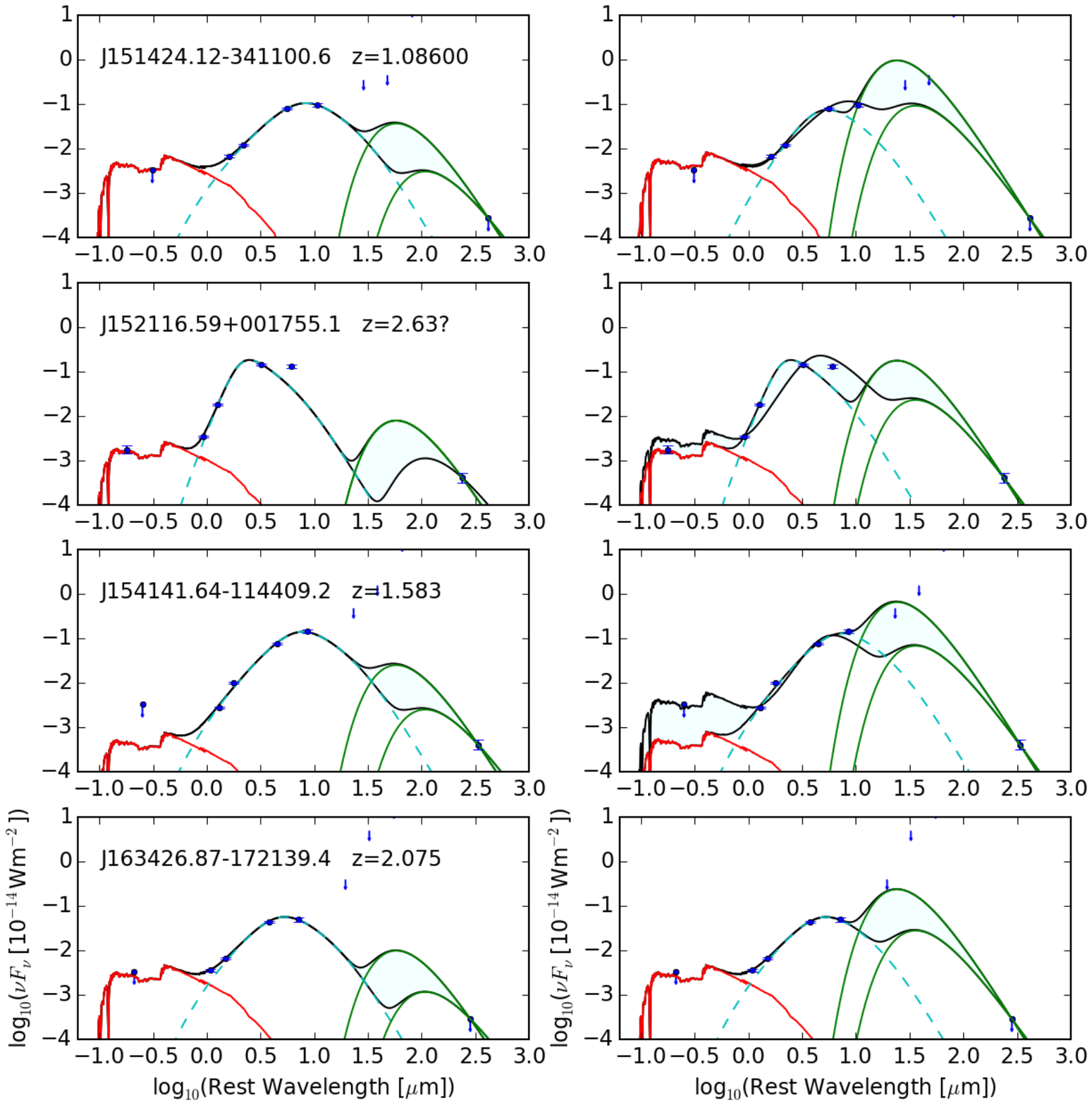}
	\caption{continued.   SED plots.}
\end{figure}
\addtocounter{figure}{-1}
\begin{figure}[!ht]
	\centering
	\includegraphics[trim=35 130 50 100,clip,width=0.6\textwidth]{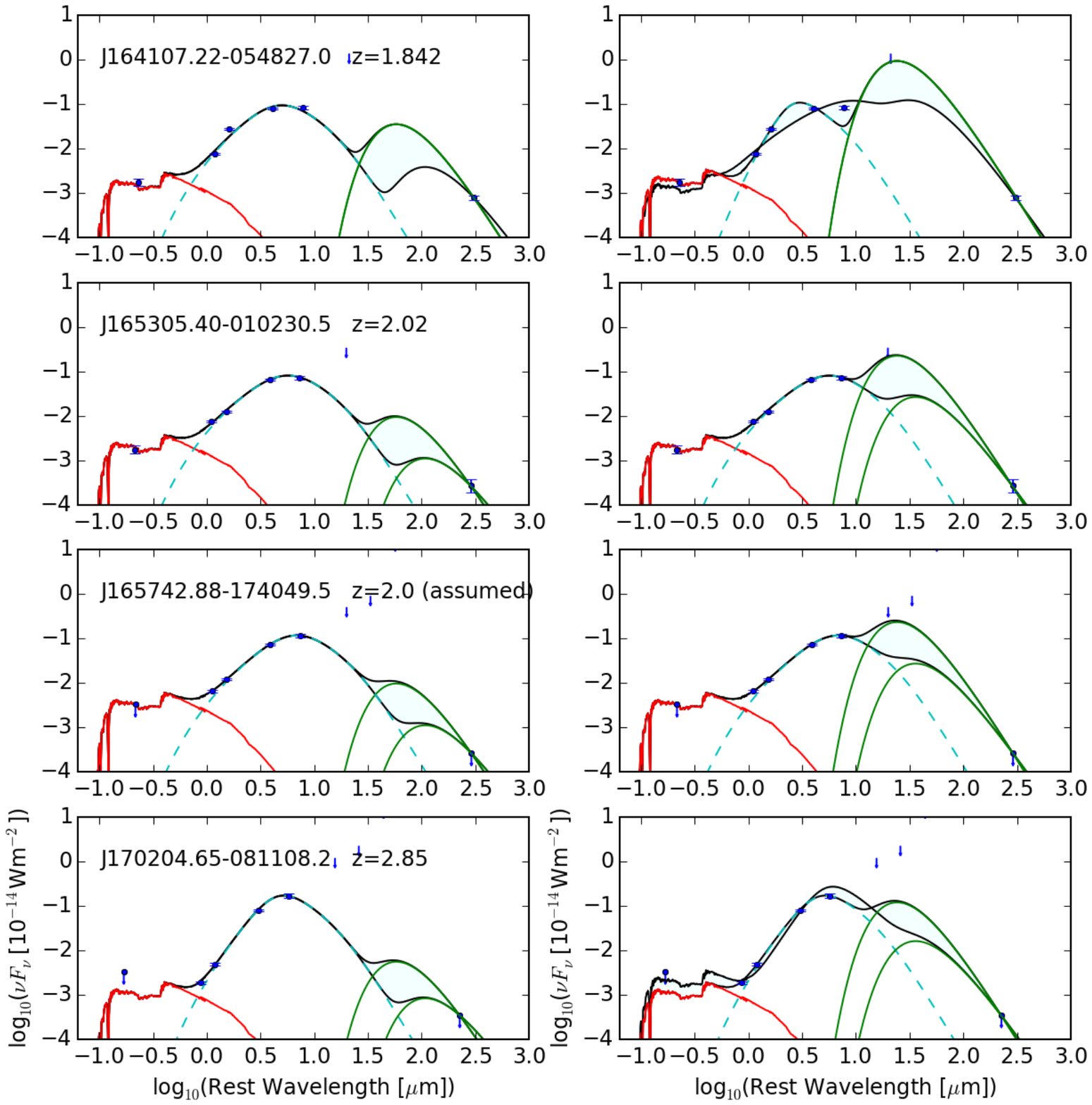}
	\includegraphics[trim=35 130 50 100,clip,width=0.6\textwidth]{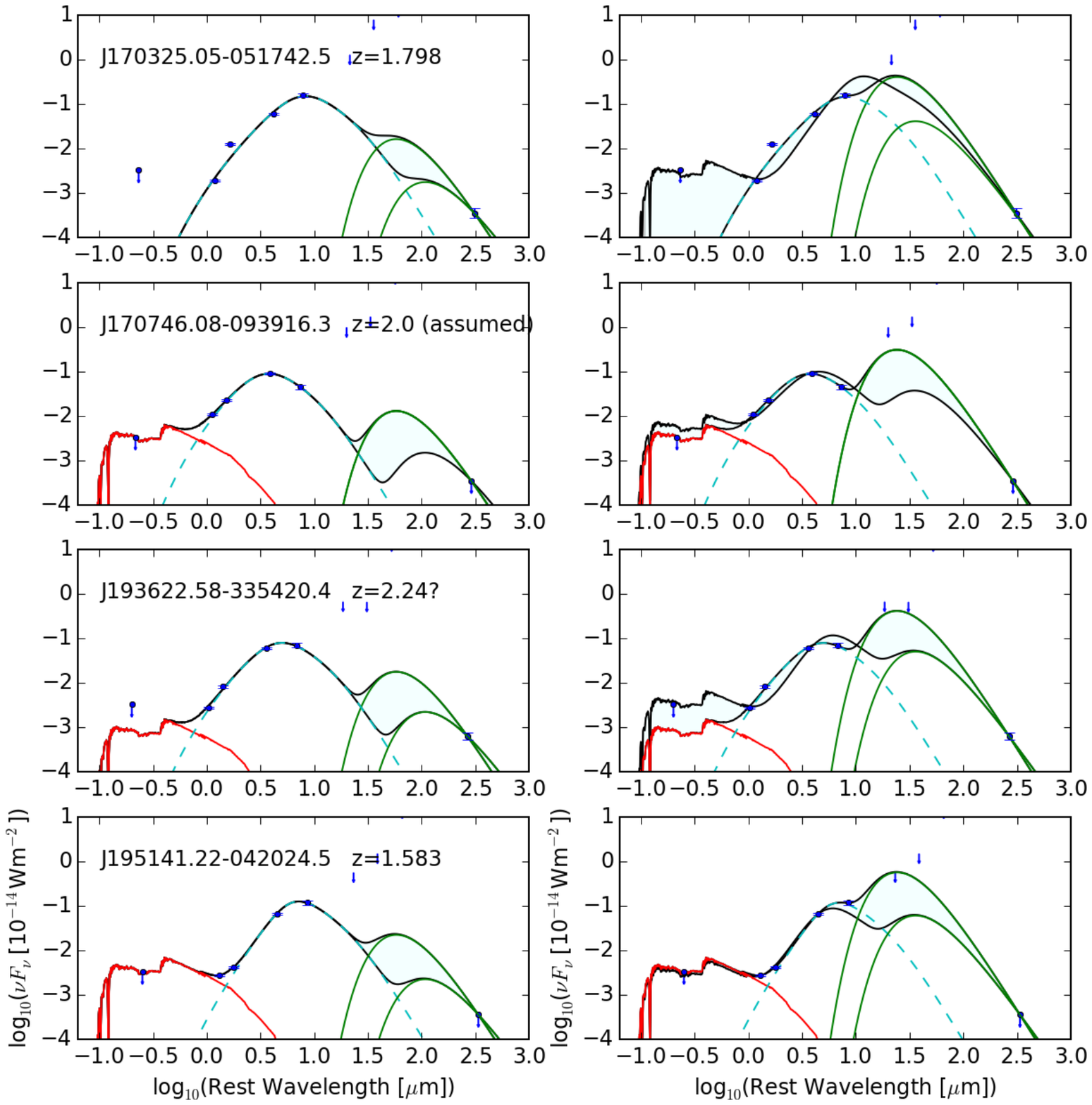}
	\caption{continued.   SED plots.}
\end{figure}
\addtocounter{figure}{-1}
\begin{figure}[!ht]
	\centering
	\includegraphics[trim=35 130 50 100,clip,width=0.6\textwidth]{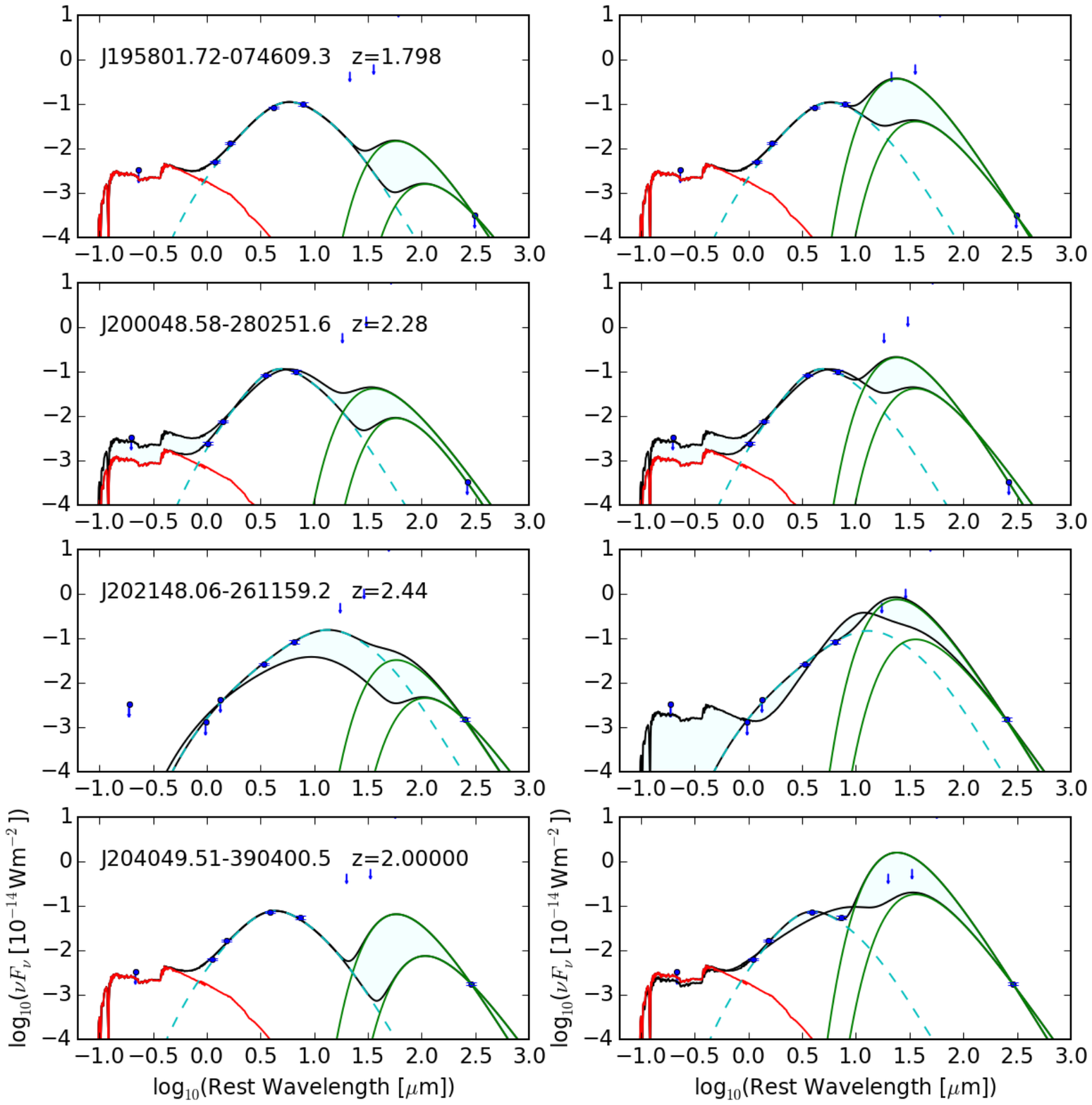}
	\includegraphics[trim=35 300 50 300,clip,width=0.6\textwidth]{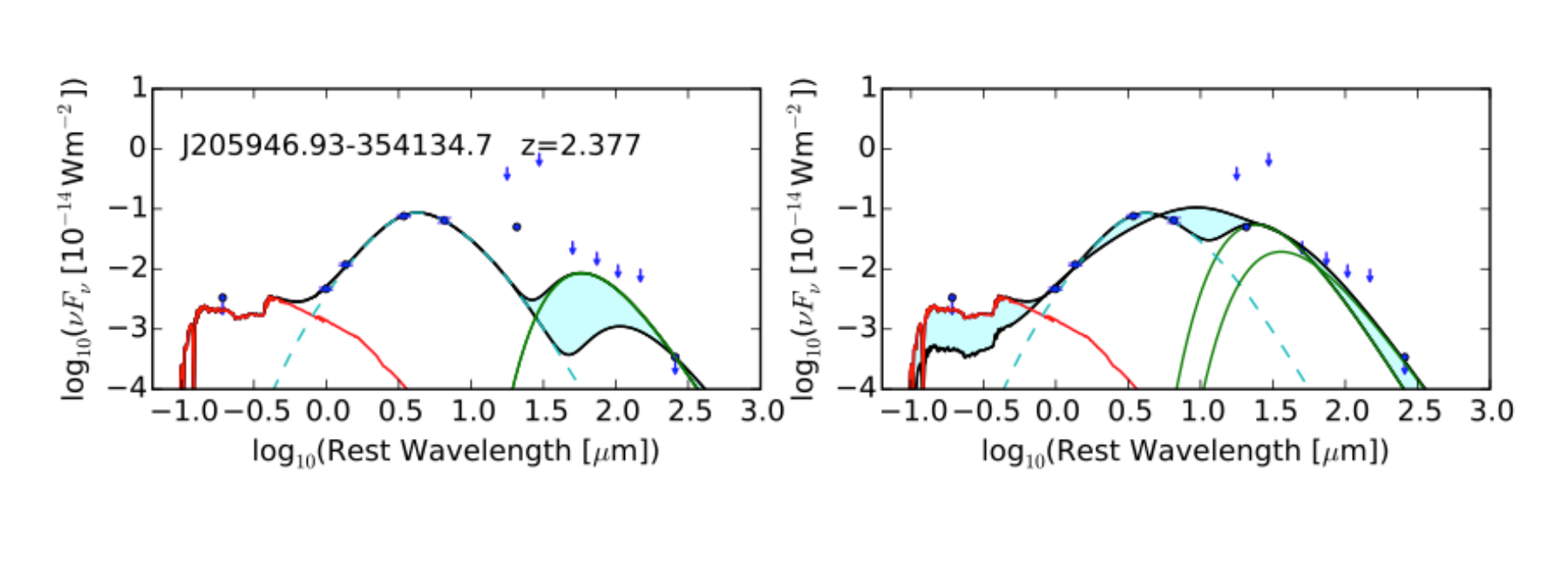}
	\caption{continued.   SED plots.}
\end{figure}

\end{document}